\documentclass{tADP2e}
\usepackage{graphicx}   % need for figures
\usepackage{color}      % use if color is used in text
\usepackage{hyperref}   % use for hypertext links, including those to external documents and URLs
\usepackage{dcolumn}
\usepackage{tocstyle}

\def\lsno{La$_{2-x}$Sr$_x$NiO$_4$}
\def\lco{La$_2$CuO$_4$}
\def\lsco{La$_{2-x}$Sr$_x$CuO$_4$}
\def\lbco{La$_{2-x}$Ba$_x$CuO$_4$}

\def\lnsco{La$_{1.6-x}$Nd$_{0.4}$Sr$_x$CuO$_4$}
\def\lesco{La$_{1.8-x}$Eu$_{0.2}$Sr$_x$CuO$_4$}
\def\ybco{YBa$_2$Cu$_3$O$_{6+x}$}
\def\lcod{La$_2$CuO$_{4+\delta}$}
\def\bscco{Bi$_2$Sr$_2$CaCu$_2$O$_{8+\delta}$}
\def\hbco{HgBa$_2$CuO$_{4+\delta}$}

\def\newr{\color{black}}

\begin{document}

\title{Cuprate superconductors as viewed through a striped lens}

\author{
\name{J. M. Tranquada\thanks{CONTACT J.~M. Tranquada. Email: jtran@bnl.gov}}
\affil{Condensed Matter Physics and Materials Science Division, Brookhaven National Laboratory, Upton, New York 11973, USA}
}

\maketitle

%\date{\today} 

\begin{abstract}
Understanding the electron pairing in hole-doped cuprate superconductors has been a challenge, in particular because the ``normal'' state from which it evolves is unprecedented.  Now, after three and a half decades of research, involving a wide range of experimental characterizations, it is possible to delineate a clear and consistent cuprate story.  It starts with doping holes into a charge-transfer insulator, resulting in in-gap states.  These states exhibit a pseudogap resulting from the competition between antiferromagnetic superexchange $J$ between nearest-neighbor Cu atoms (a real-space interaction) and the kinetic energy of the doped holes, which, in the absence of interactions, would lead to extended Bloch-wave states whose occupancy is characterized in reciprocal space.  To develop some degree of coherence on cooling, the spin and charge correlations must self-organize in a cooperative fashion.  A specific example of resulting emergent order is that of spin and charge stripes, as observed in \lbco.  While stripe order frustrates bulk superconductivity,  it nevertheless develops pairing and superconducting order of an unusual character.  The antiphase order of the spin stripes decouples them from the charge stripes, which can be viewed as hole-doped, two-leg, spin-$\frac12$ ladders.  Established theory tells us that the pairing scale is comparable to the singlet-triplet excitation energy, $\sim J/2$, on the ladders.  To achieve superconducting order, the pair correlations in neighboring ladders must develop phase order.  In the presence of spin stripe order, antiphase Josephson coupling can lead to pair-density-wave superconductivity.  Alternatively, in-phase superconductivity requires that the spin stripes have an energy gap, which empirically limits the coherent superconducting gap.  Hence, superconducting order in the cuprates involves a compromise between the pairing scale, which is maximized at $x\sim\frac18$, and phase coherence, which is optimized at $x\sim0.2$.  To understand further experimental details, it is necessary to take account of the local variation in hole density resulting from dopant disorder and poor screening of long-range Coulomb interactions.  At large hole doping, kinetic energy wins out over $J$, the regions of intertwined spin and charge correlations become sparse, and the superconductivity disappears.  While there are a few experimental mysteries that remain to be resolved, I believe that this story captures the essence of the cuprates.
\end{abstract}

\tableofcontents

\section{Introduction}

The discovery of high-temperature superconductivity in \lbco\ by Bednorz and M\"uller in 1986 \cite{bedn86} came as a tremendous and stupefying shock.  Transition-metal oxides were out of fashion at the time, and they were certainly not viewed by most researchers as a good place to search for new superconductors.  In terms of theoretical concepts that would become relevant, Mott's conclusion that NiO is an insulator because of strong onsite Coulomb repulsion experienced by Ni $3d$ electrons \cite{mott49} was still hotly debated \cite{tera84,sawa84}.  Thus, it should not be surprising that it has taken us a long time to make sense of the layered cuprates.

The success of solid-state physics in the mid-20$^{\rm th}$ century was largely due to the band theory of solids.  With the assumption that atomic states combine to form Bloch states \cite{bloc29}, coherent in every unit cell and characterized by wave vector {\bf k},  one obtains a set of levels that can be filled, two at a time, with fermionic electrons.  In a metal, the top of the distribution defines a Fermi surface.  With Landau, the picture is modified to allow interactions that modify the Bloch states, but one nevertheless ends up with well-defined quasiparticles close to the Fermi energy.  The resulting Fermi liquid is the starting point for the Bardeen-Cooper-Schrieffer theory of superconductivity \cite{bard57}, which, in turn, is the inevitable starting point for the consideration of any new superconductor.

Of course, not all solids are metals.  If the highest-energy occupied band of states is completely filled and there is an energy gap to the next (empty) band, one has a band insulator.  In contrast, there are also cases of insulators that cannot be explained by having an even number of electrons to fill a band.  As emphasized by Mott \cite{mott49}, there are cases where local Coulomb repulsion overwhelms kinetic energy.  This is most easily pictured in a tight-binding model based on atomic orbitals; if the highest-energy, partially-filled orbital has a strong Coulomb repulsion $U$ between two electrons on the same site, it is possible to end up with an insulator where band theory would predict a metallic conductor.

Hubbard \cite{hubb64} provided a convenient parameterization of this picture involving competing kinetic energy, represented by intersite hopping energy $t$, and onsite repulsion $U$.   The Hubbard model is especially useful for treating insulating transition-metal oxide compounds, many of which exhibit antiferromagnetic order.  Anderson \cite{ande59} showed that the superexchange interaction between nearest-neighbor magnetic moments can be understood by applying second-order perturbation theory to a Hubbard-like model.  For the case of a single half-filled orbital, the kinetic energy is lowered when one electron hops to a neighboring site and back home again; however, because of the Pauli exclusion principle, this can only happen if the neighboring spins are antiparallel.  The net magnetic interaction, termed superexchange, is given by $J = 4t^2/U$.  Hence, antiferromagnetism can be a symptom of strong onsite Coulomb repulsion.

When Bednorz and M\"uller discovered high-temperature superconductivity, the specific compound that was superconducting came as a surprise even to them.  They were attempting to explore the perovskite LaCuO$_3$ doped with Ba; however, follow-on work determined that the superconducting compound is \lbco\ (LBCO), a layered perovskite with the K$_2$NiF$_4$ structure \cite{taka87,bedn87}.  It took further investigations to establish that the parent compound, La$_2$CuO$_4$, is an antiferromagnetic insulator \cite{vakn87}.\footnote{La$_2$CuO$_4$ can easily pick up a small amount of excess oxygen during synthesis, and phase separation can result in an impurity phase that is metallic and superconducting \cite{well96}.}

Anderson quickly recognized that La$_2$CuO$_4$ should be a Mott-Hubbard insulator with antiferromagnetic correlations due to superexchange \cite{ande87}.  This set up the initial theoretical quandary.  The doped compounds are superconductors, and the extremely successful BCS theory of superconductors is based on Fermi liquid theory.  At the same time, the parent compounds are antiferromagnetic insulators due to strong electron-electron interactions, as eventually confirmed by experiment (see Sec.~\ref{sc:parent}).  Something would have to give.

There is a great deal of mathematical machinery that has been developed within the context of the Fermi liquid model. All types of ordering (magnetism, charge density waves, etc.) are typically described in terms of interactions between quasiparticles near the Fermi level.  From this perspective, antiferromagnetism (or spin-density-wave order) should be described by an interaction that causes a scattering between quasiparticles at the Fermi surface that are separated by the AF wave vector.  In the absence of actual order, such scattering should result in ``hot'' spots on the Fermi surface.  It took quite some time before this proposal could be tested by experiments using angle-resolved photoemission spectroscopy (ARPES).  Now we have many results {\newr on hole-doped cuprates} \cite{dama03}, and {\newr\it none of them show evidence of hot spots}.  To be clear, there are strong anomalies, but they are not restricted to unique points on the Fermi surface.

As experiments on cuprates have shown, doping a sufficient density of holes $p$ (normalized per Cu) into the parent antiferromagnetic insulator leads to unusual metallic and superconducting states, with a density of states near the Fermi energy that is strongly suppressed compared to predictions based on conventional band calculations (see Sec.~\ref{sc:hole}).  Within the Fermi-liquid approach, the depressed density of states is typically analyzed in terms of some sort of competing order.  With increasing hole concentration, the effects attributed to the competing order decrease, eventually disappearing at an assumed quantum critical point (QCP) below the superconducting dome, at the critical hole density $p^*$.  This perspective draws parallels to a number of experimental systems where superconductivity is observed around a QCP associated with ferromagnetism \cite{mont07,vonl07,aoki19}, charge-density-wave order \cite{moro06,zhu16,li17}, or even antiferromagnetism due to RKKY coupling of $f$-electron moments \cite{mont07,vonl07}.  It is the soft quantum fluctuations of the dying order parameter that drive the electron pairing in the vicinity of a QCP \cite{aban20}.  For cuprates, there have been theoretical proposals for a QCP associated with antiferromagnetic \cite{sach03b}, charge-density-wave \cite{cast96}, current-loop \cite{varm06}, $d$-density-wave \cite{chak01} and nematic \cite{maie14,lede15} orders.

A problem with the QCP scenario is that it tends to imply a large density of coherent carriers at high temperature, with a transition at a doping-dependent temperature $T^*$ to a ``pseudogapped'' state.  There is no experimental support for such a high-temperature phase; {\newr\it the depressed density of states occurs over a large energy scale and involves largely incoherent states}, as will be discussed.  Furthermore, {\newr \it there is no evidence for a phase-transition-like onset of a pseudogap on reducing temperature}.

As emphasized by Anderson \cite{ande97a}, the superexchange interaction $J$ underlying the antiferromagnetism in the parent cuprates is a very short-range effect driven by $U$; it has nothing to do with Bloch states.  He also proposed that the two large interactions $J$ and $U$ should dominate the electronic and magnetic properties of cuprates \cite{ande06a}.  While I agree with this idea, there is still the trick of how to explain the details.  Anderson proposed that the two-dimensionality and minimal spin $S=1/2$ of the Cu moments would enable quantum fluctuations to overwhelm the usual ordering tendencies, resulting in a resonating valence bond (RVB) state \cite{ande87}.  The pairing of Cu moments in singlets would somehow lead to pairing of mobile carriers when holes were introduced.  While this concept has motivated much work by many theorists \cite{ande04}, some features of the mean-field analysis disagree with experiment.  For example, while the idea that singlet-triplet excitations provide the pairing scale  \cite{lee92} is a good one, {\newr\it this energy does not continuously rise to its maximum value at zero hole doping.}

\begin{figure}
 \centering
    \includegraphics[width=10cm]{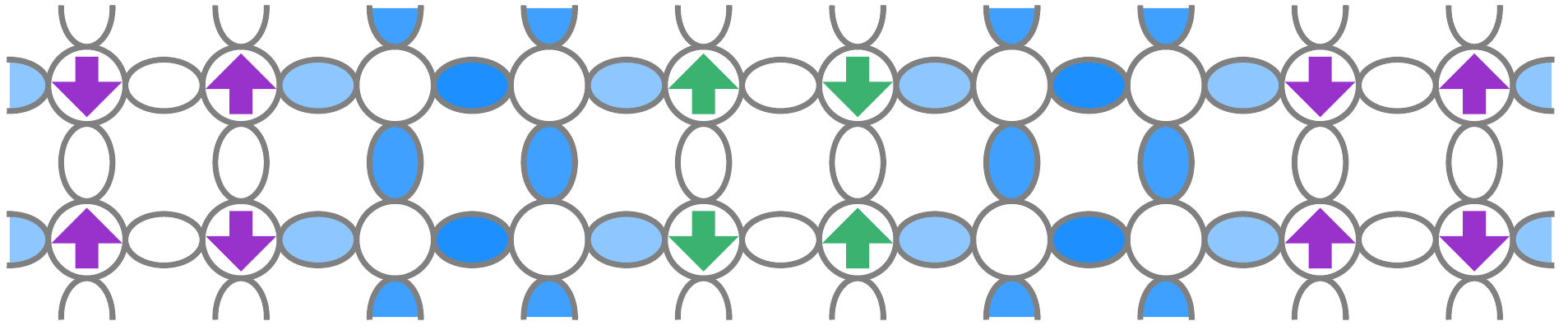}
    \caption{\label{fg:stripe0} Cartoon of bond-centered charge and spin stripes as alternating doped and undoped 2-leg ladders.  Arrows indicate ordered magnetic moments on Cu atoms (circles), with color changing between antiphase domains.  Blue intensity indicates hole density (with white $=0$) on O sites (ellipses).   The empty circles indicate the lack of magnetic order on the charge stripes; the magnetic moments on the charge stripes are dynamic.
    }
\end{figure}

To make the connection between superexchange and pairing, we need to take account of another insight.  Emery and Kivelson (and coworkers) \cite{emer90} pointed out a tendency towards phase separation (and real-space segregation of holes and spins) when holes are introduced into the antiferromagetic insulator, as described in a simplified variant ($t$-$J$ model) of the Hubbard model; such large-scale segregation is possible because of the neglect of long-range Coulomb interactions.  Inclusion of those interactions in an effective model provided evidence for periodic patterns of hole-rich and hole-poor (antiferromagnetic) patches \cite{emer93,low94,chay96}, including charge and spin stripes (see Fig.~\ref{fg:stripe0}).\footnote{Earlier Hartree-Fock and related calculations had provided solutions of charge and spin stripes \cite{zaan89,schu89,poil89,mach89}; however, the charge stripes were always insulating.}  With the experimental discovery of stripe order in one specific cuprate family \cite{tran95a}, a model of superconductivity based on stripes was proposed \cite{emer97}.  An assumption of this model was that the magnetic spin gap needed to induce pairing correlations within the charge stripes would come from the hole-poor spin stripes \cite{arri04}; that assumption appears to be incompatible with static spin stripes (which exhibit no significant spin gap, as I will discuss), so that stripe order would then appear to compete with superconductivity \cite{tran97a}.\footnote{Another challenge in the superconducting stripe model of \cite{emer97} is that charge-density-wave order within the stripes would compete with pairing.  To avoid this possibility, the idea of fluctuating stripes in the form of nematic and smectic orders was introduced \cite{kive98}.  Again, this has a bias of static stripe order being bad for hole pairing.} 

The situation changed a decade later, when experiments exploring the anisotropic response of LBCO single crystals provided evidence that {\newr\it stripe order is compatible with pairing and two-dimensional (2D) superconductivity} \cite{li07,li19a}.   This discovery has been rationalized in terms of a proposed pair-density-wave superconducting state \cite{berg07,hime02,agte20}.  It shifted the narrative from one of competing orders to a picture of intertwined orders \cite{frad15}.

The concept of intertwined orders, while providing a framework for interpreting experiments on cuprates, does not provide a specific explanation for the pairing within charge stripes.  This is where I propose a new synthesis that exploits pre-existing ideas (see Sec.~\ref{sc:pairing}).  {\newr\it The pairing is a consequence of singlet correlations of Cu moments, but these singlets are on hole-doped stripes}, not the spin stripes.  Here I invoke the theoretical analysis \cite{dago92,dago96} and experimental evidence \cite{ueha96} that hole-doped $S=1/2$ two-leg ladders superconduct.  The limit on the energy scale for pairing is the singlet-triplet excitation gap of the ladder.  {\newr\it The spin stripes have an antiphase order that decouples them from the excitations of the charge stripes} (for excitations below the pairing scale).  Superconducting order results from Josephson coupling between the charge stripes \cite{emer97}.

The antiphase character of the spin stripes makes them incompatible with spatially-uniform superconductivity.  It immediately follows that {\newr\it the only way to develop 2D superconductivity in the state with spin-stripe order is to have the pair wave function avoid the spin stripes by forming a sinusoidally-modulated pair-density wave} \cite{berg07}.  {\newr\it In order to have spatially-uniform in-phase superconducting order, it is necessary to avoid spin-stripe order and to gap the incommensurate spin-stripe excitations}.\footnote{This is supported by a recent phenomenological analysis \cite{dahm18}.}  Indeed, experiments indicate that the coherent superconducting gap is limited by the incommensurate spin gap \cite{li18}{\newr, as discussed in Sec.~\ref{sc:spatial}}.  

To appreciate why there is a coherent superconducting gap that is smaller than the maximum of the {\newr commonly-accepted {\bf k}-dependent}
%$d$-wave 
gap function, we have to consider the role of a second form of inhomogeneity, which is quenched dopant disorder that results in local variations in the average charge density on a scale coarser than that of the stripes.  This leads to behavior similar to that of a granular superconductor \cite{imry12}, such that the development of superconducting order tends to be limited by phase fluctuations \cite{emer95a,li21c}.

The normal state from which the stripes and pairing develop involves both antiferromagnetic correlations based on superexchange and a low density of doped holes that want to minimize their kinetic energy by delocalizing.  {\newr\it The superexchange interaction is only nearest-neighbor, so it essentially lives in real space, whereas electronic quasiparticles live in reciprocal space}.  The doped holes strongly damp the antiferromagnetic spin correlations, resulting in a short spin-spin correlation length.  Where antiparallel spin correlations overlap with Bloch states, they cause a large {\bf k}-dependent electronic self energy, corresponding to a pseudogap.  The effects of superexchange dominate at low doping, but with increasing hole concentration, the energy-scale of the pseudogap decreases.  The holes and spins initially find a compromise in the form of stripes, but at higher hole concentrations, the pseudogap eventually disappears, indicating that regions with strong local AF correlations no longer percolate across the CuO$_2$ planes.  Since the Cu singlet correlations are crucial to pairing, {\newr\it the strong decrease in spin-correlation weight in highly-overdoped samples leads to a rapid decrease in superfluid density}.

\begin{table}[t]
\tbl{List of frequently mentioned cuprate (and nickelate) families and associated acronyms.}
{\begin{tabular}[l]{@{}ll}\toprule
  Formula & Acronym \\
  \colrule
  \lbco & LBCO \\
  \lsco & LSCO \\
  \ybco & YBCO \\
  Bi$_2$Sr$_{2-x}$La$_x$CuO$_{6+\delta}$ & Bi2201 \\
  \bscco & Bi2212 \\
  \hbco & Hg1201 \\
  \null \\
  \lsno & LSNO \\
\botrule
\end{tabular}}
\label{tb}
\end{table}

I have made some bold claims in this introduction {\newr (in {\it italics} above)}.  The rest of this perspective is aimed at backing these up, largely based on experimental evidence, relying on relevant theoretical results when available.  We start in Sec.~\ref{sc:phenom} by reviewing the experimental observations on the normal and superconducting states.  Section~\ref{sc:stripes} covers the evidence for stripe order and associated superconductivity, followed by a description of the stripe-based pairing mechanism in Sec.~\ref{sc:pairing}.  This mechanism is used to explain the spatially-uniform superconducting state in Sec.~\ref{sc:spatial}.  Some remaining experimental aspects are discussed in Sec.~\ref{sc:disc}, with a brief conclusion in Sec.~\ref{sc:conc}.

\section{Cuprate phenomenology}
\label{sc:phenom}

To appreciate the challenges in understanding the cuprates, it is necessary to start with the parent compounds, such as \lco.  We then look at what happens when holes are introduced, considering first the correlated evolution of electronic and magnetic responses, and then measured features of the superconducting order.

\subsection{Parent correlated insulators}
\label{sc:parent}

Within the CuO$_2$ planes, the states near the chemical potential are dominated by contributions from the Cu $3d_{x^2-y^2}$ and O $2p_\sigma$ orbitals (where the $\sigma$ orbitals are the ones that have their lobes pointing at the neighboring Cu sites) \cite{emer87}.  If one considers only the nearest-neighbor hopping energy $t_{pd}$, then, since there is just one hole per Cu site, one would expect to find a half-filled hybridized band.  From experiment, we know that the parent cuprates are insulators with an optical gap  of 1.5--2~eV \cite{toku90}, and x-ray spectroscopies confirm that the Cu $3d$ hole is of $x^2-y^2$ character \cite{tran87c,heal88,bian88}, with the other $3d$ orbitals separated by at least 1.5 eV \cite{more11}.\footnote{There is some hybridization of the $3d_{x^2-y^2}$ orbital to the neighboring O $2p_\sigma$ orbitals \cite{walt09}, where the degree of hybridization can vary among cuprate families \cite{rybi16}.} To capture this behavior, one must take account of the strong Coulomb repulsion $U$ that is experienced by two electrons in the same $3d_{x^2-y^2}$ orbital, as treated by the  theory of Mott-Hubbard insulators \cite{mott49,hubb64}.  The optical gap does not correspond to $U$, however; the orbital energies are such that the valence band has strong O $2p$ character, while the lowest excited state corresponds to filling the $3d_{x^2-y^2}$ orbital (upper Hubbard band).  This arrangement corresponds to a charge-transfer insulator \cite{zaan85}.

\begin{figure}
 \centering
    \includegraphics[width=9cm]{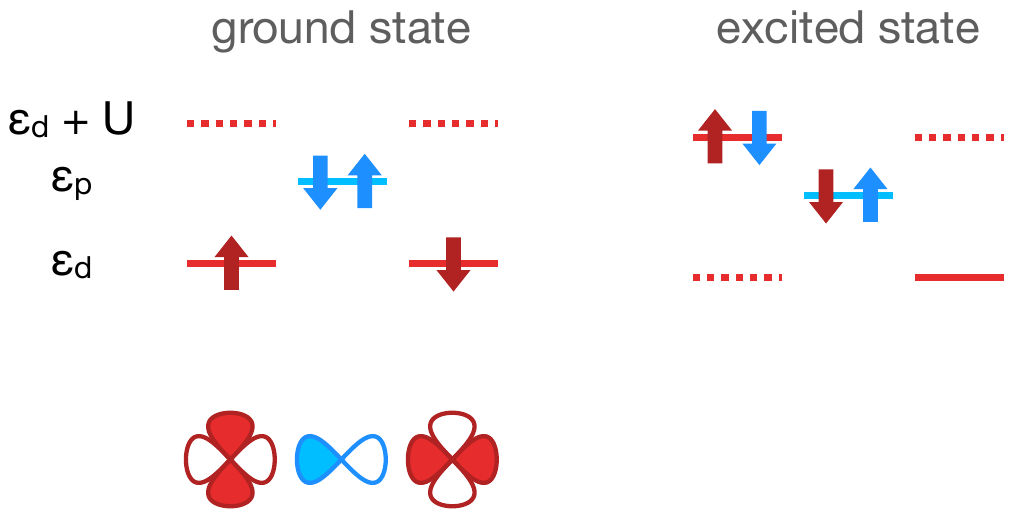}
    \caption{\label{fg:J}  Energy level diagrams illustrating states relevant to the superexchange mechanism.  In the ground state, there is one electron in each Cu $3d_{x^2-y^2}$ orbital, and the O $2p$ is filled (with orbitals illustrated schematically below).  In an intermediate excited state, one electron has hopped from O to fill the Cu level, raising its energy by $U$, which also allows the electron on the other Cu site to hop and refill the O site.  Because of the Pauli exclusion principle, these hops are only possible when the ground state involves antiparallel spins on neighboring Cu atoms.  (Obviously, electrons do not have a color; the color is used to indicate schematically how hops between sites can occur.) }
\end{figure}

As illustrated in Fig.~\ref{fg:J}, the electron in a half-filled $3d_{x^2-y^2}$ acts as a local moment, even though it sits below the filled O $2p_\sigma$ level, since filling it with an electron from O raises its energy by $U$, to above $\varepsilon_{p}$.  The large $U$ acts to frustrate the kinetic energy in the ground state; nevertheless, the energy can be lowered if virtual hops from O to Cu and back occur.  Of course, due to the Pauli exclusion principle, hops to both neighboring sites can only occur if the spins on neighboring Cu sites are antiparallel.  Thus, as shown by Anderson \cite{ande59,ande87}, these effects lead to a net superexchange interaction $J$ between neighboring Cu spins that drives antiferromagnetic (AF) correlations.\footnote{If one ignores the O sites and uses a single-band Hubbard model, then $J=4t^2/U$, where $t$ is the Cu-Cu hopping parameter.  Quantitative agreement with experiment requires consideration of the O, which leads to a more complicated formula, proportional to $t_{pd}^4$, where $t_{pd}$ is the hopping energy from O to Cu \cite{zaan87,emer88}.}

The superexchange mechanism of the Hubbard model translates to an effective Heisenberg Hamiltonian for spins $S=\frac12$.  {\newr The antiferromagnetic order within the CuO$_2$ planes is characterized momentum transfer ${\bf Q}_{\rm AF}=(\pi,\pi)$,\footnote{\newr In this notation, the nearest-neighbor Cu spacing $a$ is set equal to 1.  In reciprocal lattice units based on $a^*=2\pi/a$, one would write ${\bf Q}_{\rm AF}=(\frac12,\frac12)$.} as determined by neutron diffraction \cite{vakn87,tran88a}. } If the AF correlations were purely two dimensional, then the system would be disordered at finite temperature due to the Mermin-Wagner theorem \cite{merm66}.  There is also the fact that the ordered moment is reduced by almost 40\%\ due to zero-point fluctuations \cite{sand97b}.\footnote{Anderson speculated that the quantum spin fluctuations for the 2D $S=\frac12$ Heisenberg model might be great enough to prevent order, resulting in an RVB state \cite{ande87}.  This stimulated a great deal of theoretical effort on quantum spin liquids, and speculation that it was the RVB character of the initial state that led to superconductivity on doping \cite{sach00,ande04,lee06}.  In contrast, as we will see, competition been electronic kinetic energy and AF order is a key factor in the present story.} Nevertheless, the spin-spin correlation length is observed to grow exponentially with cooling  \cite{shir87,birg99,imai93} (theoretically, $\xi\sim \exp(\alpha J/k_{\rm B} T)$ \cite{chak88,hase93}), and only a small interlayer exchange interaction is necessary to produce three-dimensional order \cite{yasu05}.  The excitations of the ordered antiferromagnet are spin waves, with a bandwidth of $2J$; neutron scattering measurements on various layered cuprate insulators find $J$ in the range of 100 to 150 meV \cite{kive03,birg06,head10}.  

Detailed studies of the spin-wave dispersion in \lco\ \cite{cold01,head10} reveal modest deviations from the predictions based on linear-spin-wave theory for the nearest-neighbor Heisenberg model.  The superexchange interaction is the lowest-order spin-spin coupling term obtained in a perturbation expansion of the Hubbard model \cite{ande59}, and one can find corrections with higher-order terms, such as 4-spin cyclic exchange \cite{more06}, as well as a multimagnon continuum \cite{sand01}, that can account for much of the deviation.   A recent analysis in terms of a single-band Hubbard model found that fitting the spin-wave dispersion of \lco\ \cite{cold01} gave $U\approx 8t$ \cite{dela09}.  Since $8t$ is the electronic bandwidth in a noninteracting square lattice, one can see that the cuprates have a strong competition between Coulomb and kinetic energies.  

Anderson \cite{ande97a} argued that the superexchange interaction cannot be described in terms of a conventional band structure picture, {\newr which would involve an} interaction between quasiparticles.  Starting with a mean-field approach such as density-functional theory (DFT) or Hartree-Fock theory, it is certainly possible to find a solution with antiferromagnetic order and an associated insulating gap \cite{lane18,laug14}; however, that gap disappears in the absence of AF order \cite{coma08}, in contrast to experiment.  It is necessary to account for the local dynamical electronic correlations, and calculations on the Hubbard model using a version of Dynamical Mean Field Theory (DMFT) confirm that the insulating gap does not require AF order \cite{mouk01}.   It was only in 2016 that a first-principles calculation, employing dynamical mean-field theory (DMFT), was finally able to reproduce the charge-transfer gap in the paramagnetic phase of \lco\ \cite{choi16}.

\subsection{Hole-doped antiferromagnet}
\label{sc:hole}

To obtain superconductivity, one must dope charge carriers into the CuO$_2$ planes.  For a given compound, one generally can dope either holes or electrons.  Switching from one to the other requires a large jump in chemical potential, and I know of only one compound, Y$_{0.38}$La$_{0.62}$(Ba$_{0.87}$La$_{0.13}$)$_2$Cu$_3$O$_y$, in which ambipolar doping has been reported \cite{sega10}.  There is considerable asymmetry between electron and hole doping, with much higher $T_c$'s achieved with hole doping.  Here we consider only the case of holes.

The character of the dopant-induced holes has been characterized by x-ray absorption spectroscopy and electron-energy-loss spectroscopy at the O $K$ edge (probing $1s \rightarrow 2p$ transitions) and Cu $L_3$ edge (probing $2p \rightarrow 3d$) \cite{chen91,nuck89,bian88}.  The unanimous conclusion is that the holes are dominantly associated with O $2p_\sigma$ states, as expected for the case of a large $U$ on Cu \cite{eske88}.

\begin{figure}
 \centering
    \includegraphics[width=8cm]{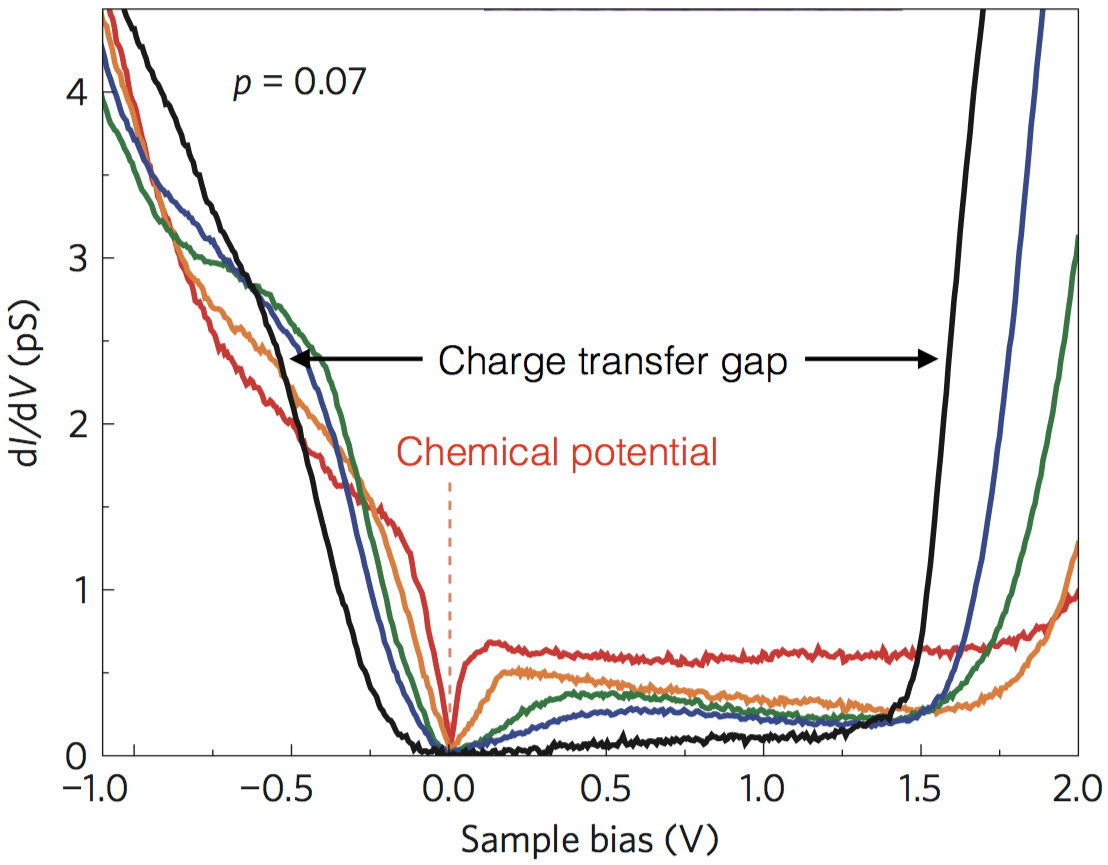}
    \caption{\label{fg:cai}  Tunneling curves at various points on the surface of Bi$_2$Sr$_{2-x}$La$_x$CuO$_{6+\delta}$ with $p = 0.07$ (charge-ordered insulator) measured by STM.   Reprinted with permission from \cite{cai16}, Springer Nature \copyright 2016. }
\end{figure}

In a conventional, mean-field band model, one would expect doping to cause the chemical potential to drop into the top of the valence band.  On the other hand, if the doped charges segregate into stripes, the chemical potential should be pinned within mid-gap states induced by the doping \cite{braz81,salk96}.  The presence of mid-gap states was initially inferred from optical conductivity\footnote{\newr For a review of techniques for measuring optical conductivity and applications to high-$T_c$ superconductors, see \cite{baso05,taji16}} measurements on LSCO \cite{uchi91}.  Their existence has been confirmed by a recent scanning tunneling microscopy (STM)\footnote{\newr For a review of scanning tunneling spectroscopy and applications to high-temperature superconductors, see \cite{fisc07}.} study of lightly-doped  Bi$_2$Sr$_{2-x}$La$_x$CuO$_{6+\delta}$ (Bi2201) \cite{cai16}.  Figure~\ref{fg:cai} shows tunneling conductance measurements made at several different points within 200~\AA\ of one another on the same atomically-flat sample surface \cite{cai16}.  (The hole concentration is estimated to be $p=0.07$, but the sample is not superconducting.)  A crucial feature of these spectra is that the range of bias voltage spans the charge-transfer gap \cite{uchi91,vanh09}.  Looking at the black curve, we see an example where, at that particular spot, there are essentially no states within the gap and the charge-transfer gap is about 2~eV.  For the other curves, we see that, locally, weight from both the lower and upper edges of the gap has moved into the gap, and the chemical potential is pinned within the gap.  Furthermore, the transferred weight is spread over a large energy scale ($\sim0.5$~eV), and there is a pseudogap centered at the chemical potential that is relatively particle-hole symmetric.

The variation in the local weight of the mid-gap states suggests a significant role for the long-range Coulomb interaction{\newr, and we will return to the impact of disorder and long-range charge inhomogeneity in Sec.~\ref{sc:disorder}}.  In the present case, a hole is introduced by replacing a La$^{3+}$ ion with a Sr$^{2+}$.  The hole goes into a neighboring CuO$_2$ plane, but it is not free to move away.  It feels a substantial Coulomb attraction to the dopant site, and in the limit of very small doping, the screening comes mainly from phonons \cite{kast98}.   As reviewed in \cite{kast98}, a single hole may be localized on a scale of a few lattice spacings.  Evidence for variable-range hopping in a Coulomb potential has been provided by transport measurements on \lsco\ with a hole concentration $p\lesssim0.05$  \cite{kast98,shkl84}.  

\begin{figure}
 \centering
    \includegraphics[width=12cm]{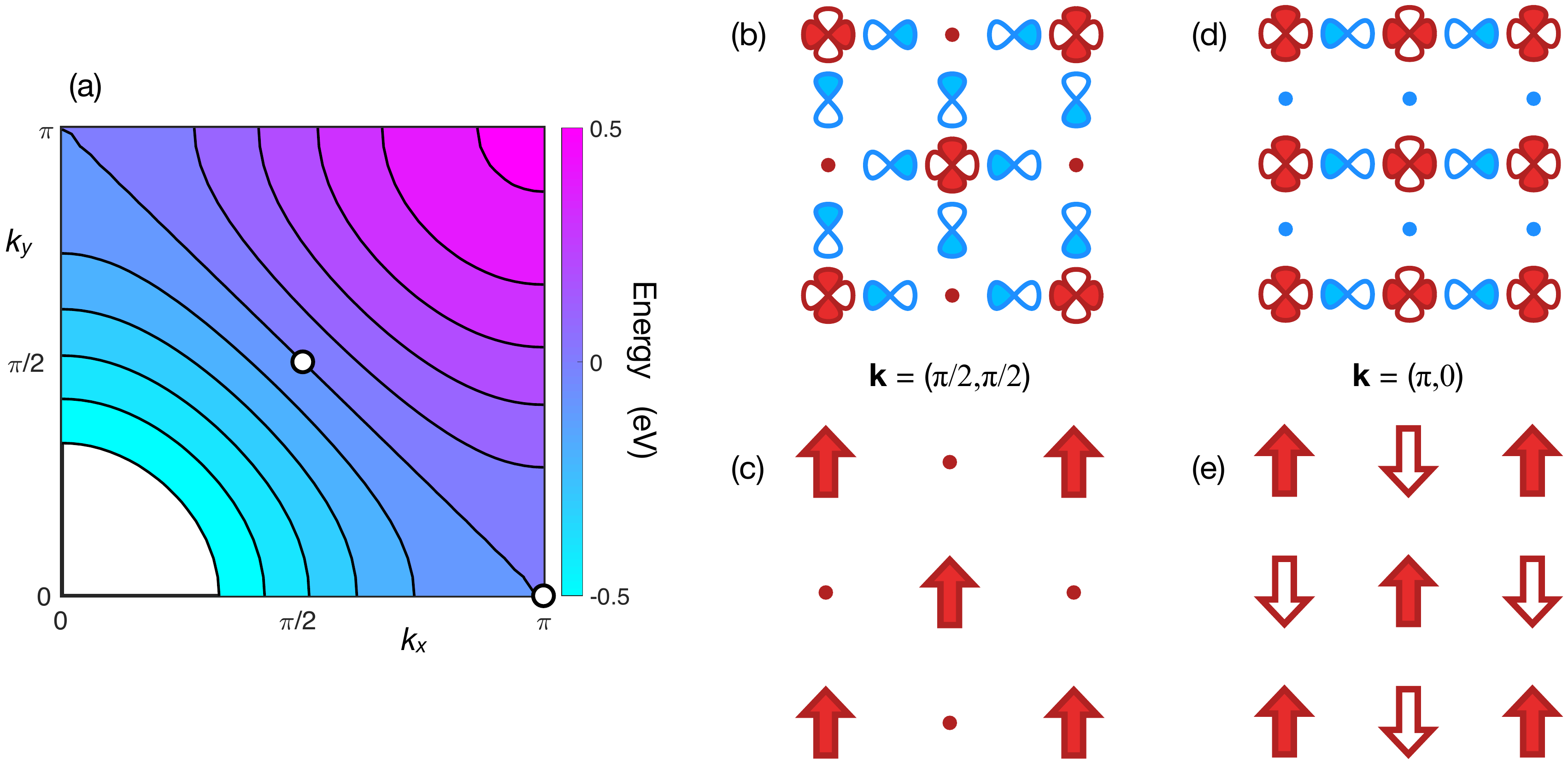}
    \caption{\label{fg:es}  (a) Dispersion of the antibonding band for a non-interacting 3-band model with only nearest-neighbor hopping, projected onto a quadrant of the first Brillouin zone with energy referenced to the chemical potential at half-filling.  Circles denote the points on the Fermi surface at $(\pi/2,\pi/2)$ and $(\pi,0)$.  (b) Schematic wave function for ${\bf k} = (\pi/2,\pi/2)$ \cite{ande95}.  (c) Antiferromagnetic spin structure with spin size proportional to the $d_{x^2-y^2}$ weights in the $(\pi/2,\pi/2)$ wave function.  (d) Schematic wave function at ${\bf k} = (\pi,0)$ \cite{ande95}, and (e) the corresponding weighted spin components. }
\end{figure}

It is clear that the real materials are quite different from the predictions of conventional band theory \cite{matt87,frie89}; nevertheless, from comparison of such predictions with the antiferromagnetic structure of the undoped system, one can gain some appreciation for how the electronic quasiparticles and the antiferromagnetic correlations compete.  For this purpose, we return to the ``3-band'' model involving the Cu $3d_{x^2-y^2}$ and O $2p_\sigma$ orbitals with only nearest-neighbor hopping, and look at a tight-binding calculation.\footnote{This is essentially the Emery model \cite{emer87}, but with the onsite Coulomb repulsion $U$ set to zero.}  The relevant formulas have been given in detail by Andersen {\it et al.}\ \cite{ande95}.  Using common parameter values (3~eV for the energy separation between the Cu and O levels and hopping energy $t_{pd}=1.6$~eV \cite{ande95}), the antibonding band, which should cross the chemical potential $\mu$ and be half-filled for an undoped CuO$_2$ layer, is easily calculated.  The dispersion, projected onto a quadrant of the first Brillouin zone, is shown in Fig.~\ref{fg:es}(a).  In this simple model, the Fermi surface at half filling runs diagonally from $(\pi,0)$ to $(0,\pi)$ through $(\pi/2,\pi/2)$, the position of the node in the case of the $d$-wave superconducting gap (the nodal point).\footnote{Note that, for this section, the lattice parameter $a$ is set equal to 1.}

The schematic wave functions at two of these points [indicated by circles in Fig.~\ref{fg:es}(a)] are also given in \cite{ande95}, and we reproduce them in Fig.~\ref{fg:es}(b) and (d).  At each wave vector, the variation of the real part of the Bloch wave in real space follows $\cos({\bf k}\cdot {\bf r}_n)$, where ${\bf r}_n$ is the center of the $n^{\rm th}$ unit cell.\footnote{The imaginary part of the Bloch wave forms a degenerate state, decoupled from the real part.}  For ${\bf k} = (\pi/2,\pi/2)$, we see that the phase of the $d$ orbitals is identical for all sites along the diagonal, but is staggered for second neighbors along the Cu-O bond directions; for half of the Cu sites, the amplitude is zero.  For ${\bf k}=(\pi,0)$, the $d$ orbitals are in-phase along [0,1] and antiphase along [1,0].  Again, these states have identical energies in this noninteracting model, where there is no magnetic order.

Now consider the impact of antiferromagnetism.  At the nodal wave vector, the Cu sites that contribute to the wave function all have weight on a sublattice of the AF order with unique spin direction, as shown in Fig.~\ref{fg:es}(c), so that the electronic state is compatible with the spin correlations.  On the other hand, at the antinodal (AN) wave vector $(\pi,0)$, neighboring Cu sites correspond to antiparallel spins of the AF state.  A Bloch state must have a unique spin, so the wave function at $(\pi,0)$ is incompatible with AF spin correlations.  Note that the incompatibility does not depend on static order; if the spins on neighboring sites are instantaneously antiparallel, then they cannot be part of the same extended state.  \footnote{For a proper analysis of the problem of one hole in a two-dimensional (2D) antiferromagnet, see \cite{lau11,ebra14}.}

The electronic spectral function can be probed directly by ARPES.\footnote{\newr For a review of angle-resolved photoemission and applications to cuprates, see \cite{dama03}.}  While much of the discussion of ARPES results on cuprates has focused on the difference between nodal and antinodal responses (the ``nodal-antinodal dichotomy'' \cite{fu06}), we are interested in the points in between, as well.  For a particular {\bf k} along the noninteracting Fermi surface, a practical measure is the weight of the wave function on nearest-neighbor Cu sites,
\begin{equation}
  w_{\rm AF}({\bf k}) = \frac12 \left(|\cos(k_xa)|+|\cos(k_ya)|\right).
\end{equation}
In fact, this has the same {\bf k} dependence as the absolute magnitude of the $d$-wave gap,
\begin{equation}
  \Delta({\bf k}) = \frac12 \Delta_0\left[\cos(k_xa)-\cos(k_ya)\right].
\end{equation}
Both functions go to zero at the nodal point and have extrema at the antinodes.

To estimate the energy scale of the pseudogap, note that, in order to transform the AF state in Fig.~\ref{fg:es}(e) to a single-spin state, it is necessary to flip half of the Cu spins.  The energy to locally flip a spin corresponds to the maximum spin-wave energy, $\sim2J \sim 300$~meV.  Hence, it is plausible that, for a hole moving in an AF, the energy difference between the antinodal (AN) and nodal points is $\sim 2J$.\footnote{The picture of an antinodal pseudogap defined by scattering from spin fluctuations is supported by a variety of advanced numerical calculations applied to the Hubbard model \cite{dago94,kyun06,gunn15,wu17}.}  

\begin{figure}
 \centering
    \includegraphics[width=9cm]{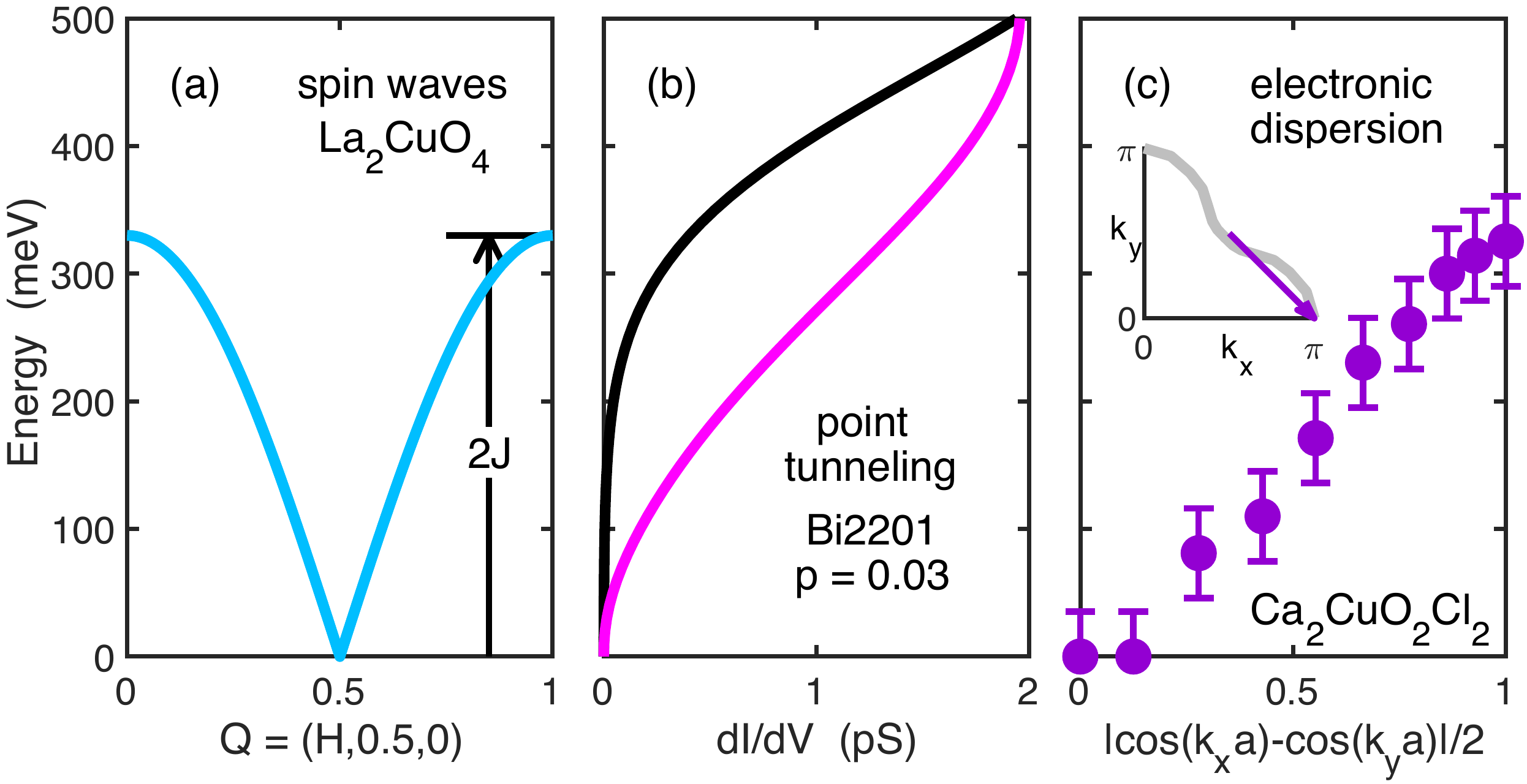}
    \caption{\label{fg:big}  (a) Spin-wave dispersion in La$_2$CuO$_4$ measured by neutron scattering \cite{cold01,head10}.  (b) Schematic version of typical conductance curves measured by STM at typical locations in Bi$_2$Sr$_{2-x}$La$_x$CuO$_{6+\delta}$ with $p = 0.03$ \cite{cai16}, plotted as binding energy vs.\ conductance; black: region with no in-gap states; magenta: region with significant in-gap states.  (c) Effective gap or dispersion along the nominal weak-coupling Fermi surface measured by ARPES in Ca$_2$CuO$_2$Cl$_2$ \cite{ronn98}. }
\end{figure}

Figure~\ref{fg:big} compares the energy scales of spin fluctuations in a parent insulator, in (a), with the effective dispersion of one hole in an antiferromagnet, in (c).  Here, the spin-wave results are from neutron scattering\footnote{\newr  For reviews of neutron scattering, with applications to cuprates and related systems, see \cite{tran13a,zali15i}.} measurements on \lco \cite{cold01}, while the ARPES were obtained on Ca$_2$CuO$_2$Cl$_2$ \cite{ronn98}.  It may at first seem surprising that one can measure electronic dispersion in an insulating compound; however, the photoemission process involves the absorption of a photon and emission of an electron, leaving a final state with one hole, whose energy varies with wave vector.  The results shown are roughly along the direction of the anticipated Fermi surface for noninteracting electrons, so that the dispersion corresponds to the pseudogap {\newr  for one hole in an antiferromagnet}.  The pseudogap has a {\bf k} dependence similar to $\Delta({\bf k})$ and an energy scale essentially equal to $2J$ \cite{naza95}, as proposed above.

STM effectively measures the density of states.  For noninteracting band structure, there is a Van Hove singularity at the $(\pi,0)$, $(0,\pi)$ points \cite{frie89}, so we expect the largest contributions to the density of states to come from the antinodal regions.  Figure~\ref{fg:big}(b) shows representative tunneling conductance curves for Bi2201 with small-but-finite $p\sim0.03$ \cite{cai16}.  (Such data are usually plotted as conductance vs.\ bias voltage, {\newr as in Fig.~\ref{fg:cai},} but here the axes are reversed to allow direct comparison of the energy scales.)  At this low doping, there are some points on the sample that appear undoped, such as indicated by the black line. Where dopant-induced states appear (magenta line), their relative weight (on the positive binding-energy side) has a maximum that is comparable to $2J$, consistent with the pseudogap seen in ARPES.  Of course, we have already seen that the STM conductance curves in Fig.~\ref{fg:cai} show similar gapping effects for states both above and below the Fermi level, which is also consistent with the pseudogap due to AF correlations.\footnote{Another factor for the STM results  concerns the nature of the tunneling process from the probe tip to the sample surface and along the $c$ axis, typically through an apical O site, to a Cu site in the CuO$_2$ plane nearest the surface.  The $2p$ orbital on the apical site has $s$ symmetry relative to the Cu, so that it cannot couple to the $3d_{x^2-y^2}$ but may couple to the in-plane O $2p_\sigma$ states.  From Fig.~\ref{fg:es}(b) and (d), one can see that no coupling is possible at ${\bf k}=(\pi/2,\pi/2)$ because the O orbitals are all in phase with the nearest Cu $3d_{x^2-y^2}$ orbital, but there is a finite coupling at $(\pi,0)$ as the phasing is different.  Such effects were originally noted in analyses of $c$-axis conduction and planar tunneling \cite{chak93,xian96}, and are discussed for STM in \cite{bala06}.}

\begin{figure}
 \centering
    \includegraphics[width=4cm]{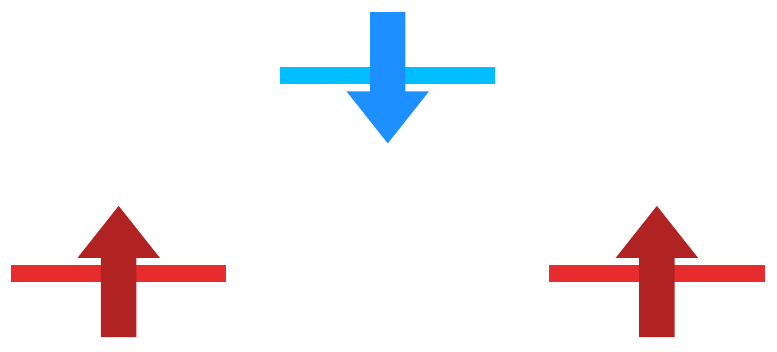}
    \caption{\label{fg:Emery}  Hole on O favors parallel alignment on neighboring Cu sites. }
\end{figure}

While the AF correlations have a big impact on the states associated with the doped holes, we also have to consider the back action of the holes on the AF correlations, which is drastic.  For example, as one introduces holes into \lsco\ (LSCO) by increasing $x$ from zero, the AF ordering temperature, $T_{\rm N}$, decreases rapidly, with commensurate order disappearing at $x\approx0.02$ \cite{birg06}.  In the case of \lcod, where interstitial oxygen atoms grab electrons and induce holes, there is actually phase separation into distinct antiferromagnetic and doped superconductor phases \cite{well97}.  The doped holes at low temperature only become delocalized as the static magnetic order disappears \cite{ando01}; nevertheless, dynamic AF correlations survive \cite{fuji12a}, as we will discuss.  It is this competition between the kinetic energy of the holes and superexchange between local Cu moments that is the dominant feature for understanding the cuprates.

While superexchange tends to cause neighboring Cu spins to be antiparallel, a hole on an O site will have the best chance to hop and reduce its kinetic energy if its Cu neighbors have parallel spins \cite{emer88}, as indicated in Fig.~\ref{fg:Emery}.  This alignment clearly frustrates commensurate AF order.\footnote{This is different from the proposal of the Zhang-Rice singlet \cite{zhan88}.  That picture assumes that a hole is bound symmetrically about a Cu site, resulting in no net moment.  This would tend to act as a dilution of the AF lattice.  But we know from experiment that dilution with Zn only destroys long-range order at the percolation limit (41\%\ Zn concentration) \cite{vajk02}. }  The motion of a hole through the lattice causes further disruption.  This can be seen easily in the case of a single hole introduced into an AF array of Cu atoms, ignoring the O sites.  For the hole to move from one Cu site to the next, a Cu spin must move in the opposite direction.  This means that as the hole moves, a string of sites develops with ferromagnetic correlations to nearest neighbors \cite{trug88,weng97}. Such behavior in a model of one hole in an Ising AF has been simulated with cold atoms, confirming the string-like defects \cite{chiu19,grus18}.

\subsection{Intertwined spin and charge fluctuations}
\label{sc:intertwined}

While superexchange interactions among Cu spins and the kinetic energy of doped holes compete with one another, the charge and spin fluctuations of hole-doped cuprates display a correlated evolution with doping and with temperature.  Here we will first consider fluctuations on the scale of $J$ and then on the scale of $k_{\rm B}T$.

One useful measure of the high-energy AF excitations is given by Raman scattering.\footnote{\newr For a review of Raman scattering and applications to cuprates, see \cite{deve07}.}  While a photon does not couple directly to a spin, it can cause a pair of spins to flip; the corresponding inelastic scattering feature is known as two-magnon scattering \cite{fleu68}.  Recent theoretical analysis for 2D antiferromagnets \cite{weid15} gives a peak energy of $2.44J$ and an asymmetric peak shape that is in quantitative agreement with experimental measurements of cuprates \cite{chel18}.  

\begin{figure}
 \centering
    \includegraphics[width=7cm]{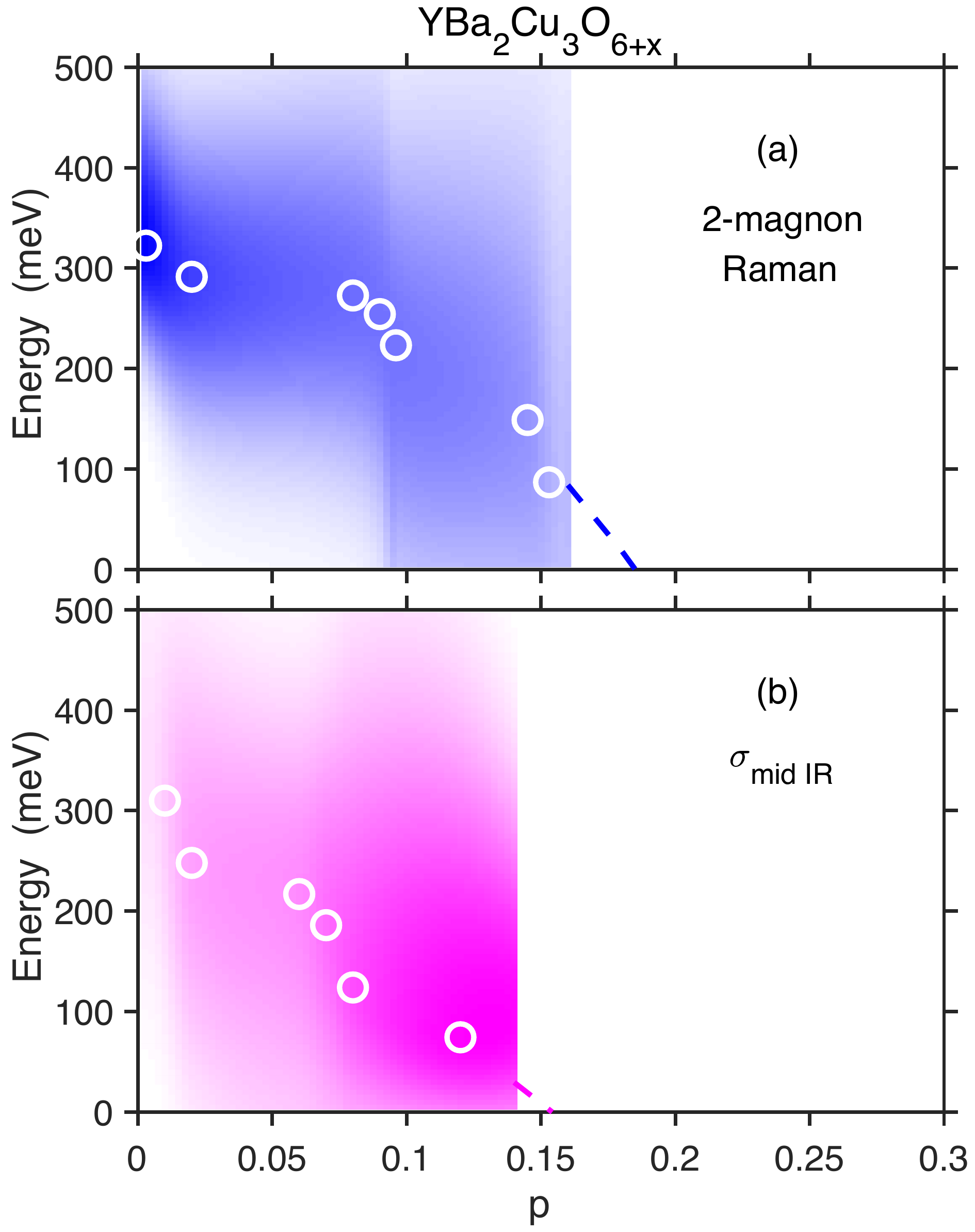}
    \caption{\label{fg:mYBCO}  (a) Two-magnon response measured by Raman scattering in \ybco\ as a function of estimated hole concentration $p$ \cite{suga03}. (b) Mid-infrared optical conductivity in \ybco\ \cite{lee05}, with the energy scale divided by 2.  In both (a) and (b), the data have been interpolated, and circles denote measured peak positions.  Dashed lines are quadratic extrapolations of the peak positions.}
\end{figure}

Figure~\ref{fg:mYBCO}(a) shows how the two-magnon scattering evolves with doping in \ybco\ \cite{suga03}.  As one can see, the well-defined peak in the lightly-doped system softens in energy as holes are added.  Sugai {\it et al.} \cite{suga03} have similar results for  \lsco\ (LSCO), \bscco\ (Bi2212), and Bi2201, extending to higher doping, and in all cases the two-magnon peak appears to become overdamped in the vicinity of $p\sim0.2$.  If we assume that the excitation mechanism remains the same in the doped systems as in the parent AF, then the peak energy will depend both on the strength of $J$ and the degree to which a finite patch of Cu sites retains AF correlations.  While the AF correlations restrict the hole motion and spatial distribution, the increasing density of holes must reduce the spatial areas in which AF correlations among Cu spins can survive.  Neutron scattering measurements have shown that the instantaneous spin correlation length is reduced to approximately one lattice spacing near optimal doping \cite{thur89b,xu07}, so a decrease in correlation length can describe much of the decay in the peak energy with $p$.  Nevertheless, it can also be convenient to think of an average reduction in $J$.  For example, Johnston \cite{john89} pointed out that the temperature dependence of the bulk magnetic susceptibility in \lsco\ can be scaled by doping-dependent values of $J$ and average moment per Cu, with both trending toward zero at $p=x\sim0.2$, as confirmed by others \cite{naka94}.  This is consistent with neutron-scattering observations of increasing damping of high-energy spin excitations with doping \cite{stoc10,fuji12a}.\footnote{\newr Magnetic excitations can also be probed by resonant inelastic x-ray scattering (RIXS) at the Cu $L_3$ edge.  For example, measurements of magnetic excitations for {\bf Q} from zero to the AF zone boundary performed on thin films of LSCO for doping extending out to the highly overdoped regime suggest less doping dependence of the measured excitations \cite{dean13}.  However, there are several issues to consider in evaluating these results, the most important being that the measurements cannot reach the vicinity of the ${\bf Q}_{\rm AF}$, where inelastic neutron scattering studies clearly demonstrate large changes with doping in the definitive AF correlations that theory considers most relevant to superconductivity \cite{huan17b,jia14,kung15}.}

The charge fluctuations are probed by measurements of the optical conductivity, $\sigma_1(\omega)$ \cite{baso05}.   At energies below that of the charge-transfer gap, optical measurements probe all direct transitions from filled to empty mid-gap states; from the STM results in Fig.~\ref{fg:cai}, we already saw that, for a very underdoped cuprate, the mid-gap states extend over a significant energy range with a pseudogap about $\mu$.  For moderately underdoped cuprates, it is possible at low temperature to distinguish the corresponding broad ``mid-IR" conductivity from the more coherent Drude peak, centered at zero energy and corresponding to the quasiparticles that contribute to the metallic transport \cite{lee05}.  Figure~\ref{fg:mYBCO}(b) shows a rough version of the mid-IR feature observed in YBCO at temperatures close to but above the superconducting $T_c$ \cite{lee05}.  Here the energy scale has been divided by two to make it comparable to electron binding energies.   The energy scale is virtually the same as, and decreases in a fashion similar to, the two-magnon energies in Fig.~\ref{fg:mYBCO}(a), while the spectral weight grows with doping.

\begin{figure}
 \centering
    \includegraphics[width=4cm]{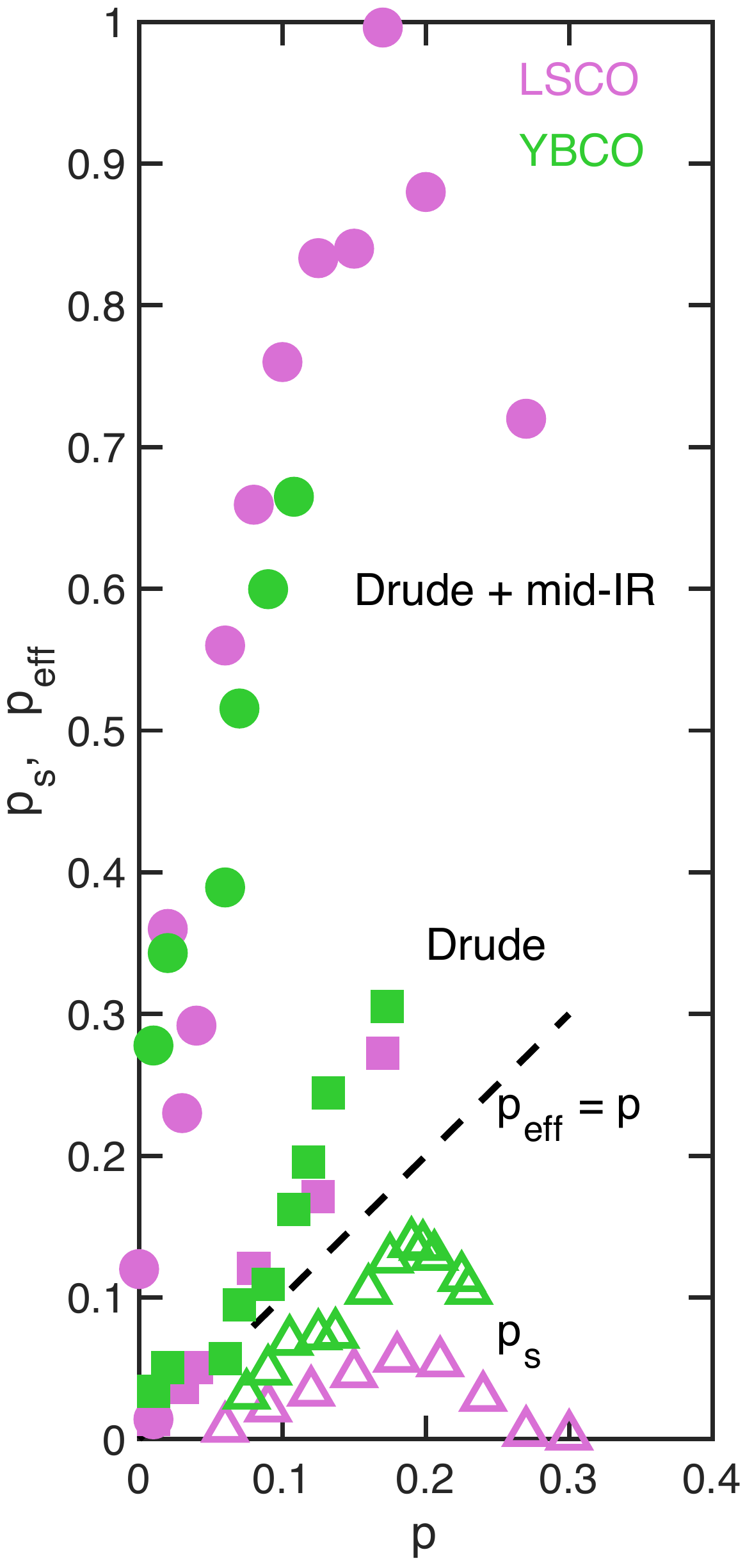}
    \caption{\label{fg:p} Comparison of various estimates of carrier density in \lsco\ (violet) and \ybco (green).  Triangles: superfluid density, $p_s$, from measurements of the magnetic penetration depth on Y$_{0.8}$Ca$_{0.2}$Ba$_2$Cu$_3$O$_{6+x}$ by muon spin rotation \cite{bern01} and on LSCO films by mutual inductance \cite{lemb11}; squares: effective carrier density, $p_{\rm eff}$, from integrating in-plane optical conductivity to 80 meV \cite{padi05}; circles: $p_{\rm eff}$ from integrating to 1.5 eV \cite{padi05,uchi91}.  In evaluating $p_{\rm eff}$, the effective masses of $4m_e$ and $3m_e$ for LSCO and YBCO, respectively, determined in \cite{padi05} were used. }
\end{figure}

The variation of the effective carrier concentration is considered more quantitatively in Fig.~\ref{fg:p}.  Integrating the optical conductivity from zero to a cutoff energy gives a result proportional to the effective hole density $p_{\rm eff}$ divided by an effective mass $m^*$.  Padilla {\it et al.}\ \cite{padi05} compared optical data with measures of $p$ from the Hall coefficient to obtain $m^*/m_e\approx4$ for LSCO and 3 for YBCO, where $m_e$ is the electron mass.\footnote{A study using terahertz spectroscopy in pulsed magnetic fields has determined a cyclotron mass of $4.9\pm0.8\ m_e$ for LSCO with $x=0.16$ \cite{post20}.}  (A later study has suggested that $m^*$ decreases by $\sim50$\%\ with doping in the range $0.1 <p<0.22$ \cite{vanh09}.)  To estimate the carrier density associated with the Drude peak, they integrated $\sigma_1(\omega)$ data for LSCO and YBCO to 80 meV, which yields the $p_{\rm eff}$ values indicated by filled squares in Fig.~\ref{fg:p}, plotted against the estimated dopant-induced carrier density $p$ \cite{padi05,lee05}.\footnote{Related data for Bi2212 are reported in \cite{hwan07b}.}  At small $p$, the effective Drude carrier density is close to $p$, while it begins to rise above it for $p\gtrsim0.1$; such behavior in LSCO was originally noted based on measurements of the Hall coefficient \cite{taka89b}, and similar behavior is seen in the nodal weight detected by ARPES \cite{yosh03}.  Integrating up to 1.5~eV, an energy comparable to the charge-transfer gap of the parent insulators, one captures the $p_{\rm eff}$ associated with both the Drude peak and the mid-IR signal, indicated by the filled circles.  This value is several times larger than the Drude weight alone, and approaches a maximum of $\sim1$ for LSCO at $p\approx0.18$.

The maximum in $p_{\rm eff}$ is consistent with other measurements.  The $T$-linear electronic specific heat coefficient for the normal state\footnote{This was estimated by suppressing the superconductivity by Zn substitution \cite{momo94}.} of LSCO shows a maximum at $p\sim0.2$ \cite{momo94}.  Similarly, ARPES studies on LSCO \cite{yosh07,miao20} indicate that the nominal Fermi surface crosses through the $(\pi,0)$, $(0,\pi)$ points in the range $0.17 < p < 0.21$; as already mentioned, a noninteracting system would have a Van Hove singularity that hits $E_F$ at this point \cite{frie89}. That crossing would also represent a Lifshitz transition, where the Fermi surface changes from hole-like to electron-like.\footnote{In the system \lnsco, measurements of the Hall effect at low temperature and high magnetic field (to suppress superconductivity) have been interpreted as indicating a rapid rise of the carrier concentration to a level of $1+p$ at $p^* \approx 0.23$ \cite{coll17} (with analogous behavior in YBCO at $p^*\approx0.19$ \cite{bado16}).  The analysis here is not clear cut, as ARPES measurements \cite{matt15} indicate that a Lifshitz transition, along with the closing of the antinodal pseudogap, occur at or near $p^*$.  These latter features also appear to be consistent with the observation of a peak in the electronic specific heat at $p^*$ \cite{mich19} (where analysis is complicated by a Schottky anomaly due to the magnetic Nd ions \cite{xie12}). Hall effect measurements on LSCO and Bi2201 yielded sharp cusps in the vicinity of $p^*$ \cite{bala09}.  Theoretical analyses indicate that anomalous behavior can occur near a Lifshitz transition \cite{maha17}, especially when strong-correlation effects are important \cite{shas93,wang20}.}

In a complementary fashion, one can obtain the superfluid density $p_s$ from measurements of the magnetic penetration depth $\lambda$ at $T\ll T_c$, using \cite{tink75} 
\begin{equation}
\frac{1}{\lambda^2} = \frac{4\pi p_se^2}{m_ec^2}.
\end{equation}
Two ways to measure $\lambda$ are by muon spin relaxation ($\mu$SR) in an applied magnetic field \cite{uemu89} and by mutual inductance \cite{lemb11}.  The open triangles in Fig.~\ref{fg:p} indicate $p_s$ values for Y$_{0.8}$Ca$_{0.2}$Ba$_2$Cu$_3$O$_{6+x}$ (determined by $\mu$SR \cite{bern01}) and for LSCO thin films (from mutual inductance \cite{lemb11}).\footnote{A related figure of $p_s$ in LSCO, including detailed data from \cite{bozo16}, is presented in \cite{pelc19}.}  Much of the Drude weight from $T>T_c$ goes into the superfluid, while $p_s$ in these materials reaches a maximum near $p\sim0.2$.

Returning to the normal state, the electronic spectral function, as measured by ARPES on Bi2212, also evolves significantly with doping.  We saw in Fig.~\ref{fg:big}(c) that there is a large pseudogap for one hole in an antiferromagnet, going to zero at the nodal point and rising to its maximum at the antinodes.  With doping, the gapless behavior of the nodal point {\newr in the normal state} spreads to a finite arc, with the arc length growing with doping \cite{norm98,kani06,kami15}.  As the arc length grows, the antinodal gap decreases.  

\begin{figure}
 \centering
    \includegraphics[width=8cm]{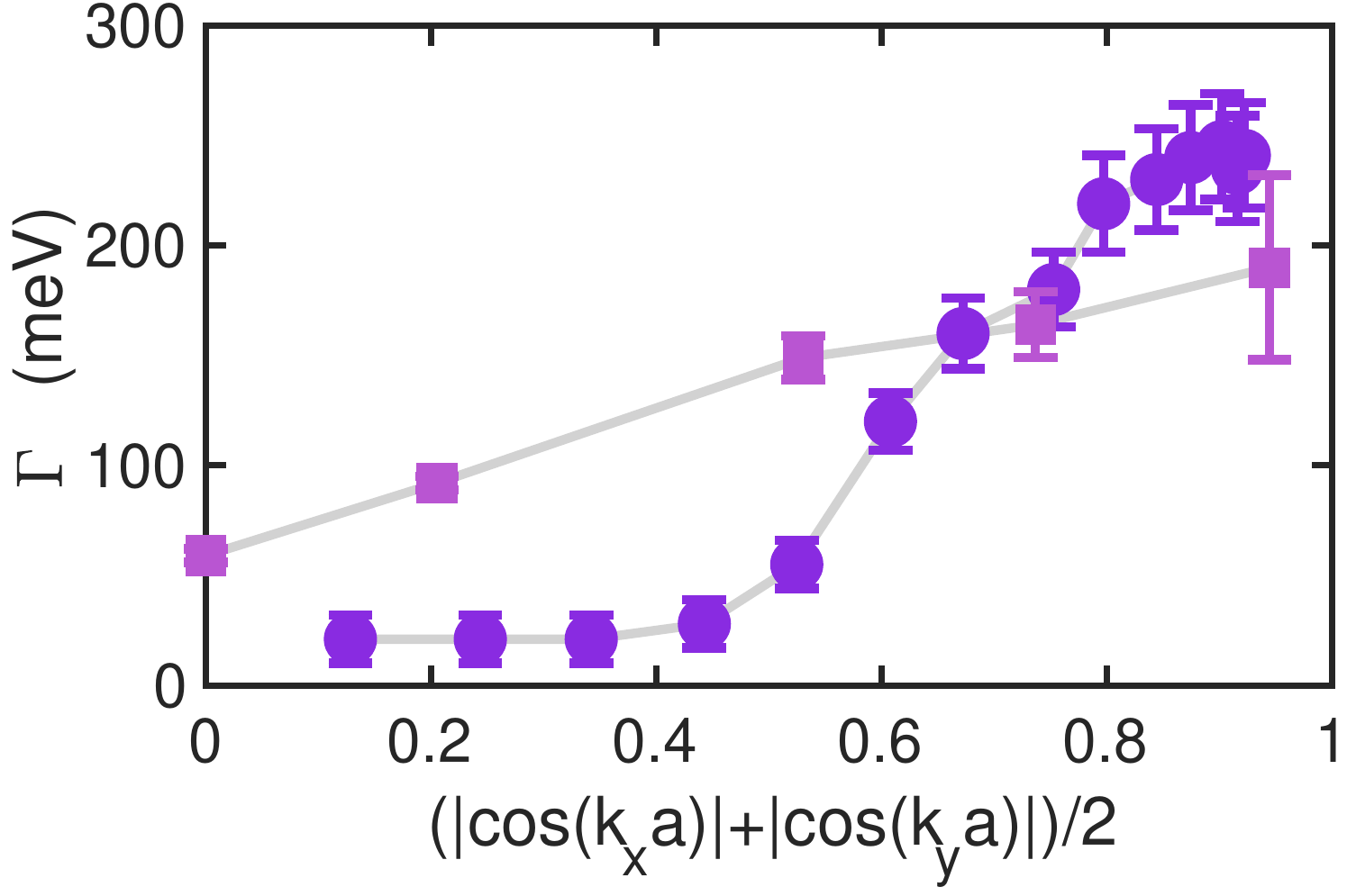}
    \caption{\label{fg:gamma}  Scattering rate $\Gamma$ for electron spectral function obtained on optimally doped Bi2212 from ARPES measurements at 100~K (violet squares) \cite{vall00}; at 140~K (purple circles) \cite{kami05}.  {\newr  These results are from separate analyses of distinct data by different groups.  The common feature is that the scattering rate grows continuously from the nodal to the antinodal regions.}}
\end{figure}

While the antinodal pseudogap has decreased substantially by optimal doping, the scattering rate is still large there.  As shown in Fig.~\ref{fg:gamma}, the scattering rate also shows a continuous variation from the nodal to antinodal points, reaching its maximum at the latter \cite{vall00,kami05}.  This variation is consistent with the impact of short-range AF correlations of Cu moment.\footnote{Note that there is no sign of enhanced scattering at any unique wave vector, as proposed in ``hot spot'' models, where the interaction is assumed to be restricted to {\bf k} points nested by ${\bf Q}_{\rm AF}$.}

An exception to this trend occurs when incommensurate AF order is present.  For example, neutron scattering shows static or quasi-static incommensurate spin correlations in LBCO with $x=1/8$ for $T\lesssim 50$~K \cite{fuji04,tran08}, and ARPES finds a pseudogap consistent with Eq.~(1), reaching an antinodal gap of $\sim 20$~meV \cite{vall06,he09}.  A similar relationship exists for quasi-static spin correlations in LSCO with $x=0.07$ \cite{jaco15} and ARPES on LSCO with $x=0.08$ \cite{razz13}. 

Given the minimum in scattering rate in the near nodal region, it is plausible that the near-nodal states make the dominant contribution to the in-plane conductivity and the Drude peak in optical conductivity.  Indeed, an analysis of the self energy of near-nodal states as a function of temperature and doping in Bi2212 supports such a conclusion \cite{rebe19}.  On the other hand, the conductance between layers depends on antinodal states \cite{chak93}, so that the evolution of the pseudogap is reflected in the $T$ dependence of the $c$-axis resistivity \cite{naka93,komi02} and $c$-axis polarized optical conductivity \cite{home93b,baso94}.

\subsection{Temperature dependence}
\label{sc:temp}

We have already seen that the magnetic and electronic excitations have correlated energy scales that vary with doping in a parallel fashion.  There are related correlations in behavior with temperature.  From the temperature dependence of various physical quantities such as the bulk spin susceptibility $\chi$ and the in-plane Hall coefficient $R_{\rm H}$, a temperature $T^*$ (which decreases with $p$) is commonly identified which behaves as a crossover to the pseudogap phase \cite{emer97,keim15}.  It is instructive to consider the measurements and analysis by which $T^*$ has been determined.  

\begin{figure}
 \centering
    \includegraphics[width=6cm]{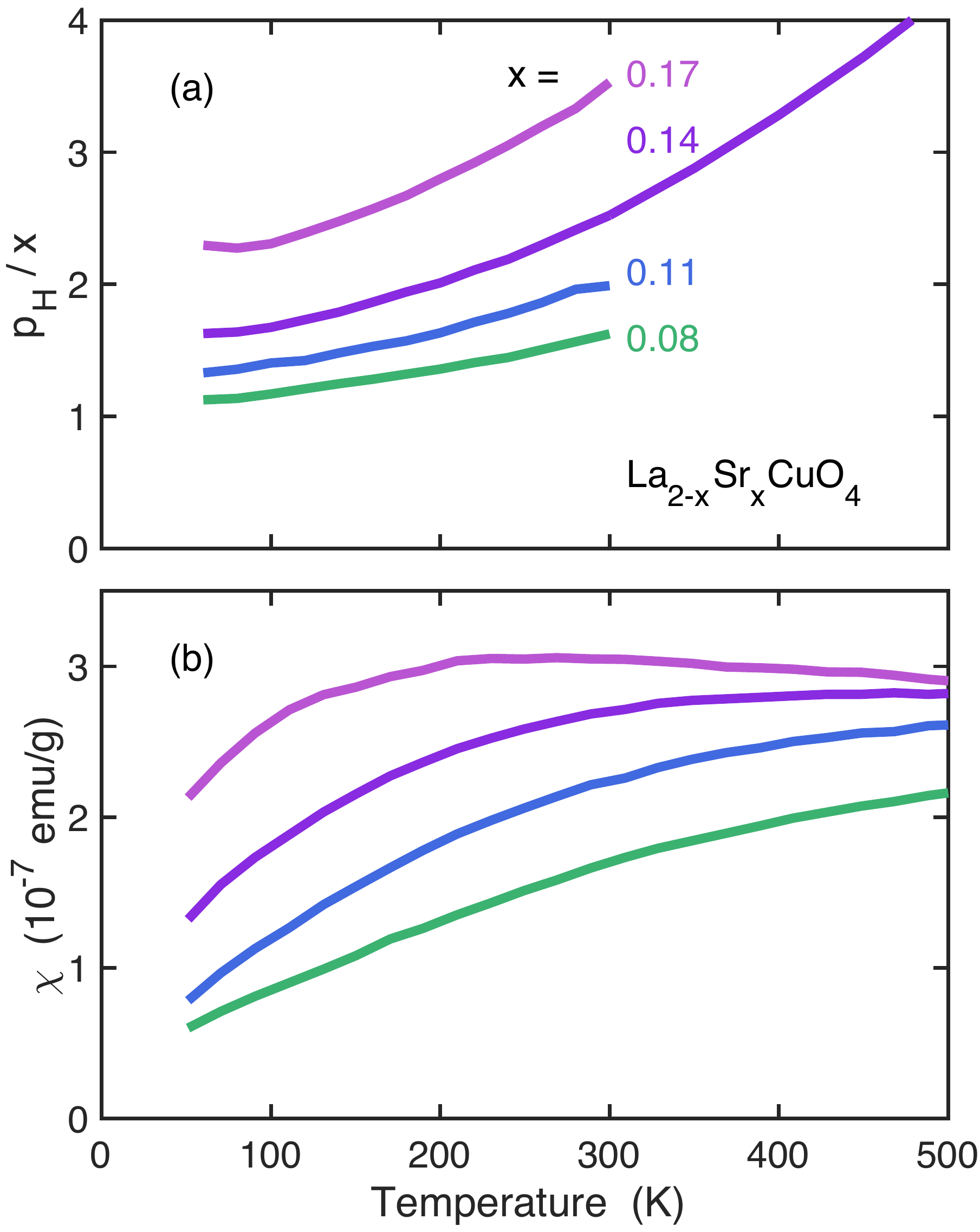}
    \caption{\label{fg:Hallchi}  (a) Ratio of the hole concentration $p_{\rm H}$ determined from the Hall effect to the Sr fraction $x$ in \lsco; data from \cite{ando04} have been interpolated.  $T^*\sim410$~K and 535~K for $x=0.17$ and 0.14, respectively \cite{hwan94}. (b) Bulk magnetic susceptibility in \lsco; data from \cite{naka94} have been interpolated.  $T_{\rm max}\sim280$~K and 500~K for $x=0.17$ and 0.14, respectively.}
\end{figure}

$\chi(T)$ evolves continuously with doping from the antiferromagnetic parent state.  For a $S=\frac12$ 2D Heisenberg model, as appropriate to a system such as \lco, $\chi$ is predicted to have a maximum at $k_{\rm B}T_{\rm max} \approx J$ \cite{yama19}.  The maximum occurs when the spin-spin correlation length reaches $\sim2.5a$ \cite{taka89}, and $\chi$ decreases as the AF correlation length grows further.  $T_{max}$ for \lco\ is estimated to be comparable to the crystal melting temperature; nevertheless, $\chi(T)$ has been measured to decrease on cooling below 1200~K \cite{yosh90}.  For \lsco, the $\chi(T_{\rm max})$ becomes observable for $x\gtrsim0.09$ as $T_{\rm max}$ drops below 800~K \cite{yosh90,naka94}.  It has been demonstrated that the temperature dependence of $\chi$ can be scaled with doping, with the temperature scaled by $T_{\rm max}$ \cite{john89,naka94}.  Representative (interpolated) results for several values of $x$ are shown in Fig.~\ref{fg:Hallchi}(b).\footnote{One of the first identifications of the pseudogap was based on Knight-shift measurements using Y nuclear magnetic resonance (NMR) in YBCO \cite{allo89}.  The Knight shift is generally proportional to the bulk spin susceptibility, and the results for YBCO are similar to bulk susceptibility results for the CuO$_2$ planes, after correction for a chain contribution \cite{john88,yama89}.  In the original interpretation of the temperature dependence of the Knight shift data \cite{allo89}, the role of Cu moments (clearly detected by neutron scattering \cite{sham93}) was ignored; instead, it was interpreted as the response of electronic quasiparticles, with a decrease in carrier density on cooling.  While other measurements, such as the Hall effect, do indicate a temperature-dependent carrier density, the spin susceptibility is dominated by the Cu moments (at least for $p\lesssim0.2$).  Perhaps the clearest demonstration of this is from anisotropic susceptibility measurements on single crystals of LBCO \cite{huck08}.  }

It happens that $R_{\rm H}$ is also temperature dependent \cite{hwan94,ando04}, and it can be scaled in terms of a temperature $T^*$, which, within error bars, is consistent with $T_{\rm max}$.  One can estimate the hole concentration using $p_{\rm H} = 1/(R_{\rm H}ec)$, and the temperature dependence of $R_{\rm H}$ suggests that $p_{\rm H}$ decreases on cooling, as shown in Fig.~\ref{fg:Hallchi}(a).  There are no sharp changes in the $T$ dependence of $p_{\rm H}$ near $T^*$ (estimated as 410~K and 535~K for $x=0.17$ and 0.14, respectively \cite{hwan94}], so that $T^*$ should not be viewed as a  phase transition.  Instead, Gor'kov and Teitel'baum \cite{gork06} found that the temperature dependence of $R_{\rm H}(T)$ is well described by a thermally-activated carrier density.  The excitation gap has a magnitude and doping dependence similar to that observed for the mid-infrared conductivity, shown in Fig.~\ref{fg:mYBCO}.

In Fig.~\ref{fg:Hallchi}(a), $p_{\rm H}$ has been normalized to the dopant concentration $x$.  We see that $p_{\rm H}\approx x$ for small $x$, but ratio $p_{\rm H}/x$ grows as $x$ reaches optimal doping.  This is consistent with the doping dependence of the Drude weight shown in Fig.~\ref{fg:p}.  Studies of optical conductivity vs.\ temperature in underdoped LSCO show that the frequency-dependent conductivity becomes completely incoherent at high temperature \cite{take02b,take03}), so that the $T$ dependence of $p_{\rm H}/x$ reflects a crossover from a larger density of incoherent carriers at high $T$ to a smaller density of quasiparticles on cooling.  In terms of $\chi$, we have to remember that strong AF correlations result in a small $\chi$.  Hence, we see that, as AF correlations grow on cooling, the effective carrier concentration decreases.  With doping, the strength of AF correlations weakens, and $p_{\rm H}/x$ grows.

\subsection{Pairing and Coherence}

We know from experiment that the superconducting state involves pairing of holes \cite{goug87}, and 
there is broad acceptance that the superconducting gap of optimally-doped cuprates has $d$-wave symmetry.  The amplitude of the gap, with zeros at the nodal wave vectors and maxima in the antinodal regions, was first detected in ARPES studies \cite{shen93,ding96c}.  The phase factor, corresponding to a sign change across each node, was determined with various ingenious Josephson junction devices \cite{vanh95,tsue00}.

The gap symmetry differs from the $s$-wave gap of BCS theory \cite{bard57}, the latter being a consequence of an attractive interaction between quasiparticles associated with exchange of a phonon.  The possibility of replacing the phonon with an AF magnon leads to a repulsive interaction that can still lead to pairing if the gap has the $d$-wave form \cite{scal86,mont91,mori90}.  The challenges for applying this concept to cuprates are that 1) we do not have a Fermi liquid with sharp quasiparticles all around the nominal Fermi surface \cite{norm98}, and 2) we do not have sharp AF paramagnons to be exchanged \cite{fuji12a}.

\begin{figure}
 \centering
    \includegraphics[width=9cm]{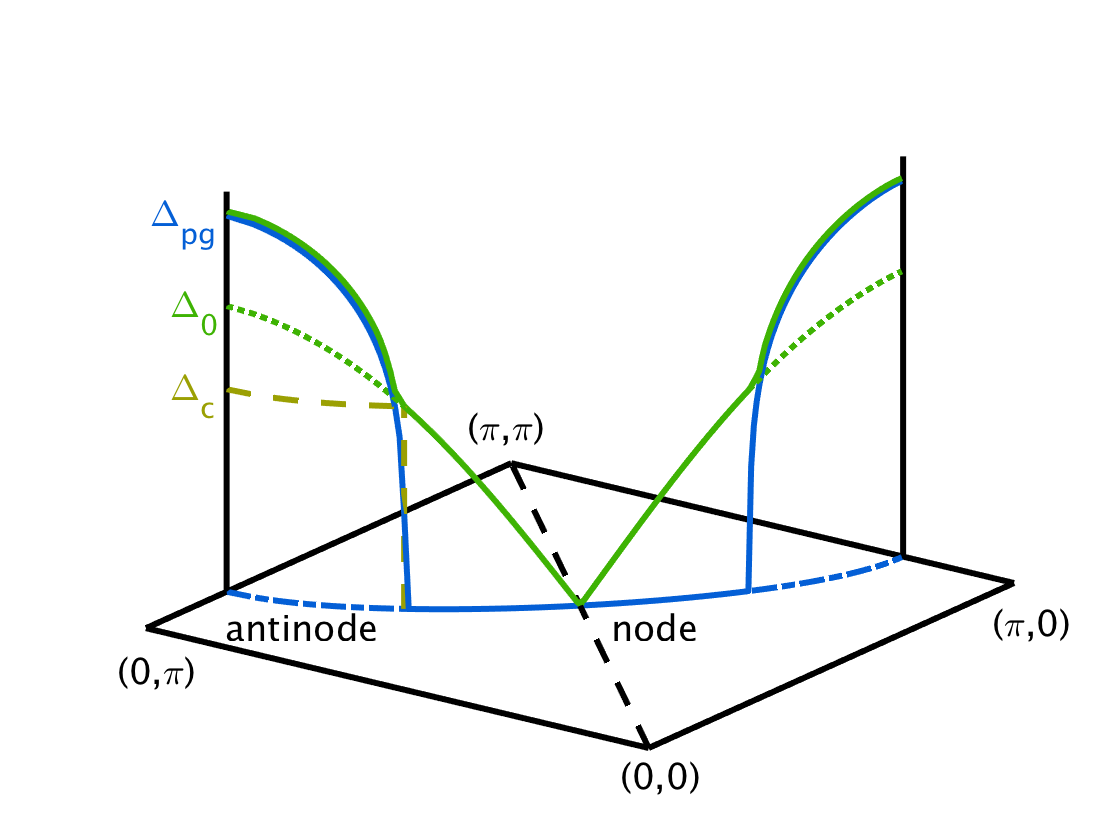}
    \caption{\label{fg:dc} Schematic of electronic gaps around a quadrant of reciprocal space for a square CuO$_2$ plane with lattice constant $a=1$.  Green line: below $T_c$, a $d$-wave gap is found near the node, extrapolating to $\Delta_0$.  For underdoped samples, the dispersion observed by ARPES tends to deviate upwards in the antinodal region \cite{vish12}, rising to the pseudogap energy $\Delta_{\rm pg}$.  Blue line: above $T_c$, the coherent gap (with maximum energy of $\Delta_c$) closes to form a Fermi arc. }
\end{figure}

A simple picture of a $d$-wave gap is also challenged by ARPES measurements below $T_c$ in underdoped cuprates.  As indicated in Fig.~\ref{fg:dc}, the proper $d$-wave gap shape is only observed over a finite arc of the nominal Fermi surface \cite{lee07,kond10}, up to a coherent gap scale $\Delta_c$, with the spectral function broadening and deviating at higher energies \cite{tana06,vish12,anza13}.  The scale $\Delta_c\sim 6k_{\rm B}T$ has also been identified in Raman scattering studies \cite{munn11,sacu13}.  An interpretation of this behavior will be discussed in Sec.~\ref{sc:spatial}.  An important piece of the story is the charge disorder already indicated by the variation among STM conductance curves on the same sample shown in Fig.~\ref{fg:cai}.

\section{Stripes}
\label{sc:stripes}

The most extreme form of intertwined orders occurs with stripe order.  Charge and spin stripes provide an energetic compromise between the competing superexchange and kinetic energies, as we will discuss here.  In related oxides, stripe order can result in an insulating ground state, as occurs with \lsno.  While it is true that stripe order in the cuprates tends to frustrate bulk superconductivity, there is good evidence that pairing, along with a distinct superconducting order, occurs.  Furthermore, ordered stripes provide a simple way to connect pairing with a model system that is better understood, both in terms of theory and experiment.  Using a consistent interpretation of the incompatibility of the incommensurate AF spin correlations and spatially-uniform superconducting coherence then leads to a systematic explanation of the correlation between the AF spin gap and the coherent superconducting gap.

\subsection{Stripe order}

Charge order turns out to be quite common among underdoped cuprates \cite{comi16,fran20}; however, it frequently comes along with a substantial gap in the spin excitations \cite{huck14}.  The focus here is on combined charge and spin stripe orders.  The best examples occur in \lbco\ \cite{fuji04,huck11}, \lnsco\ \cite{tran95a,ichi00}, and \lesco\ \cite{klau00,fink11}.  The maximum amplitude of stripe order occurs at $x\approx1/8$ \cite{huck11}, corresponding to a minimum in the bulk superconducting transition \cite{mood88,craw91}.  The orientation of the charge-stripe order is pinned by the structural anisotropy of a crystallographic phase that is stable only at low temperature \cite{axe94,tran95a}.

The spin order is readily detected by neutron scattering as incommensurate peaks split about the antiferromagnetic wave vector, ${\bf Q}_{\rm AF}$; in terms of reciprocal lattice units for a square CuO$_2$ plane, the spin-stripe peaks appear at $(0.5\pm\delta,0.5)$ and $(0.5,0.5\pm\delta)$, due to the presence of orthogonal stripe domains in nearest-neighbor layers.\footnote{For an extended discussion of scattering measurements of stripe order, see \cite{tran13a}}  For $p=x=1/8$, $\delta\approx1/8$.  The charge order peaks appear about fundamental Bragg peaks, split by $2\delta$.  They can be inferred from neutron scattering measurements, but the signal in that case is due to modulation of the lattice in response to the charge stripes, since neutrons do not couple to charge.  The modulation of the O $2p$ hole density is directly detected by resonant soft-x-ray scattering measurements at the O $K$ edge \cite{abba05}.  It should be noted that the spin-stripe order develops at a temperature below that of the charge order.

\begin{figure}
 \centering
    \includegraphics[width=10cm]{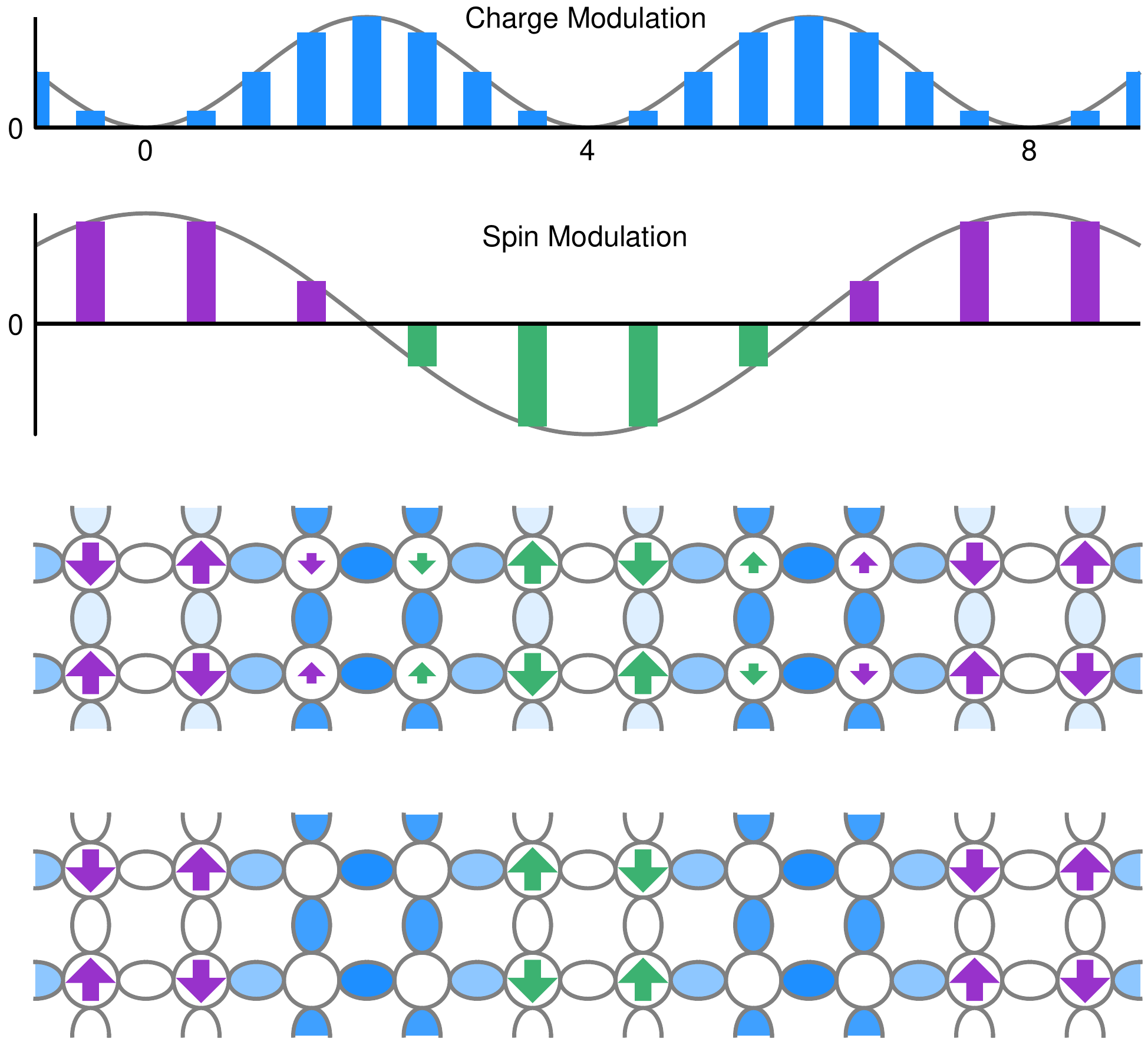}
    \caption{\label{fg:stripe} Bottom shows a cartoon of bond-centered charge and spin stripes as alternating doped and undoped 2-leg ladders.  Arrows indicate the size of the ordered magnetic moments on Cu atoms (circles), with color changing between antiphase domains.  Blue intensity indicates hole density (with white $=0$) on O sites (ellipses).  Panel above shows what such stripes would look like in the case of purely sinusoidal modulations, with the modulations of the charge and spins indicated in the top two panels; the charge amplitude (indicating hole density) is always positive, in contrast to the spin modulation, which changes sign. % (Only a single harmonic is detected experimentally for the charge and spin modulations, so there is no evidence for modulations beyond simple sinusoidal forms.)  
    }
\end{figure}

Based on the scattering measurements, we infer a picture of combined spin and charge stripe orders\footnote{\newr  There were early theoretical considerations of a spin-spiral order in a mean-field description of the hole-doped antiferromagnet \cite{shra89,kane90,sush05}.  Such a spiral might, in principle, yield the incommensurate magnetic peaks seen with neutrons; however, it was soon shown that the spiral state is unstable to charge modulation \cite{auer91,arri91,zhou95}, which is confirmed by the direct detection of charge-stripe order.} as shown in Fig.~\ref{fg:stripe}.  The period in real space is inversely proportional to the modulation wave vector, which means that the period of the charge order ($\sim4a$ for $p=1/8$) is half that of the spin stripes.  Experiments have not yet determined the registry of the stripes with respect to the lattice; in the figure the charge stripes are suggestively shown as bond-centered,\footnote{The sinusoidal modulation shown is consistent with the experimental observation of a single modulation wave vector.  Deviations from sinusoidal require higher harmonics, with consequent higher-harmonic superlattice peaks.  Examples of stripe order with multiple harmonics occur in the case of La$_2$NiO$_{4.133}$ \cite{woch98}.  As the diffraction intensity is proportional to the square of the amplitude, the relative intensities tend to be quite weak, at best, and would be reduced by disorder and fluctuations.  In any case, bond-centered stripes should minimize harmonic content.} for reasons that will be discussed shortly.   The locally-antiferromagnetic spin stripes flip their phase on crossing a charge stripe, resulting in a doubled period of $8a$.  (The phase difference between the charge and spin modulations is based on theoretical predictions, discussed below.)

{\newr
Real-space images showing atomic-scale modulations of the local density of states consistent with charge-stripe order have been obtained by STM on underdoped Bi2212 and Na-doped Ca$_2$CuO$_2$Cl$_2$ \cite{kohs07}.  The results look quite similar to local conductance curves obtained on LBCO (see Supplementary Online Material for \cite{vall06}).  The doping dependence of the stripe-like modulations in Bi2212 has also been reported \cite{park10}.  A theoretical analysis has modeled these modulations in terms of pair density waves \cite{chou20}, a concept that we will discuss in Sec.~\ref{sc:sc_stripes}.
}  

\begin{figure}
 \centering
    \includegraphics[width=5cm]{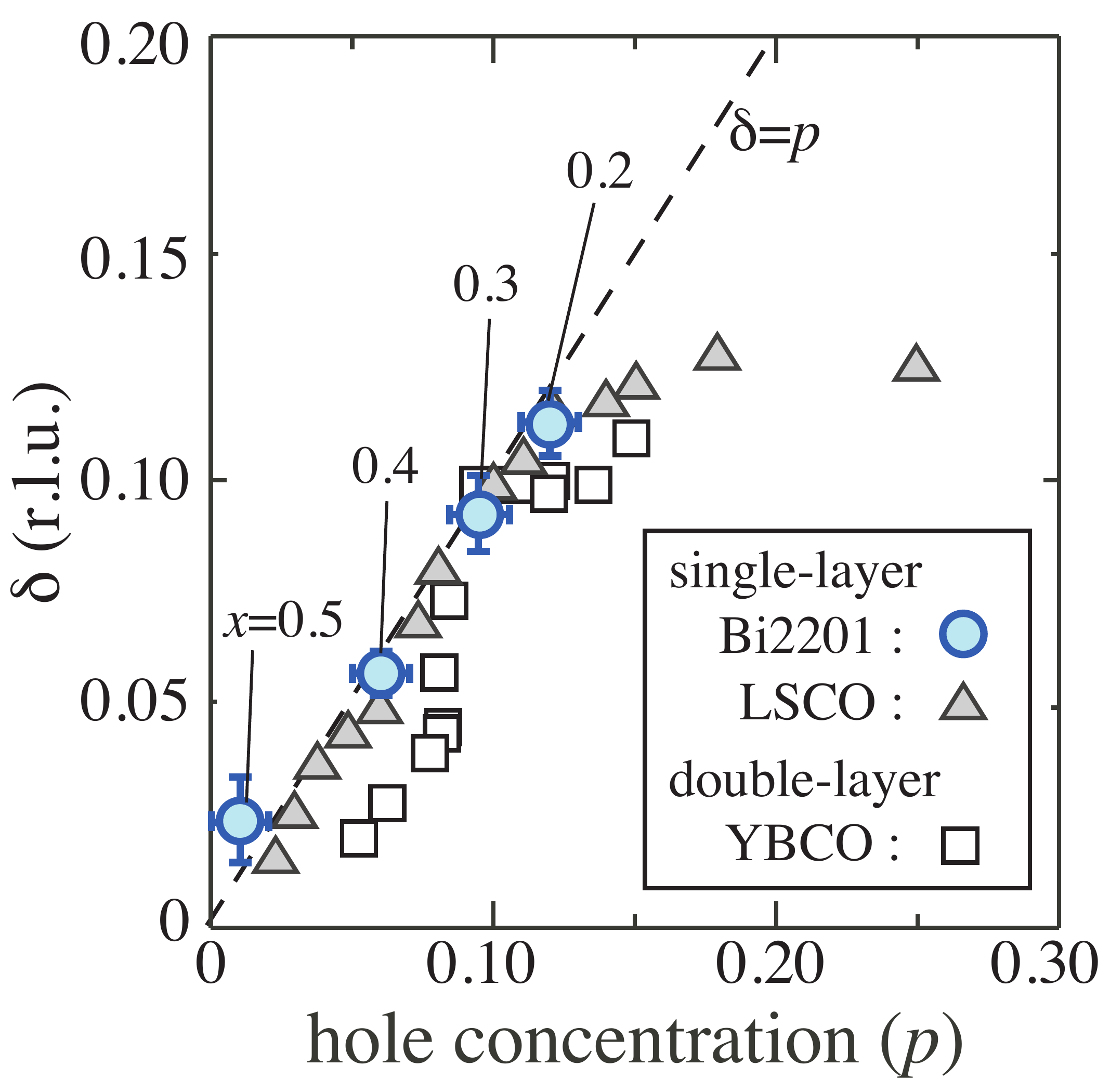}
    \caption{\label{fg:eno}  Hole concentration dependence of the incommensurability $\delta$ of low-energy spin fluctuations in Bi$_{2+x}$Sr$_{2-x}$CuO$_{6+y}$ (blue circles) compared with results for LSCO (gray triangles) \cite{yama98,waki00,fuji02c} and YBCO (open squares) \cite{dai01,haug10} (with $p$ estimated from $T_c$ via \cite{lian06}).  The dashed line represents $\delta=p$.  Reprinted with permission from \cite{enok13}, \copyright2013 by the American Physical Society.}
\end{figure}

The stripe order evolves with doping: for $p<1/8$, the incommensurability follows $\delta\approx p$, while $\delta$ saturates at higher doping \cite{birg06,huck11}.  Charge-stripe correlations of weaker amplitude are also observed in LSCO over a broad range of doping \cite{crof14,wen19,miao20}.  The occurrence in LSCO has been rationalized by detection of a reduced crystallographic symmetry, favorable to stripe pinning \cite{jaco15}.  Evidence of the spin-stripe order is also provided by neutron scattering measurements of the low-energy inelastic magnetic excitations that rise from the wave vectors of the spin-stripe order \cite{birg06,yama98}; these excitations are gapless (in zero magnetic field) for $x\lesssim0.13$ \cite{chan08}.  At low doping, where LSCO becomes insulating ($x<0.06$), the stripes rotate from bond-parallel to running diagonally with respect to the Cu plaquettes \cite{birg06}.{\newr\footnote{\newr For LSCO, commensurate order disappears at $x=0.02$, with neutron scattering detecting only incommensurate peaks at low energy for $x>0.02$ \cite{fuji02d}.  Optical conductivity measurements on detwinned crystals with $x=0.03$ and 0.04 suggest that the spin order is accompanied by modulated charge order \cite{dumm03}.  For $x<0.02$, phase separation occurs at $T\lesssim 30$~K, with loss of commensurate intensity and corresponding appearance of incommensurate peaks characteristic of $x=0.02$ \cite{mats02}.}}  The incommensurability $\delta$ measured in LSCO is plotted (gray triangles) for a large range of doping in Fig.~\ref{fg:eno}.  Static or dynamic spin stripes are also observed over a significant doping range in Bi$_{2+x}$Sr$_{2-x}$CuO$_{6+y}$ (blue circles) with the same $\delta\approx p$ relationship \cite{enok13}.  For YBCO, looking at the lowest-energy spin excitations leads to a similar trend, even though a substantial spin gap develops for $p\gtrsim0.09$ \cite{dai01,haug10,hink10}.{\newr\footnote{\newr For \ybco, there is a jump in the hole concentration in the planes from $p\lesssim0.02$ to $\sim0.05$ at the tetragonal-to-orthorhombic transition which occurs at $x\approx0.3$ \cite{bozi16}.  Hence, magnetic incommensurability appears \cite{hink08} as soon as there is a meaningful hole density in the planes.}}

The existence of spin and charge stripes in hole-doped cuprates is well supported by theory \cite{vojt09}.  In fact, stripes were initially predicted in calculations on the Hubbard model using the Hartree-Fock approximation \cite{zaan89,schu89,poil89,kato90}; however, these mean-field calculations yielded insulating charge stripes.\footnote{The insulating stripes have an increased hole density (one per lattice spacing along the length of a stripe) compared to metallic stripes, with a corresponding increase in the stripe period.  In particular, the charge-stripe period predicted by Hartree-Fock is inconsistent with experiment.}  An alternative approach, motivated by evidence for phase separation in the hole-doped $t$-$J$ model\footnote{If one starts with a single-band Hubbard model, calculations at large $U/t$ indicate that it is very unlikely that two electrons will sit on the same site, while electron spins on neighboring sites are coupled by $J=4t^2/U$.  An approximate version of this, called the $t$-$J$ model, keeps the hopping kinetic energy $t$, includes the superexchange $J$, and excludes any states with electron double-occupancy \cite{zhan88}.  To tune the calculated electronic structure near the chemical potential, variants may include next- or next-next-nearest-neighbor hoppings $t'$ and $t''$, respectively.} \cite{emer90}, considered the role of the extended Coulomb interaction, which tends to frustrate the phase separation \cite{emer93}.  Monte Carlo calculations on an effective model yielded both stripe and checkerboard phases \cite{low94}; however, such modeling was not able to address the electronic character of such phases.

A new approach of numerical variational techniques applied to correlated-electron Hamiltonians allowing general inhomogeneous solutions began with the application of the density matrix renormalization group (DMRG) by White and Scalapino \cite{whit98a}.  Evaluating the $t$-$J$ model on finite clusters at $p=1/8$, they found evidence for metallic charge stripes and antiphase spin stripes with a form very similar to that shown in Fig.~\ref{fg:stripe}.   Extending the calculations to other hole concentrations provided theoretical evidence for vertical charge and spin stripes over a large range of $p$ \cite{whit98c}.

A variety of numerical techniques appropriate to this problem have now been developed.  A comparison study applying 4 techniques to the single-band Hubbard model (with only nearest-neighbor hopping) with $U/t=8$ and $p=1/8$ found that vertical filled (insulating) stripes have the lowest energy, but partially filled stripes are close in energy, followed by diagonal stripes and uniform $d$-wave order \cite{zhen17}.  The key messages are that the various techniques yielded consistent results and a variety of inhomogeneous states are close in energy.  For a model that yields mean-field electronic structure that is close to experimental results of ARPES studies \cite{dama03}, it is important to include longer-range hoppings \cite{ande95}.  Recent DMRG studies of the Hubbard model with a broad range of $t'$ and $p$ found that vertical stripes are the ground state for plausible values of $t'$ \cite{jian19,jian20}.  Monte-Carlo results for a three-band spin-fermion model yield half-filled, bond-centered stripes \cite{huss19}.

{\newr
Given all of the computational evidence for stripes in variations of the Hubbard model with intermediate coupling, is the concept of electronic phase separation (and the role of frustration by extended Coulomb interactions) ruled out?  That is an interesting question, but it is too early to say.  The numerical calculations are generally done on systems with anisotropic boundary conditions (periodic in one direction, open in the other), and that anisotropy tends to favor unidirectional modulations.  Calculations on symmetric clusters tend to involve a limited number of sites.  Two very recent reviews on the Hubbard model conclude that there is evidence for various competing states, but that more work is needed to reach a consensus on the ground state for a given set of parameters \cite{arov21,qin21}.  Of course, that still leaves the open question of whether that ground state gives a good description of the real materials.
}

\subsection{Fluctuating stripes}

Static stripes are the exception rather than the rule.  Experimentally, they are associated with a crystalline phase in which the rotational symmetry is reduced from 4-fold to 2-fold \cite{tran95a,jaco15}.  Similarly, the calculations that yield stripes are typically done on clusters with anisotropic boundary conditions, which impact the results \cite{whit98a}.  When rotational symmetry is restored, the stripe correlations in experimental systems tend to be purely dynamical.

A good example is LBCO with $x=1/8$.  Static stripe order is absent above the structural transition at 56~K, but the magnetic excitation spectrum remains virtually unchanged across the transition \cite{xu07}, except at low energies ($\lesssim10$~meV), where the imaginary part of the {\bf Q}-integrated dynamical susceptibility changes from being roughly constant in the ordered state (as for antiferromagnetic spin waves) to varying linearly with $\hbar\omega$ in the disordered state \cite{fuji04}.  The low-energy incommensurability of the spin-stripe excitations remains well-resolved at 100~K, but the incommensurability decreases with temperature and evolves to a single broad peak at 200 K.  (A more complete study of temperature and frequency dependence of the low-energy spin response has been reported for LSCO $x=0.16$ \cite{aepp97}; for that composition, one obtains a spin gap, rather than spin order, at low temperature \cite{lake99,chan08}.)  The charge stripe fluctuations in the disordered phase have now been detected by resonant inelastic x-ray scattering at the Cu $L_3$ edge \cite{miao17}.\footnote{The incommensurability of the charge-stripe fluctuations appears to increase with temperature, opposite to the spin correlations \cite{miao17}.  This might be a consequence of the apparent strong sensitivity of the RIXS measurements to electron-phonon coupling \cite{miao19,peng20,chai17}, which, for the relevant Cu-O bond stretching response, increases greatly toward the Brillouin-zone boundary \cite{deve16}.}

The experimental evolution of the spin and charge correlations with temperature is qualitatively consistent with recent theoretical results.  Calculations on the one-band Hubbard model for $p=1/16$ using the minimally-entangled typical thermal states (METTS) method find well-defined spin and (half-filled) charge stripe correlations at low temperature, but a disordered inhomogeneous state at a significantly higher temperature \cite{wiet20}.  Calculations of the spin dynamics have been performed for a 3-band Hubbard model using the determinant quantum Monte Carlo (DQMC) technique \cite{huan17}.  Here the calculation is done at finite temperature, with the lowest temperature limited by the ``sign problem''; the minimum temperature achieved for $p=1/8$ is about $J/4$.  One can see damped AF spin-wave-like excitations at high energy, which have a broad {\bf Q}-width near the AF wave vector \cite{huan17}, consistent with the high-temperature neutron scattering results on LBCO $p=1/8$ \cite{fuji04}.

Returning to experiment, it is important to note that the thermal evolution of fluctuating charge and spin stripe correlations underlies the charge and spin responses probed by the transport, optical, and electron-spectroscopic techniques discussed in Sec.~\ref{sc:intertwined}.  As the stripe correlations grow, the charge and spin responses of the collective system develop some coherence.  The appearance of partially coherent responses is the consequence of cooperative organization.  

\subsection{Superconducting stripes}
\label{sc:sc_stripes}

Before we discuss a model of superconductivity based on stripes, we need to consider the evidence that charge stripes can be superconducting.  Much of the evidence has been reviewed previously \cite{frad15,agte20}, but a few words are appropriate here to provide a complete story.

The charge carriers and interactions that drive the superconductivity live within the CuO$_2$ planes.  To achieve 3D superconducting order, it is necessary to lock together the superconducting phases of the planes through interlayer Josephson coupling \cite{frie89}.  Once the superconducting correlations within the planes become substantial, a very small interlayer coupling is enough for the order to become 3D.  

If the interlayer Josephson coupling is absent or frustrated, then one might observe 2D superconductivity without 3D order.  That turns out to happen in LBCO $x=1/8$.  On cooling through the temperature where the spin-stripe order develops, the in-plane resistivity drops an order of magnitude and the planes develop a weak diamagnetism, while the $c$-axis resistivity remains large (and continues to grow with cooling); the susceptibility remains positive when a weak magnetic field is applied parallel to the planes, demonstrating the absence of Josephson currents between planes, which would be needed to shield the field \cite{li07,tran08}.  With further cooling, nonlinear voltage vs.\ current relations provide evidence \cite{li07} for a 2D ordering of the superconducting phase through a Kosterliz-Thouless transition \cite{kost73}.  Only at a temperature of $\lesssim5$~K (which is quite sensitive to sample composition) does 3D bulk diamagnetism develop.

\begin{figure}
 \centering
    \includegraphics[width=10cm]{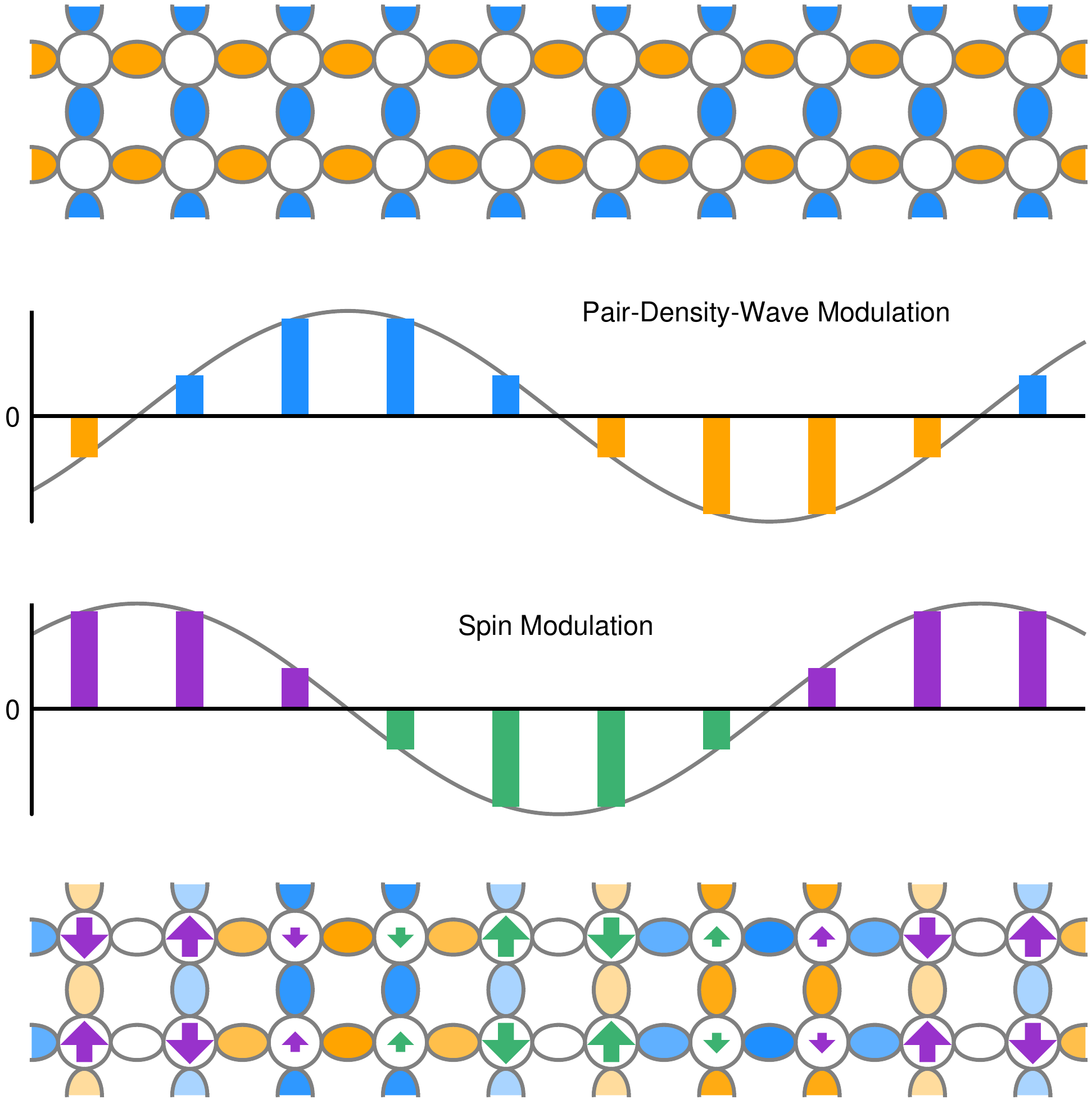}
    \caption{\label{fg:pdw} Top: schematic of uniform $d$-wave pair wave function, emphasizing that the hole density is largely on the oxygen sites and the pair wave function changes sign when rotated by $90^\circ$ (blue = positive, orange = negative).  Bottom: schematic pair wave function of the pair-density-wave order, which has the same period as the magnetic order but with a shifted phase.  Middle panels show the modulations of the PDW state and the spin-stripe order. }
\end{figure}

This behavior has been rationalized through the proposal that pair-density-wave (PDW) superconducting order develops within the charge stripes \cite{berg07,berg09b}.  In this state, shown schematically in Fig.~\ref{fg:pdw}, the superconducting wave function oscillates from positive to negative in going from one charge stripe to the next; because the pinning of the charge stripes by the lattice anisotropy rotates $90^\circ$ from one layer to the next, the interlayer Josephson coupling is frustrated.  This frustration inhibits the development of 3D superconducting order, but it puts no restriction on 2D order.

\begin{figure}
 \centering
    \includegraphics[width=6cm]{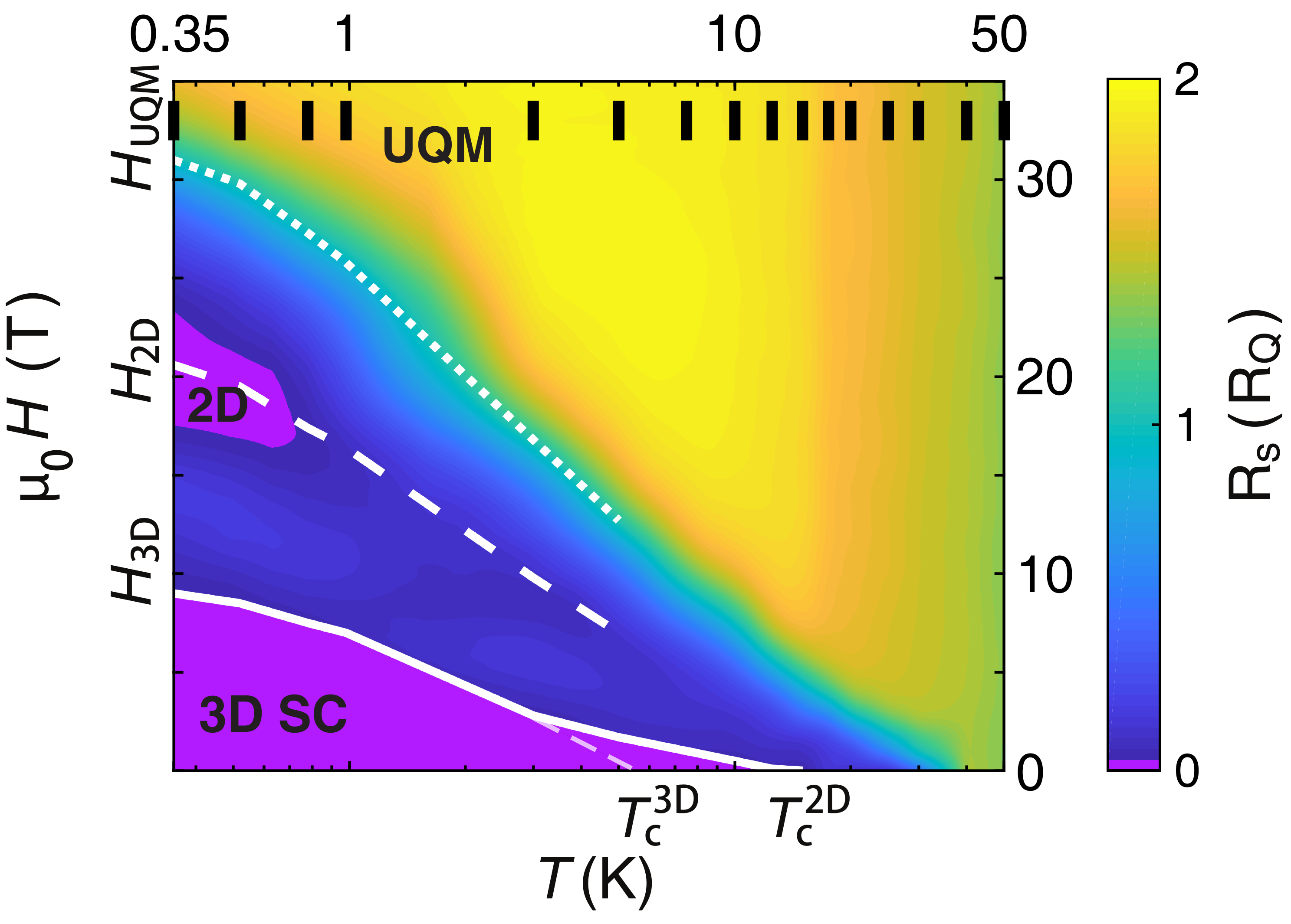}
    \caption{\label{fg:HF} Phase diagram of LBCO $x=0.125$ in terms of sheet resistance. Interpolated color contour plot of the sheet resistance $R_s=\rho_{ab}/d$ (where $d$ is the interlayer spacing) as a function of temperature and magnetic field. Black vertical marks indicate measurement temperatures. The regimes of 3D and 2D superconductivity with zero electrical resistance are labeled; the ultra-quantum metal phase occurs at fields above the dotted line. Characteristic fields $H_{\rm 3D}$, $H_{\rm 2D}$, and $H_{\rm UQM}$ are over-plotted as solid, dashed, and dotted white lines, respectively.   From \cite{li19a}.}
\end{figure}

To further test this picture, LBCO $x=1/8$ has been probed in magnetic fields up to 35 T \cite{li19a}.  Figure~\ref{fg:HF} shows a color contour map of the in-plane resistivity after converting to sheet resistance, $R_s$, where the unit is the quantum of resistance for pairs, $R_{\rm Q}=h/(2e)^2$; note that the temperature is on a logarithmic scale, while the magnetic field scale is linear.  At zero field, one can see the drop in $R_s$ at $\sim40$~K, corresponding to 2D superconductivity with phase disorder, followed by the 2D phase ordering at $\sim16$~K.  At base temperature, increasing magnetic field causes the loss of 3D superconducting order at $H_{\rm 3D}\approx 10$~T, reentrant 2D superconductivity at $H_{\rm 2D}\approx20$~T, and then a rapid rise followed by saturation at $R_s\approx2R_{\rm Q}$ in an ultra-quantum-metal (UQM) phase.\footnote{The name ``ultra-quantum metal'' is meant to convey the idea that we have a metallic phase that cannot be understood within a semi-classical model.}  {\newr The loss of 3D order is presumably due to vortex motion; the reentrant 2D superconductivity is assumed to be of PDW character with a re-pinning of vortices, as charge-stripe order has not been observed to weaken in a magnetic field.} The Hall coefficient is essentially zero over the full range of fields for $T<16$~K, indicating particle-hole symmetry and suggesting that pair correlations may survive within the charge stripes even after suppression of superconducting phase order.  Allowing for disorder, this could be evidence of a Bose metal phase \cite{ren20}.  In any case, it suggests that the charge stripes do not interfere with pairing, but do strongly impact the superconducting phase order.

Related behavior is also found in LBCO $x=0.095$.  Here, stripe order is somewhat reduced in amplitude and bulk superconductivity appears at $T_c=32$~K.  Nevertheless, perturbing the system with either a $c$-axis magnetic field \cite{wen12b,steg13} or substitution of 1\%\ Zn \cite{wen12a,loza21} enhances the stripe order and induces decoupling of the superconducting layers.  Other cuprate systems that exhibit stripe order also show evidence for layer decoupling and PDW order \cite{yang13,frad15}, such as Nd-doped LSCO \cite{ding08}, Eu-doped LSCO \cite{shi20b}, and even LSCO at $x=0.125$ \cite{kapo19}.    In an alternative direction, the 2D superconductivity in LBCO $x=0.115$ can be pushed toward 3D order by suitable application of strain \cite{gugu20}.  Going beyond single-layer cuprates, decoupling of superconducting bilayers in a $c$-axis magnetic field has been observed in crystals of La$_{2-x}$Ca$_{1+x}$Cu$_2$O$_6$ with zero-field $T_c$ as high as 54~K \cite{zhon18}; stripe order has not been directly detected there, but the spin excitations remain gapless below $T_c$ \cite{schn19}.

Numerical calculations applying various techniques (variational Monte Carlo \cite{hime02,racz07,ido18},  DMRG \cite{whit09,huan18}, infinite projected-entangled pair states (iPEPS) \cite{corb14,pons19,li21b}) to $t$-$J$ and Hubbard models at 1/8 hole doping find that PDW order is one of several competitive states that are close in energy, with the lowest-energy state varying with values of parameters such as next-nearest-neighbor hopping \cite{hime02,corb14,ido18}.  It is also found that stripe order, especially spin-stripe order, tends to compete with uniform $d$-wave order \cite{whit15,pons19,li21b}.  Given that stripe and PDW orders occur within restricted doping ranges within certain cuprates, it should not be surprising that PDW order does not stick out as a robust ground state in such calculations.  It is important to keep in mind that the models treated in these calculations are rough approximations that neglect significant factors such as the poorly-screened Coulomb interactions between neighboring sites.  The fact that the PDW state is energetically competitive provides good support for the interpretations of experimental results discussed above.

\section{Stripes and pairing}
\label{sc:pairing}

So far, we have seen that stripe order occurs in some cuprates; it tends to compete with spatially-uniform $d$ wave superconductivity, but it is compatible with 2D superconductivity that likely corresponds to PDW order.  Unrestricted variational algorithms applied to simplified models of CuO$_2$ layers containing only repulsive interactions appear to support both stripe and superconducting correlations,\footnote{It is worth noting that it has been proven that one can obtain superconductivity from a model with only repulsive interactions \cite{ragh10}; however, the proof applies in the weak-coupling limit, where $T_c$ is quite small.  The remaining challenge is to show that one can get high transition temperatures from repulsive interactions.} but they leave unclear the relationship between PDW and uniform $d$-wave orders.  Furthermore, they provide no connection with the dynamic correlations that should be relevant to the pairing scale and transition temperature.\footnote{One can get dynamic correlations from quantum Monte Carlo (QMC) calculations \cite{huan17}, as we will consider below; however, it is challenging to extend QMC calculations down to temperatures comparable to $T_c$ because of the ``sign problem'' \cite{iazz16}. }  In this section, the goal is to extract some deeper insights from experiments.

In 1997, Emery, Kivelson, and Zachar (EKZ) proposed a model for cuprate superconductivity based on stripes \cite{emer97}.  They treated each charge stripe as a one-dimensional electron gas and noted that if a spin gap were present (as in a Luther-Emery liquid \cite{luth74}), it would act as an amplitude for electron pair correlations.  Superconducting phase order would require Josephson coupling between neighboring stripes.

A challenge with this picture is that it was assumed that the spin gap on the charge stripes would be transferred from a spin gap in the neighboring spin stripes.  That assumption fails in the case of LBCO $x=1/8$, where spin-stripe order (and the absence of a significant gap on spin stripes) coexists with 2D superconductivity \cite{tran08}.  In this section, we will consider a variation on the EKZ model. 
{\newr The case involves a number of logical arguments based on a combination of experimental observations and established theoretical results:

\vspace{4pt}
1) {\it The spin degrees of freedom on a charge stripe are topologically decoupled from antiphase spin stripes.}

\vspace{4pt}
\hspace{14pt}\parbox{13.9cm}{
The evidence for this comes from studies of \lsno\ at $x=1/3$, where the 1D spin excitations of the charge stripes are clearly identified by neutron scattering \cite{boot03b,merr19}.  This is discussed in Sec.~\ref{sc:lsno}.
}
\vspace{4pt}

2) {\it The spin excitations on charge stripes in LBCO $x=1/8$ correspond to those of hole-doped 2-leg spin ladders.}

\vspace{4pt}
\hspace{14pt}\parbox{13.9cm}{
In contrast to LSNO $x=1/3$, there is no evidence for gapless 1D spin excitations in LBCO $x=1/8$ \cite{tran04}.  It then follows from point 1 that the spin excitations on the charge stripes must be gapped.  In order to have a substantial spin gap, each charge stripe must behave like an even-legged ladder.  With a charge period of $4a$, the only option is a 2-leg ladder.  This is discussed in Sec.~\ref{sc:sense}
}
\vspace{4pt}

3) {\it Hole-doped 2-leg spin ladders are the source of pairing correlations.}

\vspace{4pt}
\hspace{14pt}\parbox{13.9cm}{
The connection between doped 2-leg spin ladders and pairing correlations is provided by widely-accepted theoretical results \cite{dago92,dago96}.  Application of these results to spin excitations determined by neutron scattering provides an estimate of the charge-stripe spin gap consistent with the observed spectrum (see Sec.~\ref{sc:sense}).  This spin gap is predicted to be an upper limit on the pairing scale within the charge stripes.  As discussed in Sec.~\ref{sc:disorder}, the identified charge-stripe spin-gap scale (corresponding to the neck of the ``hourglass'' magnetic excitation spectrum) is correlated with the antinodal pseudogap energy.
}
\vspace{4pt}

4) {\it There is a confinement/deconfinement transition for the holes at the spin-gap energy on the charge stripes.}

\vspace{4pt}
\hspace{14pt}\parbox{13.9cm}{
At low temperature, the holes are confined to the charge stripes along the transverse Cu-O direction for energies below the spin gap energy by the antiphase spin-stripe environment.  At energies above the charge-stripe spin gap, there is no experimental sign of the topological constraint.  Without the constraint, unpaired holes can hop into the spin-stripe regions, causing a strong damping of the locally-antiferromagnetic spin excitations. This conclusion is based on the variation of the {\bf Q} width of the magnetic excitations as a function of energy, as discussed in Sec.~\ref{sc:sense}.  
}
\vspace{4pt}

5) {\it Within the ordered stripe phase, long-range 2D superconducting correlations among charge stripes require static order in the spin stripes.}

\vspace{4pt}
\hspace{14pt}\parbox{13.9cm}{
The onset of 2D superconductivity in LBCO $x=1/8$ is coincident with the onset of spin-stripe order \cite{tran08}.  Even without long-range 2D phase order, 2D superconducting correlations must exist over a substantial length scale in order to display diamagnetism and strongly reduced resistivity \cite{li07}, so that pairing correlations among multiple charge stripes must be phase locked.  The arguments in favor of PDW order have already been discussed.   A key point here is that low-energy spin excitations compete with superconducting phase order.  An empirical corollary is the observation that the coherent superconducting gap for spatially-uniform superconductivity is limited by the size of the spin gap that develops on the spin stripes \cite{li18}, as discussed in Sec.~\ref{sc:spin_gap}.
}
\vspace{4pt}

The rest of this section presents the details behind my interpretation of stripe-based superconductivity relevant to LBCO $x=1/8$.  The extension of these ideas to the more common case of spatially-uniform superconductivity is described in the following section.
}

% To develop this approach, we will start in a surprising place: we look at the spin excitations of insulating stripes in \lsno.  It turns out that at $x=1/3$, the charge stripes exhibit 1D spin excitations that are decoupled from the 3D spin correlations of the neighboring spin stripes.  Applying this lesson to cuprates, we reconsider the nature of the spin excitations in LBCO $x=1/8$ \cite{tran04}, and note that there must be a large spin gap on the charge stripes.  This immediately suggests that we view a charge stripe as a hole-doped 2-leg spin ladder.  The propensity for such a component to exhibit a pairing gap comparable to the singlet-triplet gap of nearest-neighbor $S=1/2$ pairs has long been recognized \cite{dago92,dago96}.  

%The hole-doped 2-leg ladder has strong similarities to the Luther-Emery liquid, so this connects us back to the EKZ model.  Given that numerical studies indicate that spin-stripe order competes with uniform $d$-wave SC \cite{li21b}, the solution is to have antiphase Josephson coupling between neighboring stripes, resulting in PDW order.  To achieve the in-phase coupling necessary for uniform $d$-wave order, we need to gap the spin excitations in the spin stripes (and avoid static order).  As we will see, experiments indicate that the spin gap limits the coherent superconducting gap in cuprates with uniform $d$-wave SC \cite{li18}.

\subsection{Coupled and decoupled spin correlations}
\label{sc:lsno}

\lsno\ has the same structure as \lsco, but it differs electronically.  While it shares a tendency to develop stripe order \cite{tran13a}, the stripes run diagonally within the NiO$_2$ planes \cite{tran94a,ulbr12b} and there is a substantial charge excitation gap below the charge ordering temperature \cite{uchi91}.  The ordering temperatures for charge and spin stripes are maximized at $x=\frac13$ \cite{yosh00,cheo94}.   Associated with the quasi-3D spin stripe order are spin excitations with a strong 2D dispersion \cite{boot03a,woo05}.  In addition, Boothroyd and coworkers \cite{boot03b} observed spin excitations with a 1D dispersion that appear below $\sim 70$~K, far below the ordering temperature for the spin stripes.

\begin{figure}
 \centering
    \includegraphics[width=0.5\textwidth]{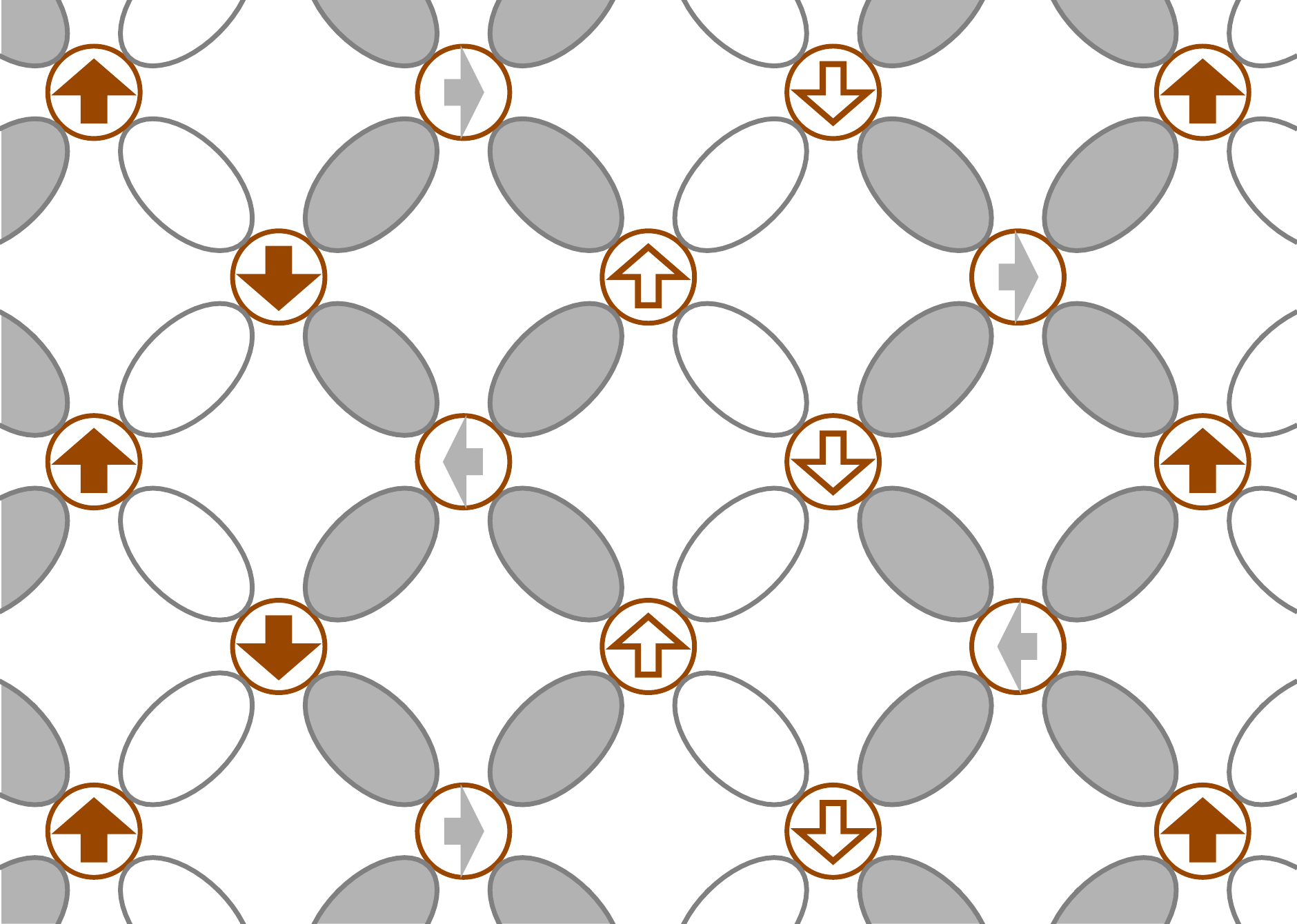}
    \caption{\label{fg:lsno}  Cartoon of the stripe order favored at low temperature in \lsno\ with $x=1/3$.  Circles represent Ni sites; ovals are bridging O; gray indicates doped holes.  Red arrows indicate $S=1$ moments in spin stripes; gray arrows correspond to low-spin $S=1/2$ moments resulting from a full Ni moment plus one hole on neighboring O sites.  Reprinted with permission from \cite{merr19}, \copyright2019 by the American Physical Society. }
\end{figure}

In a reinvestigation of the 1D spin excitations \cite{merr19}, it became apparent that they only made sense in terms of Ni site-centered charge stripes, illustrated schematically in Fig.~\ref{fg:lsno}, which become favored below a glass-like transition near 50 K.  Within the spin stripes, the Ni ions have $S=1$, with an anti-phase ordering across the charge stripes.  Within the charge stripes, there is one doped hole per Ni site.  Combining the Ni moment with the hole spin on neighboring O sites, it seems reasonable to assume a low-spin $S=1/2$ configuration.  The wave vector of the 1D spin excitations is consistent with AF correlations between Ni nearest neighbors along the charge stripe, with exchange that must occur through the intervening O atoms.  

From inspection of Fig.~\ref{fg:lsno}, one can see that the coupling of each Ni spin on a charge stripe to its neighbors in the spin stripes is geometrically frustrated.  At the same time, the 2D dispersion of the spin excitations associated with the spin stripes demonstrates coherent coupling of the antiphase spins across each charge stripe.  I cannot point to a solved model that describes this situation, but the observations essentially prove that the spin degrees of freedom within each charge stripe are decoupled from the spin stripes.  This is a lesson that we can apply to cuprates.

\subsection{Making sense of the cuprate magnetic excitation spectrum}
\label{sc:sense}

%Mean-field spin fluctuations around the stripe-ordered state: \cite{seib06}.  Problems: expect to see some dispersion away from Q-AF as well as towards it (but not see experimentally); excitations above E-cross are still sharp.  This works for diagonal stripes, where system is insulating.  For example, compare neutron results on LSCO $x=0.04$ \cite{mats13} with calculations \cite{seib14} 

\begin{figure}
 \centering
    \includegraphics[width=0.7\textwidth]{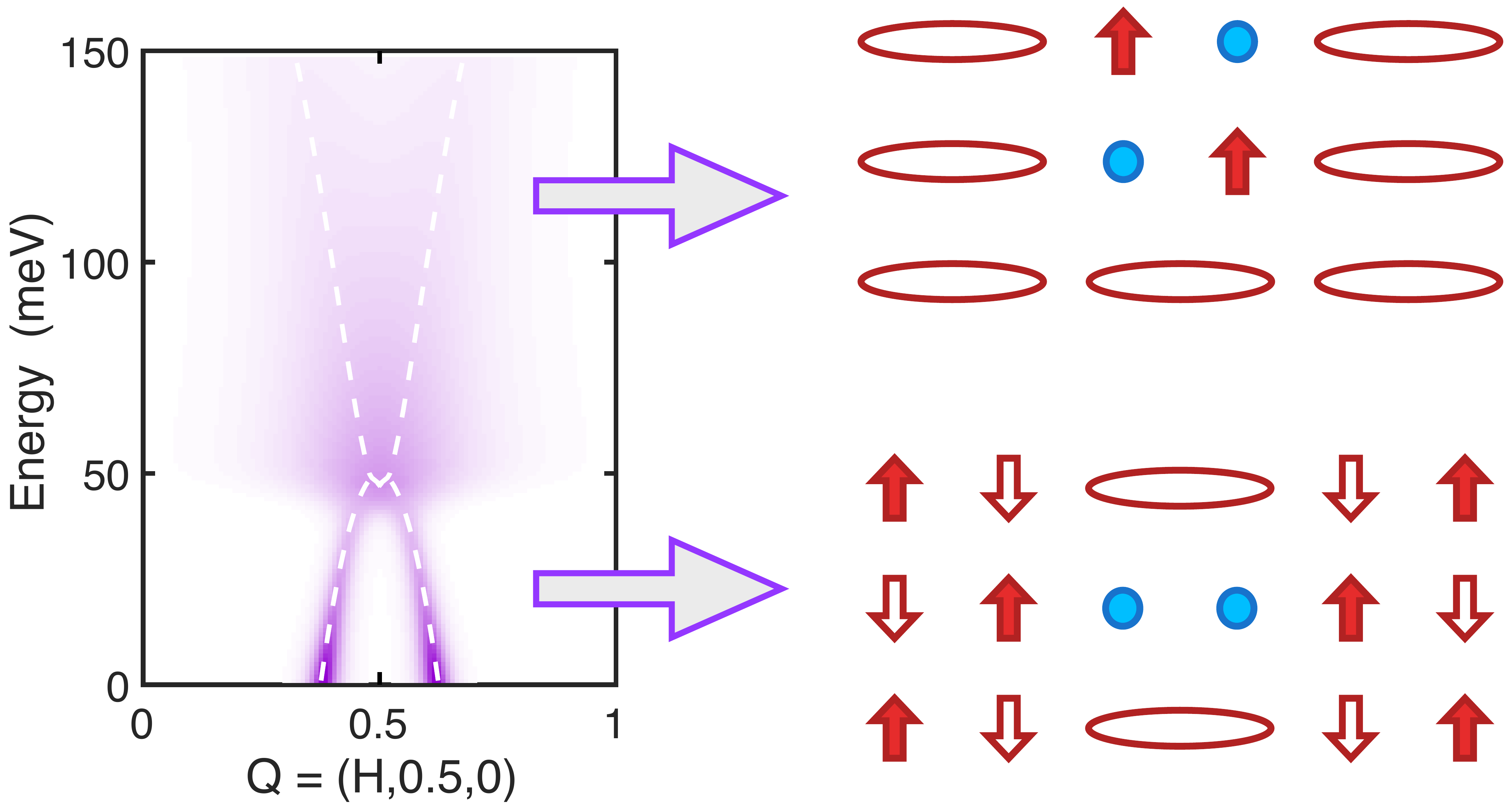}
    \caption{\label{fg:new}  Left panel shows the distribution of magnetic spectral weight, $\chi''({\bf Q},\omega)$, in LBCO \cite{tran04} and related LSCO samples \cite{vign07}. The excitations are sharp in {\bf Q} at low energy, but they are very broad at energies where commensurate scattering appears.  The white dashed line indicates the hourglass spectrum; the neck of the hourglass is labelled $E_{\rm cross}$.  The cartoons on the right indicate spin and charge correlations relevant to the observed spin dynamics, as discussed in the text.  Ellipses indicate spin singlets, blue circles indicate holes; only Cu sites are indicated.}
\end{figure}

The left-hand side of Fig.~\ref{fg:new} shows a schematic image of the magnetic spectral weight determined by neutron scattering experiments on LBCO $x=1/8$ \cite{tran04} and similar LSCO samples \cite{vign07}.  The dashed line indicates the approximate dispersion of peak features, which forms an ``hourglass'' spectrum \cite{arai99,fuji12a}, with a neck at energy $E_{\rm cross}$. 
{\newr Theoretical calculations of the spectrum are challenging.  Effective mean-field calculations based on a stripe-ordered ground state, which describe the spin excitations while the charge degrees of freedom are essentially frozen, can qualitatively describe the magnetic spectrum \cite{ande05,seib06,vojt06,yao06a,koni08,grei10,ande10}; however, there are important discrepancies.  At the lowest energies, the calculated excitations near the spin-stripe ordering wave vectors disperse both towards and away from ${\bf Q}_{\rm AF}$, whereas only the former excitations are detected experimentally \cite{tran04,chri04,lips09}.  The calculated excitations above $E_{\rm cross}$ are just as sharply defined as those below it, in contrast to experiment.  

Of course, there is crucial physics in the interactions of the holes and spins.  Techniques that can capture the excitations without constraints are promising, but they have their own challenges.  Determinant quantum Monte Carlo calculations yield intriguing images of diffuse magnetic excitations, but are presently limited to relatively high temperatures \cite{huan17}.  Zero-temperature calculations are possible, in principle, with the density-matrix renormalization group (DMRG) technique, but the challenge of keeping sufficient excited states in the calculation puts an effective limit on the number of atomic sites that can be handled, thus yielding a sampling of phase space that is far from definitive \cite{tohy20}.
}

% There is a natural bias to look for explanations of such a spectrum based on a spin-only model.  For a stripe ordered system, an obvious choice is to ignore the charge stripes and consider the excitations associated with ordered spin stripes.  Calculations based on coupled spin ladders can provide a spectrum comparable to the hourglass form \cite{yao06a,koni08,grei10}; however, the calculated excitations have comparable widths for energies above and below $E_{\rm cross}$.\footnote{An independent 2-component analysis of the spin excitations in LSCO, more in the spirit of the analysis below, is reported in \cite{sato20}.}

{\newr Returning to the schematic spectrum in Fig.~\ref{fg:new}, the excitations at $E<E_{\rm cross}$ are relatively sharp in {\bf Q}, but} those above $E_{\rm cross}$ are extremely broad.  In fact, integrating the excitations over all energies in LBCO $x=1/8$ gave a correlation length of $\sim4$~\AA\ \cite{xu07}, corresponding to one lattice spacing.   Furthermore, there is an increase in {\bf Q}-integrated magnetic spectral weight as energy increases across $E_{\rm cross}$ that is not captured by considering the spin stripes alone.
{\newr There appears to be a change in character across $E_{\rm cross}$.\footnote{An independent 2-component analysis of the spin excitations in LSCO, more in the spirit of the analysis below, is reported in \cite{sato20}.}}

%An alternative perspective is to 
{\newr To make sense of this, we can}
apply the lesson from LSNO regarding the decoupled spin degrees of freedom in charge stripes.  A difference here is that, if we take the charge stripes to be hole-doped 2-leg ladders as suggested by Fig.~\ref{fg:stripe}, then we expect the charge stripes to have a substantial spin gap.   For an undoped spin ladder with superexchange $J$  that is the same on legs and rungs, the gap is $\sim0.5J$ \cite{dago96}.  Various analyses find pairing correlations in doped $t$-$J$ \cite{dago92,troy96} and Hubbard \cite{noac94,tsve11,robi12,dolf15,gann20} 2-leg ladders.  In the limit of strong superexchange on the rungs, hole-pairing occurs in order to minimize disruption of the rung singlet correlations, and this tendency survives as one transforms towards isotropic exchange.  These analyses are backed up by experiment: neutron scattering studies of the undoped spin ladders in La$_4$Sr$_{10}$Cu$_{24}$O$_{41}$ show a spin gap of 26 meV \cite{notb07} that rises to 32 meV in Sr$_{14}$Cu$_{24}$O$_{41}$ and, under pressure, the doped ladders in Sr$_{0.4}$Ca$_{13.6}$Cu$_{24}$O$_{41.8}$ become superconducting \cite{ueha96}.

The cartoons on the right-hand side of Fig.~\ref{fg:new} indicate how we can apply these ideas to interpret the spectrum on the left.  At low energy, the incommensurability of the excitations indicates that they are fluctuations of the antiphase spin stripes{\newr; hopping of single holes transverse to the stripe direction is inhibited, effectively confining the holes to the charge stripes}.  %\footnote{Experimentally, the incommensurate spin excitations appear to disperse only inwards towards ${\bf Q}_{\rm AF}$ \cite{tran04,chri04,lips09}, whereas calculations commonly indicate a more symmetric dispersion \cite{seib06}.  One may hope that the resolution of this puzzle is connected with a proper understanding of the interstripe-couplings between spins and charges.}  
The absence of commensurate response is consistent with a spin gap on the charge stripes.  Within this picture, it is natural to associate the onset of commensurate scattering at $E_{\rm cross}$ with the singlet-triplet excitation energy of the doped ladders.  We can also associate this energy with the scale for pair-breaking within the charge stripes.

For $E\gtrsim E_{\rm cross}$, {\newr then, the holes are no longer paired, spin singlets can be excited to triplets, and we lose the self-consistent antiphase spin-stripe correlations that confined them.  With the deconfinement of the holes, the holes and spin-triplet excitations can disperse through the plane, much like the cartoon of the RVB model \cite{lee06}, as I have tried to communicate in the upper-right panel of Fig.~\ref{fg:new}.   The high-energy dispersion of spin excitations provides a measure of $J$, but with strong damping and renormalization due to the hole motion.}
%spin excitations can occur at any site in a CuO$_2$ plane.  On this energy scale, the holes are no longer confined to pairs and the spins in charge stripes are no longer confined to singlets.  Hence, when a neutron scatters and flips a Cu spin, that excitation can disperse through the AF correlations within a plane, but it will be damped by the unbound holes.   
Indeed, we already saw in Fig.~\ref{fg:mYBCO}(a) that the 2-magnon resonance is renormalized downward by $\sim1/3$ at $p\sim1/8$.\footnote{Figure~\ref{fg:mYBCO} shows results for YBCO; however, the Raman data show comparable results for other cuprates, including LSCO \cite{suga03}.}.   A fit to the peaks of the neutron spectra for LBCO $x=1/8$ gave $J\approx100$~meV, down by $\sim1/3$ from La$_2$CuO$_4$ \cite{cold01,head10}.  Applying this value of $J$ in the formula for the singlet-triplet gap of an isotropic 2-leg spin ladder gives 50 meV, consistent with the measured $E_{\rm cross}$.  The observation that all spins contribute to the excitations above $E_{\rm cross}$ (but not below) is consistent with the rise in magnetic spectral weight in that range \cite{tran04,vign07}.

In a spin-only model, achieving order in an array of 2-leg spin ladders requires a significant inter-ladder exchange coupling to overcome the large spin gap of the isolated ladders.  In the actual system (including charge stripes), there is also the energetics of the charge stripes to take into account.  Forming an antiphase order of the neighboring undoped ladders protects the charge stripe and enables a large spin gap on it.  Of course, one should also take account of the effects of spin-orbit coupling \cite{bone92}, which can help to stabilize ordered spins and which has measurable effects in terms of the Dzyaloshinskii-Moriya interaction and spin canting in La$_2$CuO$_4$ \cite{coff91,kast98}.  Electron-phonon coupling is another relevant factor that can contribute to stabilizing inhomogeneous phases \cite{han20}.

Now consider the facts that: 1) the onset of 2D superconductivity (and putative PDW order) in LBCO $x=1/8$ coincides with the static ordering of spin stripes \cite{tran08}, and 2) the strong depression of 3D superconductivity indicates a competition of uniform $d$-wave and PDW orders.  Both of these observations are consistent with the idea that local pairing coherence is incompatible with overlapping spin order or slowly fluctuating spins.\footnote{Evidence for incompatibility of SC and AF orders is also provided by a study of \lsco/\lco\  heterostructures \cite{bozo03}.}  As one can see from Fig.~\ref{fg:pdw}, the Josephson coupling between neighboring charge stripes must be antiphase so that the pair wave function can go to zero where the spin-stripe amplitude is maximum.  It is not enough to have the charge order already established, with spin stripe correlations defined and spin fluctuations virtually gapless; the spin directions have to be frozen before the PDW order develops \cite{tran08}.

In principle, uniform $d$-wave and PDW orders could locally coexist, in which case a charge modulation would be induced with the PDW period (a 1$q$ modulation compared to the 2$q$ of the observed charge-stripe order) \cite{berg09b}.  Such a $1q$ modulation in the effective density of states was measured by STM in magnetic vortex halos in near optimally-doped \bscco\ \cite{edki19}; however, that system has a large spin gap that is unlikely to be closed near the vortex cores \cite{xu09}.   In a system such as LBCO with spin-stripe order, however, careful diffraction searches for $1q$ order have been unsuccessful.  Any patches of uniform $d$-wave SC must occur where local deviations in charge density (to higher doping) correspond to gapped spin correlations.  As we will discuss below, the analysis of NMR measurements on LSCO by Imai and coworkers \cite{sing02a} provide evidence for a surprisingly broad distribution of local hole densities that is sufficient to support this concept.  Hence, the 3D superconductivity that develops in LBCO $x=1/8$ below 5~K likely results from coherence developing between dilute patches of uniform $d$-wave order.

{\newr
Before ending this section, let me note that the $E_{\rm cross}$ feature is often referred to as the ``resonance'' energy.  The latter comes from the weak-coupling interpretation of the magnetic excitations, in which they are assumed to be associated with excitations of electronic quasiparticles across the Fermi surface at wave vectors near ${\bf Q}_{\rm AF}$.  Such excitations should disappear below $T_c$ for $E < 2\Delta$, where $\Delta$ is the superconducting gap, unless interactions are able to create {\it resonant} excitations below $2\Delta$, as reviewed in \cite{esch06}.  In the case of near-optimally-doped \ybco, the commensurate excitation occurs at $\sim40$~meV and its intensity is enhanced below $T_c$ \cite{mook93} and parts of an hourglass-like spectrum are seen above and below this energy \cite{bour00}.  

The problem with this story is that a weak-coupling description is not capable of giving a consistent description of the many unusual behaviors exhibited by these materials.  As I tried to make clear in Sec.~\ref{sc:phenom}, the anomalous behaviors found in cuprates are the consequence of doping holes into a Mott-Hubbard charge-transfer insulator.  As I will discuss further in Sec.~\ref{sc:spin_gap}, the relevant scale for the magnetic excitations is not $2\Delta$, but simply $\Delta$.  The existence of magnetic correlations depends on states at high binding energy (due to a large $U$), not to states at the Fermi energy.  A consistent interpretation of neutron scattering studies of a variety of cuprates is that low-energy spin excitations are incompatible with spatially uniform superconducting coherence and must be gapped \cite{li18}.  Any magnetic excitations that exist below $\Delta$ in the normal state must be pushed above $\Delta$ in the superconducting state.  In YBCO, where the spin gap is just a bit below $E_{\rm cross}$, that weight shows up at $E_{\rm cross}$ below $T_c$.  In contrast, the spin gap in LSCO near optimal doping is far below $E_{\rm cross}$, and the enhanced intensity below $T_c$ occurs at incommensurate wave vectors \cite{tran04b,chri04}.  In a superconducting sample such as LBCO $x=0.095$, there is no obvious spin gap \cite{xu14}, consistent with expectations for a PDW superconductor \cite{chri16}.  When there is a spin gap in the normal state that is larger than the $\Delta$ that develops below $T_c$, as in underdoped Hg1201, then there is again no change in the magnetic spectral weight at $E_{\rm cross}$ \cite{chan16b}.
}

\section{Spatially-uniform superconductivity}
\label{sc:spatial}

We have seen how stripe order stabilizes doped spin ladders in which pairing is induced by the presence of a large singlet-triplet spin gap.  I have proposed that this energy corresponds to the experimental scale $E_{\rm cross}$ observed by neutron scattering, which should serve as an upper limit for the superconducting gap.  In the case of ordered stripes, the static spins inhibit uniform $d$-wave superconductivity.  But is this pairing mechanism unique to stripe-ordered samples?  Here we consider how a version of this story can also explain the behavior of cuprates with uniform $d$-wave superconductivity.

\subsection{Similar magnetic response in all cuprates}

Past reviews have already made the case that the dispersion of magnetic excitations is similar in all cuprate families where it has been studied \cite{fuji12a,tran07}.  The parent insulator phases all show commensurate antiferromagnetic order with $J$ in the range of 100--150~meV.  With hole doping, the spin waves of the antiferromagnetic phase evolve into strongly damped excitations with upward dispersion above a commensurate gap of $E_{\rm cross}$.\footnote{This behavior has been observed in LSCO \cite{vign07,mats17}, YBCO \cite{hink07,rezn08}, \bscco \cite{xu09}, HgBa$_2$CuO$_{4+\delta}$ \cite{chan16b,chan16c}, and La$_{2-x}$Ca$_{1+x}$Cu$_2$O$_6$ \cite{schn19}.}  The systematic softening of high-energy spin excitations and gradual reduction of spectral weight was clearly pointed out by Stock {\it et al.} \cite{stoc10}.  Below $E_{\rm cross}$, there are generally excitations that disperse downwards to varying degrees, the main exception being HgBa$_2$CuO$_{4+\delta}$, where there is simply a gap below $E_{\rm cross}$ \cite{chan16b}.\footnote{The dispersion in HgBa$_2$CuO$_{4+\delta}$ has been characterized as having the form of a ``wine glass'' rather than an hourglass \cite{chan16b}; however, given the large damping, the data appear roughly consistent with a parabolic upward dispersion where the lower part of the dispersion is not resolved due to the damping.  A downwardly-dispersing component that was resolvable at $T<T_c$ has now been observed in a sample with $T_c=88$~K \cite{chan16c}.  Such a change in damping across $T_c$ was previously observed in YBCO \cite{bour00,hink07}.}  We have already noted in Fig.~\ref{fg:eno} that the incommensurability of the lowest-energy spin excitations grows with $p$ in underdoped samples, before saturating.

\subsection{Role of the spin gap}
\label{sc:spin_gap}

To see how the superconducting order can change while the pairing mechanism remains the same, let {\newr us} consider LSCO.  This system exhibits at least partial charge and spin orders for $x\lesssim0.13$ \cite{chan08,jaco15}, and develops a gap in the incommensurate spin excitations for $x\gtrsim0.13$ \cite{lake99,khay05,chan08}.  Along with the loss of weight in $\chi''$ at low energy, there is a gain in weight at higher energy \cite{tran04b}.  There have been varying criteria used to define the spin gap, but it was argued in \cite{li18} that the appropriate criterion is to select the energy at which the weight transfer changes from negative to positive.  For LSCO $x=0.17$ and 0.21, it was observed that the spin gap is approximately equal to the coherent superconducting gap, $\Delta_c$ \cite{musc10,suga13}.

\begin{figure}
 \centering
    \includegraphics[width=7.5cm]{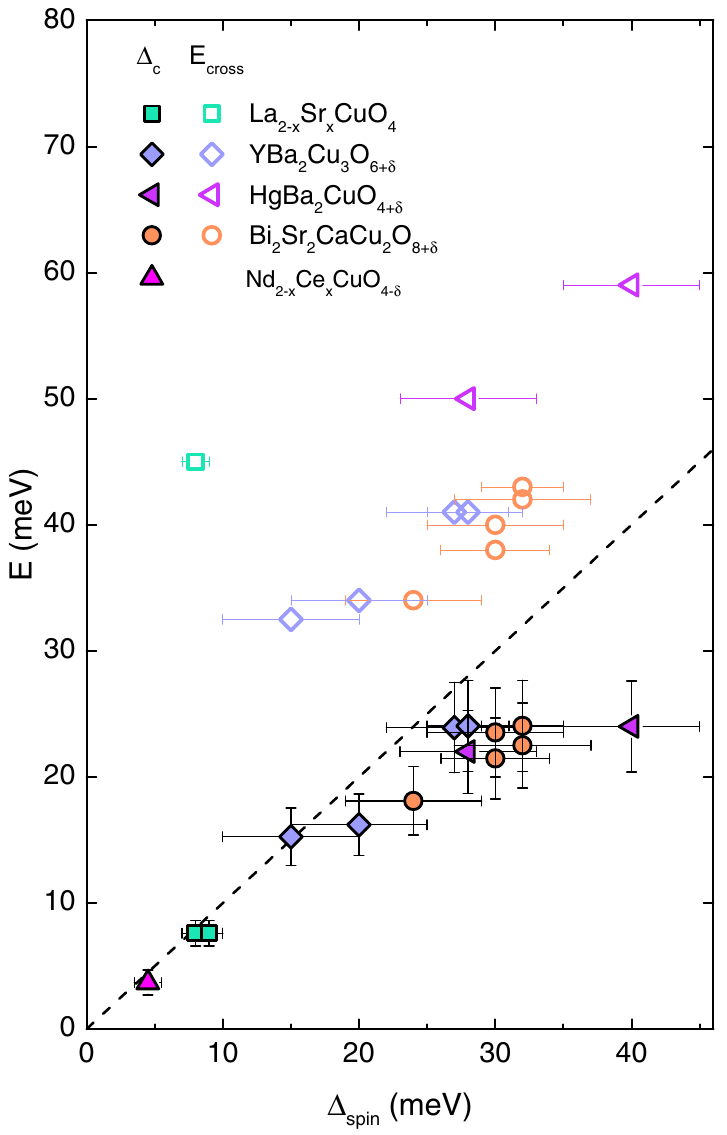}
    \caption{\label{fg:gaps}  Coherent superconducting gap, $\Delta_c$, and $E_{\rm cross}$ plotted vs.\ the spin gap in the superconducting state, $\Delta_{\rm spin}$ for a range of cuprates.   Dashed line indicates the case of $\Delta_c=\Delta_{\rm spin}$.  References to the original sources for data shown here are given in \cite{li18}.  Reprinted with permission from \cite{li18}, \copyright2018 by the American Physical Society. }
\end{figure}

Now, we have seen that stripe order locally frustrates coherent superconducting order, and spin fluctuations (in the absence of spin order) appear to inhibit the establishment of antiphase Josephson coupling between charge stripes.  It seems plausible that fluctuating spin stripes will also frustrate spatially-uniform $d$-wave order, except at energies where those fluctuations are gapped; this concept is supported by a recent analysis of contributions to pairing based on a phenomenological model for spin fluctuations in underdoped YBCO  \cite{dahm18}.  Indeed, experimental results are consistent with this picture.  Figure~\ref{fg:gaps}(a) compares values of the coherent SC gap $\Delta_c$, estimated from Raman spectroscopy, with the gap $\Delta_{\rm spin}$ determined by neutron scattering \cite{li18} for a number of cuprate families.  As one can see, $\Delta_c\le \Delta_{\rm spin}$.  Furthermore, $\Delta_c$ is always less than $E_{\rm cross}$ [plotted in Fig.~\ref{fg:gaps}(b)], which I have argued is the upper limit for pairing.

While optimally-doped LSCO has a similar normal-state carrier density to other optimally doped cuprates, it has a substantially reduced superfluid density, which correlates with its modest $T_c$ \cite{uemu89}.  We might expect all of the doped holes to be involved in pairing correlations; however, if the maximum energy scale for pairing correlations is $E_{\rm cross}$, and only pairs at energies below $\Delta_{\rm spin}$ can participate coherently in the long-range order, then it is quite understandable that the superfluid density is relatively small.  Furthermore, resonant soft x-ray scattering measurements on optimally-doped LSCO have demonstrated that charge-stripe correlations grow as the temperature is reduced towards $T_c$, but then decrease below $T_c$ \cite{wen19,miao20}.  This is consistent with the idea that the stripe correlations are good for pairing, but stripe order is bad for uniform superconductivity.  

For cuprates with a higher $T_c$, such as YBCO, Bi2212, and Hg1201, the spin gap near optimal doping tends to be much larger than in LSCO.  Within the stripe-motivated picture, a bigger spin gap means that a larger fraction of the doped holes can participate in the superfluid, as observed experimentally \cite{uemu89}.  Of course, with a bigger spin gap, one is further from the limit of spin-stripe order, so that the nature of the pairing mechanism is obscured.

It is also worth noting the impact that spin fluctuations have on the electronic self energy at energies above $\Delta_c$.  In considering the spin fluctuation spectrum in Fig.~\ref{fg:new}, we noted the impact of unpaired holes on the magnetic dispersion for $E>E_{\rm cross}$.  The converse is also significant.  Analysis of ARPES data on slightly underdoped Bi2212 show that, while the electronic scattering rate is small for energies below the SC gap, they jump to very large values for energies above the gap \cite{li18b}.  At antinodal wave vectors, the above-gap self energy is of order $2J$.  The large jump in the scattering rate is also seen in optical reflectivity studies, where the jump occurs at $\sim2E_{\rm cross}$ \cite{hwan07a}.

\subsection{Charge disorder and granular superconductivity}
\label{sc:disorder}

To understand why there is a coherent-superconducting-gap scale $\Delta_c$ that is smaller than the extrapolated $d$-wave gap maximum $\Delta_0$, we have to consider the role of charge disorder resulting from random positioning of dopants and poor screening of the long-range Coulomb interaction \cite{kres06}.  This is distinct from the form of inhomogeneity that we have already discussed: charge and spin stripes.
The (``quenched'') charge disorder forms the landscape within which stripe correlations develop, such that the stripe period depends on the local average hole density.  For now, we will focus on underdoped cuprates, and the overdoped case will be considered separately.  Besides explaining $\Delta_c$, charge disorder also provides a way to make sense of experimental observations of pairing correlations at $T\gtrsim T_c$.

While it can be easy to overlook, charge disorder in cuprates should not come as a surprise.  In the case of LSCO, the holes are introduced by partial substitution of ions that have a valence that differs by one from the host sublattice.  The holes are introduced into a correlated insulator state, which is inherently bad at screening Coulomb potentials.  In fact, the hole doped into the CuO$_2$ plane when La$^{3+}$ is replaced by Sr$^{2+}$ is the dominant channel for screening, which limits the range over which the hole can spread \cite{kast98}.  

Electronic disorder is certainly not unique to cuprates.  A new system that exhibits regimes of strong electronic correlations and superconductivity is magic-angle bilayer graphene, where impressive phase diagrams can be mapped out on a single sample by transport measurements as a function of carrier density, which is controlled by gate voltage \cite{cao18}.  Despite the quality of the transport data, nanoscale mapping of Landau levels with a scanning SQUID\footnote{SQUID = superconducting quantum interference device.} has imaged substantial disorder in the local twist angle \cite{uri20}.

Returning to cuprates, early evidence of a spatial variation of the hole concentration in \lsco\ was provided by an analysis of $^{63}$Cu nuclear quadrupole resonance \cite{sing02a} and $^{17}$O nuclear magnetic resonance \cite{sing05} measurements as a function of $x$.  For each sample, there appears to be a spread of environments that can be described in terms of a distribution of dopant concentations.  Hence, for $x\sim0.15$, the width of the distribution is $\Delta x\sim 0.05$ at low temperature, with a characteristic length scale of $\sim30$~\AA.  

In YBCO, the doping is achieved by adding O atoms to form Cu-O chains in layers intervening between CuO$_2$ bilayers.  X-ray scattering measurements with high spatial resolution ($300\times 300$~nm$^2$) on YBa$_2$Cu$_3$O$_{6.68}$ have demonstrated that the patches of ordered chains have a typical domain diameter of $\sim75$~\AA\ \cite{ricc13}, suggesting variations in the hole density transferred to the planes on this length scale.  The lack of effective charge screening is also indicated by a study of the dynamic charge susceptibility with momentum-resolved electron-energy-loss spectroscopy in Bi2212, where the plasmon features, expected for a conventional metal, are missing \cite{mitr18}.

We have already seen the evidence of substantial charge disorder provided by STM on a doped-but-insulating sample of Bi2201 shown in Fig.~\ref{fg:cai}.  This is just the latest evidence, as
from the earliest spectroscopic imaging with STM on Bi2212 at $T\ll T_c$, a common point of emphasis has been the spatial variation and disorder in apparent superconducting coherence peaks with energies in the range of 20--50~meV \cite{cren00,howa01,pan01}; such behavior is also seen in superconducting Bi2201 \cite{boye07}.  The scale of the disorder in the coherence peaks of Bi2212 is tens of \AA\ \cite{fang06}, but this is also the scale of the superconducting coherence length \cite{wang03b}.  This has led to proposals that the inhomogeneity is associated with variations in both the local hole concentration and the local pairing scale \cite{wang01,fang06}.   

When measurements were eventually performed across $T_c$, it was found that there is also a correlation between the size of the local gap and the temperature at which it closes, with larger gaps closing at higher temperatures that extend well above the bulk $T_c$ \cite{gome07,pasu08}.   The spread in gaps for $T\gtrsim T_c$ is also indicated in cuprates such as LSCO, YBCO, and Hg1201 by a study of nonlinear conductivity \cite{pelc18}.

\begin{figure}
 \centering
    \includegraphics[width=8.5cm]{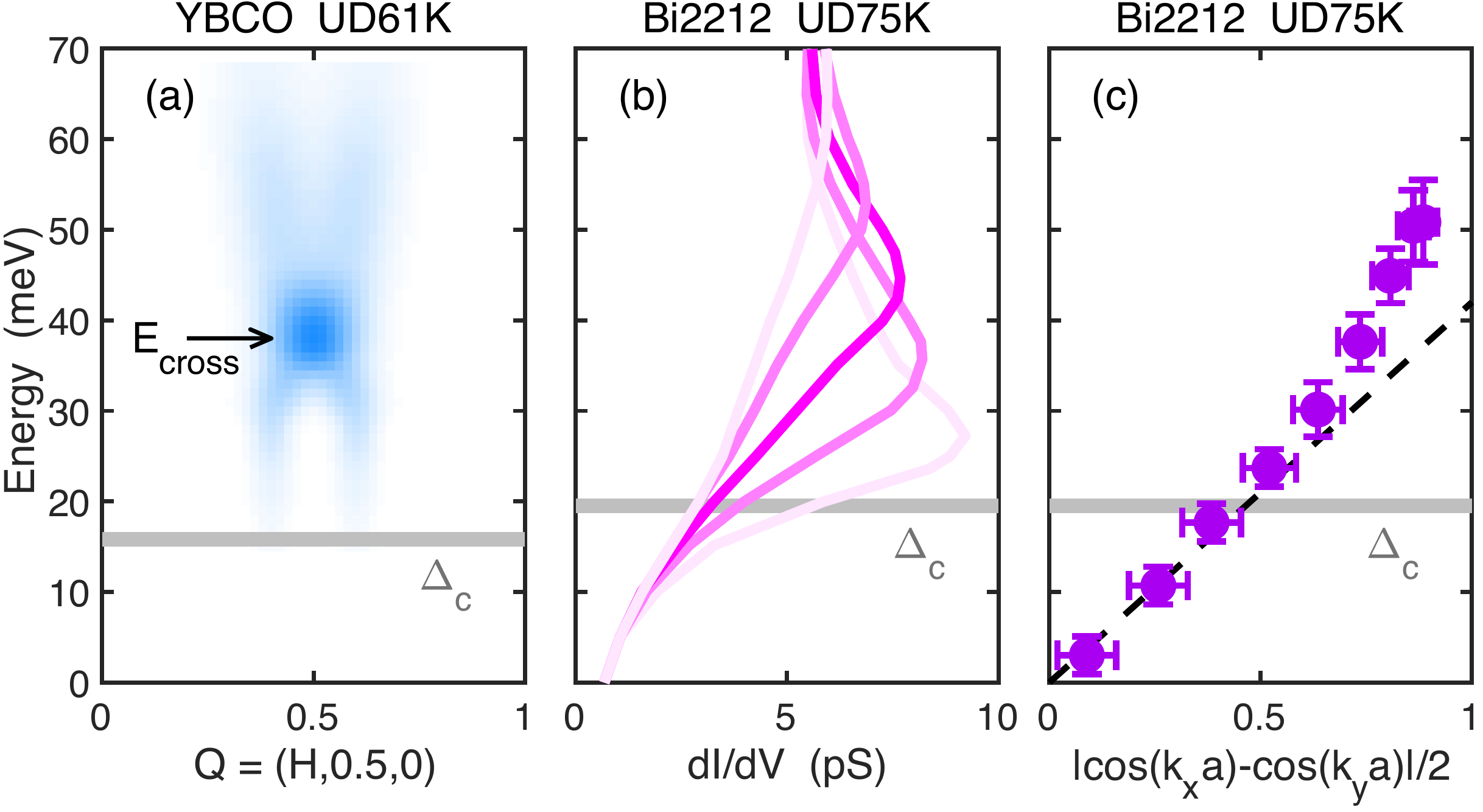}
    \caption{\label{fg:small}  (a) Magnetic spectral weight measured by neutron scattering in YBa$_2$Cu$_3$O$_{6.6}$ with $T_c=61$~K and $p\approx0.12$ \cite{hink04,hink07}.  (b) Bias voltage (binding energy) vs.\ conductance measured by STM for typical regions in Bi2212 with $T_c=75$~K and $p = 0.13(1)$ \cite{mcel05a}; shading reflects the relative frequency with which these spectra occur.  (c) Superconducting gap dispersion measured by ARPES at 10~K in a similar underdoped Bi2212 sample with $T_c=75$~K \cite{lee07}; dashed line indicates the simple $d$-wave gap form.  In each panel, the gray line indicates $\Delta_c=3kT_c$ based on Raman results \cite{sacu13,munn11}. }
\end{figure}

Figure~\ref{fg:small}(b) shows typical low-temperature conductance curves (for positive binding energy) from STM measurements at various locations on the surface of underdoped Bi2212 \cite{mcel05a}.  For a sample with $T_c=75$~K, the most frequently observed gap energy is 45~meV, with a broad spread of gaps observed around that value.  In contrast to the behavior of the coherence peaks, the conductance looks much more uniform at low energies \cite{push09,kuro10,sugi17}.  The gray line indicates the magnitude of the coherent gap $\Delta_c$ obtained from Raman scattering studies \cite{sacu13}.

Another measure of the superconducting gap is given by ARPES.  Deviations from the $d$-wave gap dispersion are observed in underdoped cuprates \cite{kond09,vish12}, as illustrated in Fig.~\ref{fg:small}(c) for Bi2212 with $T_c=75$~K \cite{lee07}.  The dashed line indicates the anticipated $d$-wave dispersion; significant deviations occur at energies above $\Delta_c$.  The ARPES measurements involve a photon beam that covers a large surface area compared to the region probed by STM, and hence should average over the distribution of coherence peaks obtained by STM.\footnote{This averaging effect was recently taken into account to explain the anomalous temperature dependence near $T_c$ of the antinodal spectral function in slightly-overdoped Bi2212 crystals \cite{zaki17}; it was used previously to simulate ARPES results for LSCO \cite{park11}.} 

I have argued that the energy $E_{\rm cross}$ determined from the spin excitation spectrum provides an upper limit for pairing, and hence a limit for superconducting coherence peaks.
To provide a comparison, Fig.~\ref{fg:small}(a) shows a schematic version of the imaginary part of the dynamic susceptibility, $\chi''({\bf q},\omega)$, in YBa$_2$Cu$_3$O$_{6.6}$ as determined by inelastic neutron scattering \cite{hink07,hink10}.\footnote{No neutron scattering studies of underdoped Bi2212 crystals have been reported yet.}  While the doping levels are not quite the same, one can see that the hourglass spectrum is characterized by $E_{\rm cross}\sim38$~meV, comparable to the energies of the coherence peaks measured with electronic spectroscopies.  

\begin{figure}
 \centering
    \includegraphics[width=5cm]{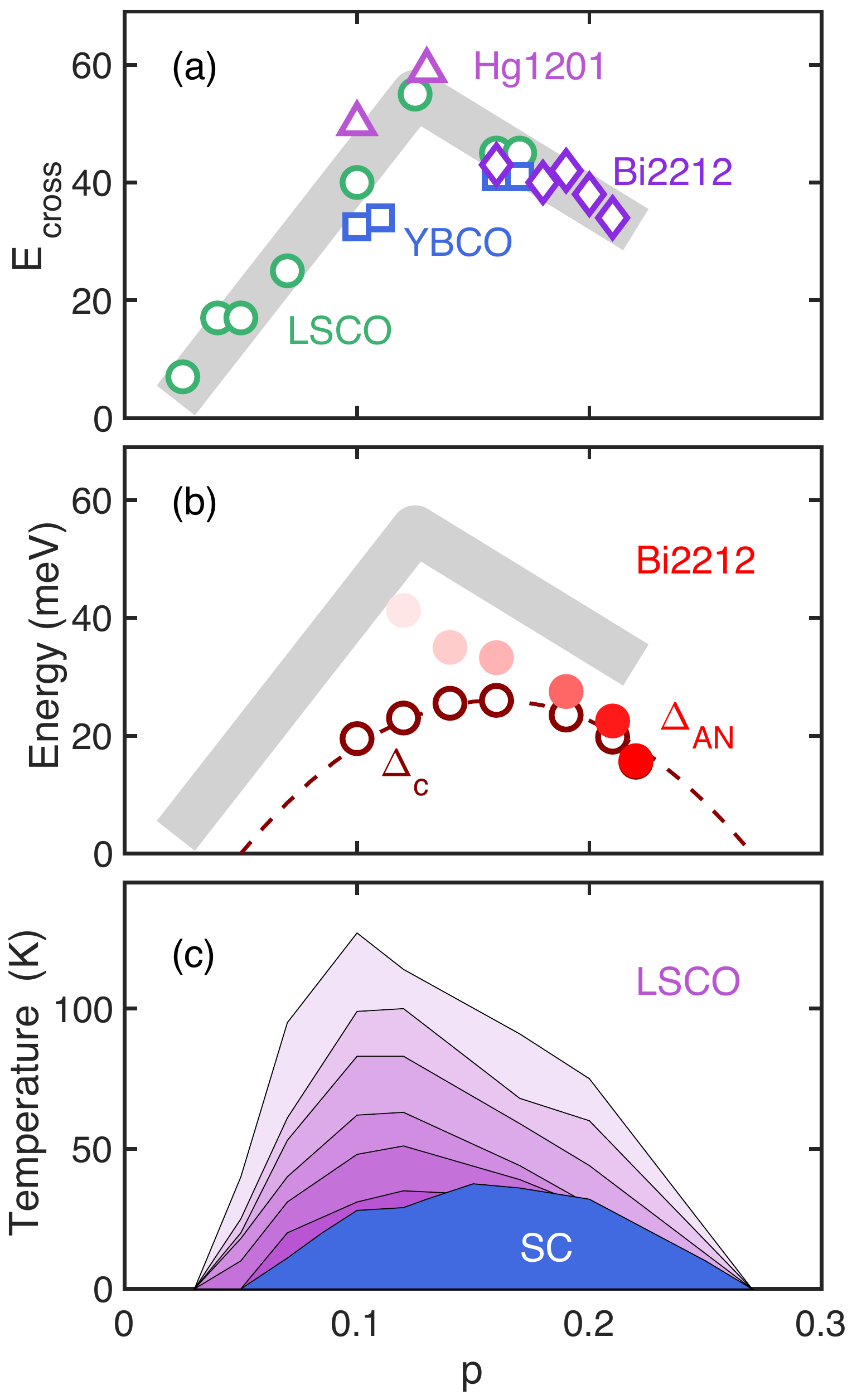}
    \caption{\label{fg:small_dope}  (a) $E_{\rm cross}$ from neutron scattering studies of \lsco\ \cite{fuji12a}, \ybco \cite{stoc05,hayd04,rezn08,woo06}, \bscco \cite{fong99,xu09,fauq07,he01,capo07}, and HgCa$_2$CuO$_{6+\delta}$ \cite{chan16b,chan16c}.  Gray line indicates the common trend among these families.  (b) Antinodal gap (red filled circles) and coherent gap (dark red open circles) energies from Raman scattering on \bscco\ \cite{blan10,sacu13}.  The shading of the $\Delta_{\rm AN}$ symbols reflects the change in peak area with doping. (c) Violet contours indicate relative strength (logarithmic scale) of superconducting fluctuations above the superconducting phase (blue) obtained from Nernst effect measurements on \lsco\ \cite{wang06}. }
\end{figure}

To appreciate why the local pairing scale may depend on hole concentration, Fig.~\ref{fg:small_dope}(a) shows the doping dependence of $E_{\rm cross}$ determined by neutron scattering in four families of cuprates.  The trends are remarkably similar in all of these compounds.  The general doping trend is approximated by the gray bars in the background.   A key point here is that, for rather underdoped cuprates, the limiting energy for pairing increases linearly with hole concentration. {\newr This means that a spatial variation in local average hole concentration will result in a local variation in the pairing scale. }

Theoretically, it has been proposed that charge disorder can cause a cuprate to behave like a granular superconductor \cite{alva05,alva08,erez10,okam10,imry12,triv12}.  In particular, Imry {\it et al.} \cite{imry12} considered a model in which superconducting grains are effectively coupled together through Josephson junctions.  Such a model is compatible with the early observation by Emery and Kivelson \cite{emer95a} that the onset of superconducting order in underdoped cuprates is limited by the development of superconducting phase coherence, rather than the onset of pairing.  Both of these approaches provide an explanation for the observation of Uemura {\it et al.} \cite{uemu89} that $T_c$ grows in proportion to the superfluid density, within the underdoped regime.

There has been quite a bit of experimental work on systems that exhibit the characteristics of granular superconductors.  A popular model system involves amorphous thin films of InO$_x$.  STM measurements reveal a gap that is associated with coherence peaks below $T_c$, but that does not close until the temperature is increased far above $T_c$ \cite{sace11}, a situation analyzed in \cite{triv12}.  Point-contact measurements reveal Andreev reflections that occur on an energy scale considerably smaller than the large energy gap obtained from tunneling conditions \cite{dubo19}.  (Related behavior is also seen in granular Al \cite{prac16}, where an oxide surface layer limits coherent coupling.)  Applying a magnetic field can destroy the Josephson coupling between grains, resulting in a superconductor-to-insulator transition (SIT) \cite{brez16}, and parallels have been drawn with the SIT behavior observed in very underdoped LSCO \cite{stei05,shi13}.

To the extent that cuprates show granular features, what limits the coherence between grains?  One proposal is that there might be a mixture of antiferromagnetic and superconducting domains, with the AF domains providing the barrier to coherence \cite{alva05,alva08}.  While there is no experimental evidence for commensurate AF order coexisting with superconductivity, we do have a varying degree of low-energy incommensurate spin fluctuations that can play a role.  We already saw in Fig.~\ref{fg:gaps} that $\Delta_c$ is limited by $\Delta_{\rm spin}$.  We can now combine that with the fact that the $\Delta_{\rm spin}$ tends to grow with doping, as does $E_{\rm cross}$, to interpret the observed behaviors of {\newr local superconducting} gap variation and appearance above $T_c$.  

Imagine, for a moment, that a sample contains regions associated with just two different hole dopings, one higher ($p_+$) and one lower ($p_-$), and that the size of these domains is much larger than $\xi_0$.  The local pairing strength should be determined by a combination of the local $E_{\rm cross}$ and $\Delta_{\rm spin}$, so that the local $\Delta_0$ should be larger in the $p_+$ domains.  The $p_+$ domains will start to develop local superconducting order at a temperature $T_c^+ > T_c$.  In order to  develop long-range order, it is necessary to achieve coherence in the $p_-$ domains, but the smaller (or possibly negligible) $\Delta_{\rm spin}$ in these domains will limit this.  Effective Josephson coupling through the $p_-$ domains will increase $\Delta_{\rm spin}^-$ to $\Delta_c$, the gap scale below which coherent, spatially-uniform superconductivity occurs, and the resulting $\Delta_c$ will determine {\newr the bulk} $T_c$.

A consequence of increasing $\Delta_{\rm spin}^-$ is that spin-fluctuation spectral weight must be pushed above $\Delta_c$.  This shift is associated with the appearance of a ``resonance'' peak\footnote{{\newr  As already discussed in Sec.~\ref{sc:sense},} the concept of a spin resonance comes from the weak coupling perspective, in which the same electrons are responsible for both the spin correlations and the superconductivity \cite{liu95,bulu96}.  In the superconducting state, where the antinodal electronic states should be gapped by $2\Delta_0$, a spin excitation at $E_r <2\Delta_0$ requires a resonant enhancement relative to the bare Lindhard spin susceptibility.  The ``resonance'' language is misleading when applied to the experimental results, because a weak-coupling description is not relevant.} {\newr (identified as enhanced intensity below $T_c$)} at energy $E_r > \Delta_c$.  Past neutron scattering studies have demonstrated a correlation between $E_r$ and $T_c$ \cite{sidi04,pail06}.  In the present interpretation, this relationship is a consequence of the connections with $\Delta_{\rm spin}$ and $\Delta_c$.

With that story in mind, consider now the results plotted in Fig.~\ref{fg:small_dope}(b).  Here the gray bars summarizing the $E_{\rm cross}$ trend are repeated to allow a comparison with measurements of $\Delta_c$ and the AN gap $\Delta_{\rm AN}$ determined by Raman scattering in Bi2212 \cite{blan10,sacu13}.\footnote{Comparisons of $2\Delta_{\rm AN}$ from a broad range of techniques have been presented elsewhere \cite{hufn08}; the other measures are consistent with the Raman results.  For the families YBCO, Bi2212, Tl$_2$Ba$_2$CuO$_{6+\delta}$, and HgBa$_2$CuO$_{4+\delta}$, it has been observed that $2\Delta_c\approx 6kT_c$ \cite{munn11,sacu13}.}\footnote{$\Delta_c$ is also detected by Andreev reflection in tunneling spectroscopy \cite{deut99}.}  The Raman result for $\Delta_{\rm AN}$ is comparable to the mean of the coherence-peak energies from STM [as in Fig.~\ref{fg:small}(b)] and the ARPES AN gap scale [as in Fig.~\ref{fg:small}(c)].  Besides indicating the energies by symbol position, the shading of the symbols is indicative of the integrated intensity of the spectral feature, which extrapolates to zero at $p\approx0.10$ \cite{sacu13}.  Note that the absence of the $\Delta_{\rm AN}$ signal for $p\lesssim0.12$ is found for other cuprate families, as well \cite{munn11}.  Also, ARPES measurements on Bi2212 with $p\lesssim0.1$ ($T_c < 70$~K) generally show an absence of legitimate quasiparticle peaks in the AN region, where the large pseudogap tends to dominate \cite{tana06,hash14,droz18,zhon18b}.  Hence, it appears that $\Delta_{\rm AN}\lesssim E_{\rm cross}$ (with the exception of LSCO, where $\Delta_{\rm AN}\ll E_{\rm cross}$ \cite{li18}).  

Figure~\ref{fg:small_dope}{b) only compares energy scales at low temperature; there are related observations at $T>T_c$.  For example, ARPES studies of Bi2212 reveal weak intensities of coherent antinodal peaks that survive for a finite temperature range above $T_c$ \cite{fedo99,feng00,ding01}.  Measurements with high resolution in energy indicate that the gap in the near-nodal region does not close uniformly at $T_c$ \cite{kond15}.  Again, these features occur within the disordered landscape indicated by STM \cite{gome07}.  Note that the coherence length $\xi_0$ is much smaller than the magnetic penetration depth $\lambda$, and as long as the grains that develop coherence above $T_c$ are small compared to $\lambda$, they will have limited impact on bulk-sensitive measurements of superconductivity, such as magnetization.

Considering the average behavior at $T> T_c$, ARPES measurements on underdoped Bi2212 and Bi2201 find that the superconducting gap closes only along a finite arc, centered on the nodal point, with a gap remaining in the AN region \cite{norm98,lee07,kond09,kond13,kond15}.  {\newr At $T<T_c$, a $d$-wave gap develops on the arc.}  The magnitude of the gap at the ends of the arc provides another measure of $\Delta_c$, as indicated {\newr schematically} in Fig.~\ref{fg:dc}, one which is quantitatively consistent with the Raman results for $\Delta_c$ shown in Fig.~\ref{fg:small_dope}(b).

Figure~\ref{fg:small_dope}(c) shows the temperature dependence of superconducting fluctuations as a function of doping in LSCO, where the measure is the Nernst coefficient, plotted with roughly logarithmic intensity contours \cite{wang06}.\footnote{The Nernst effect is the transverse voltage measured in response to a longitudinal temperature gradient in the presence of a perpendicular magnetic field; it is sensitive to vortex fluctuations for temperatures near $T_c$ \cite{wang06}.}  The maximum onset temperature occurs for $p\sim0.1$, with the response falling rapidly at lower $p$.  Confirmation that these fluctuations are associated with pairing correlations is provided by a recent study of tunneling-current noise involving LSCO films \cite{zhou19}.  The trend of the fluctuations with doping is consistent with the result in Fig.~\ref{fg:small_dope}(b) that the maximum local pairing gap is optimized near $p\sim 0.12$.  (In contrast, phase coherence and superfluid density are optimized at $p\sim p^*$, as indicated by the plot of the superfluid density in Fig.~\ref{fg:p}.)

{\newr  I have covered many points in this section, so let me try to summarize.  Indications of superconducting correlations above $T_c$, such as the antinodal ``pseudogap'' detected by ARPES and Raman scattering,\footnote{\newr This small pseudogap is distinct from the large pseudogap associated with scattering by spin fluctuations.} are a consequence of charge disorder.  The upper limit on the local AN gap is $E_{\rm cross}$.  (In Sec.~\ref{sc:sense}, I argued that $E_{\rm cross}$ is a measure of the singlet-triplet excitation energy on the charge stripes.  The charge stripes evolve with doping, and so does $E_{\rm cross}$.\footnote{\newr The stripe spacing varies with $p$.  When $p$ is small and charge stripes are farther apart, they are likely to be wider.  Correspondingly, the spin gap on a ladder with an even number of legs decreases as that number increases \cite{chak96}.  Thus, the pairing scale will be reduced for holes in wider ladders.})  The onset of long-range coherence and the bulk superconducting transition are limited by regions of reduced average hole concentration, where there is a smaller spin gap in the spin stripes.  The system effectively behaves like a granular superconductor.

To be clear, this scenario is distinct from one of ``pre-formed pairs''.   Uemura and coworkers \cite{uemu91} have emphasized that the distribution of $T_c$ vs.\ $p$ is close to that expected for Bose-Einstein (BE) condensation of a gas of bosons.  To have BE condensation in a cuprate, one would need to have a gas of hole pairs present above $T_c$.  From the stripe perspective, there may be pairing correlations within individual charge stripes well above $T_c$; however, these are quasi-1D correlations that cannot order and that are difficult to detect directly.  Superconducting correlations require phase coherence among neighboring charge stripes.  Once this local phase coherence develops, there is a superconducting grain that is detectable.  The bulk transition is limited by the development of phase coherence among many such grains.
}

\subsection{The overdoped regime}

To discuss the changes on increasing $p$ into the overdoped regime, it is convenient to start with an interpretive summary, shown graphically in Fig.~\ref{fg:sum}, and then consider the experimental results that support this picture.
In the underdoped region, as we have seen, the competition between superexchange-coupling of Cu moments and the kinetic energy of O holes leads to stripe correlations, but all good things must come to an end.  In 214 cuprates, the charge stripe period at $p=1/8$ is just 4 lattice spacings, and it saturates there for higher dopings \cite{wen19,miao19,miao20}.  It seems likely that holes added beyond $p=1/8$ form uniformly-doped regions.  Eventually, the strongly-correlated, striped regions will no longer percolate across the CuO$_2$ planes.  This crossover should correspond to the point, $p=p^*\sim 0.19$ \cite{tall01}, at which the large pseudogap due to Cu spin correlations effectively disappears  \cite{droz18,chen19}, and a coherent SC gap is observed around the entire Fermi surface, including the AN region \cite{fuji14a}.  The superfluid density reaches its maximum at $p^*$, as seen in Fig.~\ref{fg:p}.  For $p>p^*$, the superfluid density and $T_c$ decrease fairly rapidly \cite{uemu93,tana05,wang07,lemb11,bozo16,mahm19}.  While the carrier density remains substantial, the coupling of 
electrons to the excitations that drive pairing drops to zero as $T_c\rightarrow0$, as seen by both ARPES \cite{vall20} and optical spectroscopy \cite{hwan07b}; in particular, the signature of this interaction is absent in the normal state for $p>p^*$ \cite{vall20}.

\begin{figure}
 \centering
    \includegraphics[width=10cm]{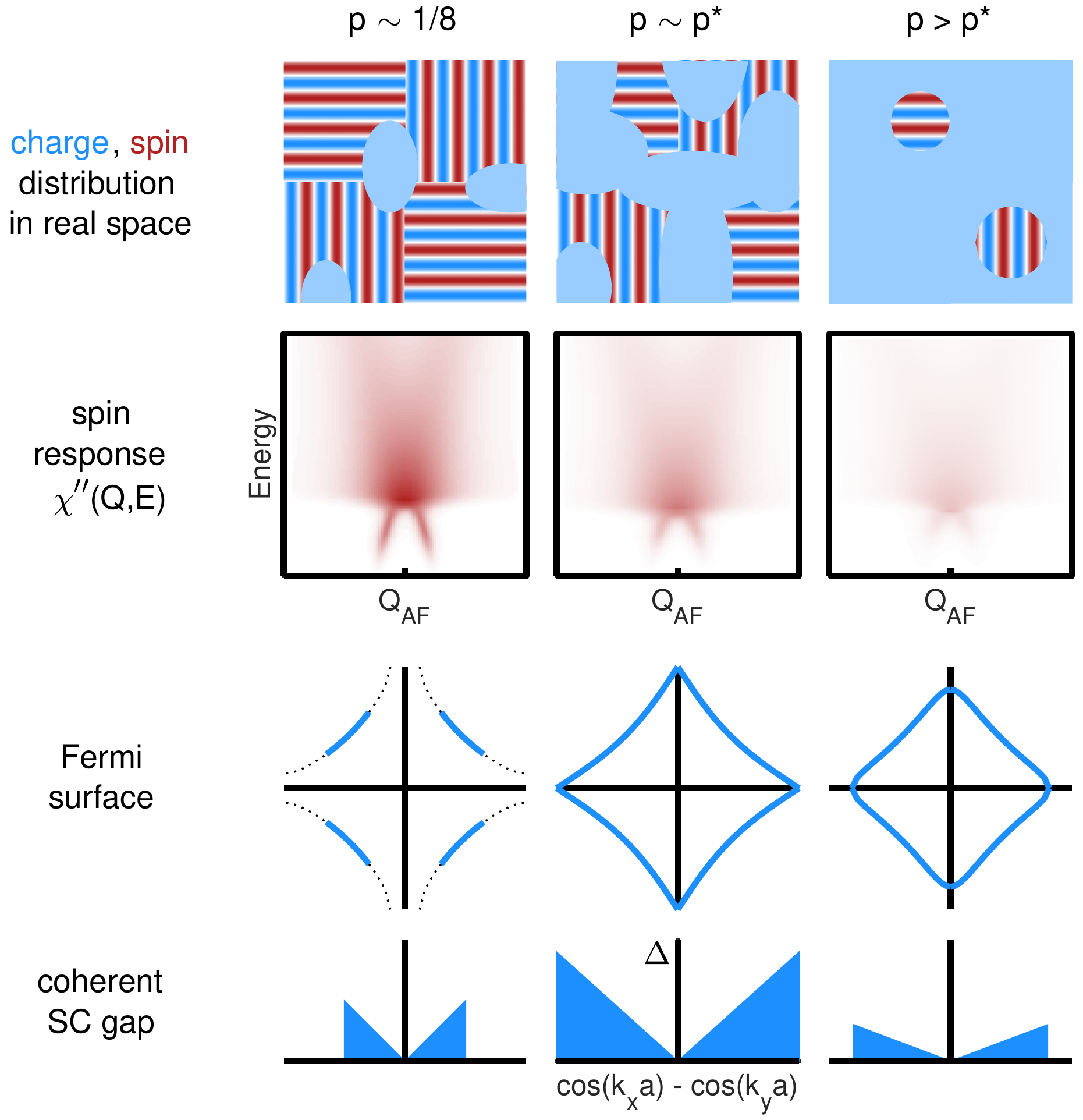}
    \caption{\label{fg:sum}  Cartoon summary of changes in various properties with overdoping.  Each column corresponds to the hole concentration listed above the top row of panels.  Top row: spatial distribution of charge (blue) and spin (red) densities in real space.  Stripes indicate strong correlations, even in the absence of static order.  Second row: spin response in the form of $\chi''({\bf Q},E)$ for ${\bf Q} = (h,0.5,0)$.  Third row: Fermi surface as indicated by $A({\bf k},E_{\rm F})$ measured by ARPES.  Bottom row: coherent SC gap along one quadrant of the Fermi surface, as a function of $\cos(k_xa)-\cos(k_ya)$. }
\end{figure}

Inelastic neutron scattering measurements on LSCO indicate that the AF spin correlations continuously weaken with doping, remaining detectable at $p>p^*$, but disappearing as the superconductivity disappears at large $p$ \cite{waki04,waki07b,li18}.\footnote{In particular, there is no evidence of a collapsing of the excitation spectrum towards zero frequency, as one might expect in the case of a quantum critical point.}  In a related fashion, the 2-magnon peak detected by Raman scattering in several cuprate families gradually shifts to lower energy [as indicated in Fig.~\ref{fg:mYBCO}(a) for YBCO], until it becomes overdamped at $p\gtrsim p^*$ \cite{suga03}.  Signatures (from neutron scattering, NMR, and ultrasound) of quasi-static AF correlations induced by high magnetic fields disappear at $p\sim p^*$ \cite{frac20,chan08}.  The characteristic temperature $T^*$ determined from quantities such as the Hall coefficient and the bulk susceptibility (discussed in Sec. \ref{sc:temp}) trends toward zero near $p^*$ \cite{hwan94,gork06}.   These results are all consistent with a loss of percolation for the regions with strong local AF correlations at $p\sim p^*$.

The data suggest that cuprates with $p>p^*$ have patches with local AF correlations, that can drive pairing, embedded within a metallic state.  Spivak {\it et al.} \cite{spiv08} have analyzed how such patches can drive bulk superconductivity through the proximity effect.  The disorder can lead to granular behavior, with $T_c$ varying with the superfluid density, rather than the pairing strength \cite{imry12,li21c}.\footnote{Alternatively, it has been argued that some of the phenomenology observed in overdoped cuprates can be understood in terms of ``dirty $d$-wave theory'', which is based on a conventional BCS picture \cite{leeh17,leeh20}. While that may be the case, it does not provide a specific understanding of the pairing mechanism, especially for $p<p^*$.  }

\section{Discussion}
\label{sc:disc}

While I have attempted to cover a large range of experimental work essential to understanding the cuprates, there are some remaining significant topics that deserve comment.  

\subsection{Nematic order}

When charge and spin stripe correlations are purely dynamic, they can still have interesting impacts on bulk properties.  An analogy to liquid crystals led to a proposal of smectic and nematic electronic phases \cite{kive98}.  Even when static stripe correlations are present, quenched disorder may destroy long-range correlations; nevertheless, vestigial effects can result in nematic order \cite{nie14,fern19}.  

Evidence of nematic order \cite{lawl10,mesa11} and its vestigial character \cite{mukh19} has been provided by STM studies\footnote{There are experimental technique issues that may qualify the identification of nematic order by STM \cite{dasi13}.} on Bi2212.  Rotational anisotropy in the Nernst effect measured on YBCO\footnote{In YBCO, ordering of Cu-O chains provides a broken symmetry that enables the detection of a nematic effect that turns on well below the chain-ordering temperature.} indicates nematic order \cite{daou10}.

Signatures of nematic responses have also been identified in Raman scattering experiments on very underdoped YBCO and LSCO \cite{capr15}.  A recent Raman study of Bi2212 proposes a nematic quantum critical point at $p^*\approx0.22$ \cite{auvr19}.   While theory shows that fluctuations at a nematic quantum critical point can, in principle, enhance superconductivity \cite{maie14,lede15,metl15}, I would argue, on the basis of nematicity being a vestigial order, that this is not of fundamental importance.  The underlying magnetic and charge correlations provide the dominant electron pairing, as discussed in Sec.~\ref{sc:spatial}.

\subsection{Charge order}

Charge-density-wave (CDW) order\footnote{Charge density waves and charge stripes are both names for a periodic modulation of the charge density.} has now been observed in most cuprate families, especially by x-ray scattering techniques \cite{comi16,fran20}.  While the modulation is always parallel to the Cu-O bonds, as with the charge stripes in  \lbco\ and other 214 cuprates, a difference is that it tends to develop in the absence of spin order.  Of course, this does not mean that there is an absence of AF correlations;  to the contrary, CDW order develops in YBCO (in zero magnetic field) over the range of doping $0.08 \lesssim p\lesssim0.18$ \cite{huck14,blan14} at temperatures comparable to the onset of the spin gap,\footnote{In bilayer cuprates, the AF coupling between neighboring CuO$_2$ layers may enhance the tendency to developing a spin gap \cite{pail06}.} as determined by the temperature dependence of the spin-lattice relaxation rate measured by Cu NMR \cite{taki91,baek12}.  The short-range and static characters of the low-field CDW are confirmed by NMR \cite{wu15}.  
{\newr  The CDW wave vector does not have a simple relationship to the wave vector of the lowest-energy spin correlations \cite{huck14,blan14}, but there is no conflict as the spin correlations are purely dynamic.
}

CDW order in Bi2212 was first detected by STM \cite{hoff02,kive03,park10} and was later connected with measurements of resonant x-ray scattering \cite{dasi14}.
The study that discovered PDW order in halos about magnetic vortices found the PDW modulation to be linked to the CDW order that was previously seen in zero magnetic field \cite{edki19}.  Follow on work, combining STM spectroscopic imaging and theoretical mean-field analysis, made the case that, even in zero field, the observed CDW order is consistent with calculations for PDW order that coexists with uniform $d$-wave order \cite{chou20}.\footnote{While there is no static spin order to conflict with the uniform SC, it may be that defects locally make the CDW/PDW orders energetically competitive, as proposed in the case of Zn-doped LBCO \cite{loza21}.} The observation of periodically-modulated Josephson tunneling demonstrates that the pairing correlations are spatially-modulated and in-phase with the CDW order \cite{hami16b}.  Complementary work on Bi2201 shows that the coherence peaks in the superconducting state are modulated at the same period as the CDW in Bi2201 \cite{ruan18,li21a}.  At minimum, these results indicate that the CDW order does not compete with pairing{\newr; however, any pairs associated with CDW order presumably do not contribute to the coherent superconductivity}.

Returning to YBCO, the zero-field CDW order involves equal and opposite atomic displacements of the CuO$_2$ bilayers immediately above and below one chain layer \cite{forg15}. {\newr  Since holes are transferred from the chains to the planes \cite{bozi16}, defects such as chain ends will cause local charge defects in the planes that are likely screened by such local CDWs. The fact that the CDW intensity decays on cooling below $T_c$ \cite{huck14,blan14} suggests that some of the holes involved in the local CDWs join the superconducting condensate.}  In contrast, a $c$-axis magnetic field of sufficient strength ($>15$~T for $p=0.12$) induces a transition to a CDW phase with modulations that are in-phase in all layers, so that there is a finite correlation length along the $c$-axis \cite{wu11,gerb15,choi20}.  The development of this phase begins slightly below the onset of finite resistivity \cite{rams12}.  Strikingly, it has been discovered that a similar CDW phase, with 3D correlations, can be induced at zero field by application of strain along the $a$ axis \cite{kim18}; it appears on cooling below the zero-strain $T_c$ and then disappears with the onset of bulk superconductivity at a reduced $T_c$ \cite{barb21}.

Given the relatively low onset field for the 3D CDW in YBCO and the association between pair correlations and CDW order in Bi2212, it would be surprising if pair correlations were not present in the 3D CDW state of YBCO.  Indeed, as discussed below, analysis of quantum oscillations in the high-field regime indicates that the ``normal-state'' response comes from only a fraction of the doped carriers \cite{tail09,seba15}, and measures of the increasing density of states with field indicate saturation in the high-field phase \cite{kacm18}.  A recent analysis suggests that the missing holes, presumably associated with antinodal states, are likely gapped \cite{hart20}.  Also, weak diamagnetism above the nominal upper critical field, $H_{c2}$, has been reported \cite{yu15,yu16}.  These effects remain to be properly understood.

\subsection{Electron-phonon coupling}

Theoretical analyses are often posed as a choice between electron interactions with phonons or with spin correlations.  In reality, both are significant; however, they are not equally important \cite{capo10}.  In particular, the charge order in cuprates benefits from, but is not driven by, electron-phonon coupling.

Given the low carrier density in cuprates, phonons make an important contribution to Coulomb screening.  The phonon branch with the biggest impact  is the Cu-O bond-stretching mode \cite{ross19}.  At ${\bf q} = (\pi,\pi)$, it forms a symmetric breathing mode of O$^{2-}$ about Cu$^{2+}$; at $(\pi,0)$ and $(0,\pi)$, we have half-breathing modes.  In cuprates such as LSCO and YBCO, neutron scattering experiments show that the strongest phonon softening is along the Cu-O bond direction, growing in strength as one moves from zone center to zone boundary \cite{mcqu99,pint04,rezn06}.  There is a strong dip near the charge-ordering wave vector \cite{rezn08,wang21}.

If charge order were driven by electron-phonon coupling and the nesting of the noninteracting Fermi surface, we would expect the charge order to be in the diagonal direction, as shown by many analyses \cite{jorg87,mach87}.  The inconsistency with experiment is direct evidence that, while phonon softening lowers the electronic energy, this occurs to complement the stronger interactions resulting from superexchange.

Resonant inelastic x-ray scattering (RIXS) is particularly sensitive to electron-phonon coupling (although with limited energy resolution, at present).  A soft-mode behavior over a large energy range in connection with the CDW wave vector has been reported for Eu-doped LSCO \cite{peng20} and Bi2212 \cite{chai17,lee21a}.  The results show an interesting anisotropy in intensity with momentum transfer {\bf q}.  The intensity rises from the CDW wave vector and connects smoothly with the bond-stretching mode at large {\bf q}.  This appears to directly illustrate the growing strength of electron-phonon coupling with {\bf q}.  The continuous variation is a consequence of poor energy resolution.  The weight must involve anticrossing transfers among the many phonon modes between zero energy and the bond-stretching branch.  For example, several non-resonant x-ray and neutron scattering studies, with much better energy resolution, have demonstrated in YBCO that the CDW is associated with phonon softening in acoustic \cite{blac13b,leta14}, bond-bending \cite{raic11}, and bond-stretching branches \cite{pint04}.

{\newr
\subsection{PDW gap}

The superconducting gap in the PDW state has been predicted to have a gapless nodal arc with the gap rising to a substantial value in the antinodal regions \cite{baru08,berg08,gran10}; however, these analyses assume that the spin-stripe order acts as a weak perturbation on the quasiparticles.  In LBCO $x=1/8$, where there is reason to believe that PDW order occurs \cite{li07,agte20}, the antiferromagnetic order within spin stripes is not a weak perturbation.  ARPES measurements on this system find that the spectral function at the nodal point has a strongly-damped peak that is shifted to finite energy \cite{he09,vall06}, and the gap grows continuously as one moves along the nominal Fermi surface, though it grows much faster near the antinodal points.

ARPES measurements on LSCO provide further evidence of the impact of spin order.  For $x=0.15$, where the spin excitations develop a gap, ARPES finds a sharp peak in the nodal spectral function \cite{razz13}; however, at $x=0.08$, a composition in the range where weak spin-stripe order is detected \cite{jaco15,nied98,juli03}, the spectral function is similar to that in LBCO $x=1/8$ \cite{razz13,he09}.

To observe a gapless near-nodal arc, one would need to consider an occurrence of PDW order in a system such as Bi2212, where spin excitations tend to be gapped.  While this system generally exhibits uniform $d$-wave order, evidence of PDW order has been observed by STM in the vicinity of magnetic vortices \cite{edki19}, as mentioned in Sec.~\ref{sc:sense}.  A magnetic vortex core is a form of defect where the superconducting order parameter must go to zero.   PDW order may be energetically favored near a vortex core, relative to uniform $d$-wave order, because it naturally incorporates zeros of the order parameter in real space.  

Zn impurities on the Cu site also suppress superconductivity locally \cite{nach96}.  Defects of this type also induce Bogoliubov quasiparticle interference that is measured in spectroscopic-imaging STM studies and analyzed to determine the form of the superconducting gap \cite{kohs08}.  Intriguingly, the gap detected in this way for underdoped Bi2212 deviates from the conventional $d$-wave form and tends toward a gapless nodal arc, as shown in Fig.~\ref{fg:qpi} \cite{kohs08}.  If a Zn impurity locally induces PDW order, then the STM measurements may selectively detect interference associated with a local PDW state, so that the observation of a gapless arc together with a large antinodal gap would be consistent with the theoretical prediction.  As a check, a recent study has looked at the impact of Zn impurities on superconductivity in LBCO $x=0.095$ \cite{loza21}, which, in the absence of Zn, exhibits 3D SC order below $T_c=32$~K.  Measurements of anisotropic resistivity and magnetic susceptibility indicate that the Zn frustrates the onset of 3D SC order while allowing 2D SC order to develop, as shown later in Fig.~\ref{fg:Zn}, consistent with expectations for the case that Zn pins PDW order.}

\begin{figure}
 \centering
    \includegraphics[width=0.5\textwidth]{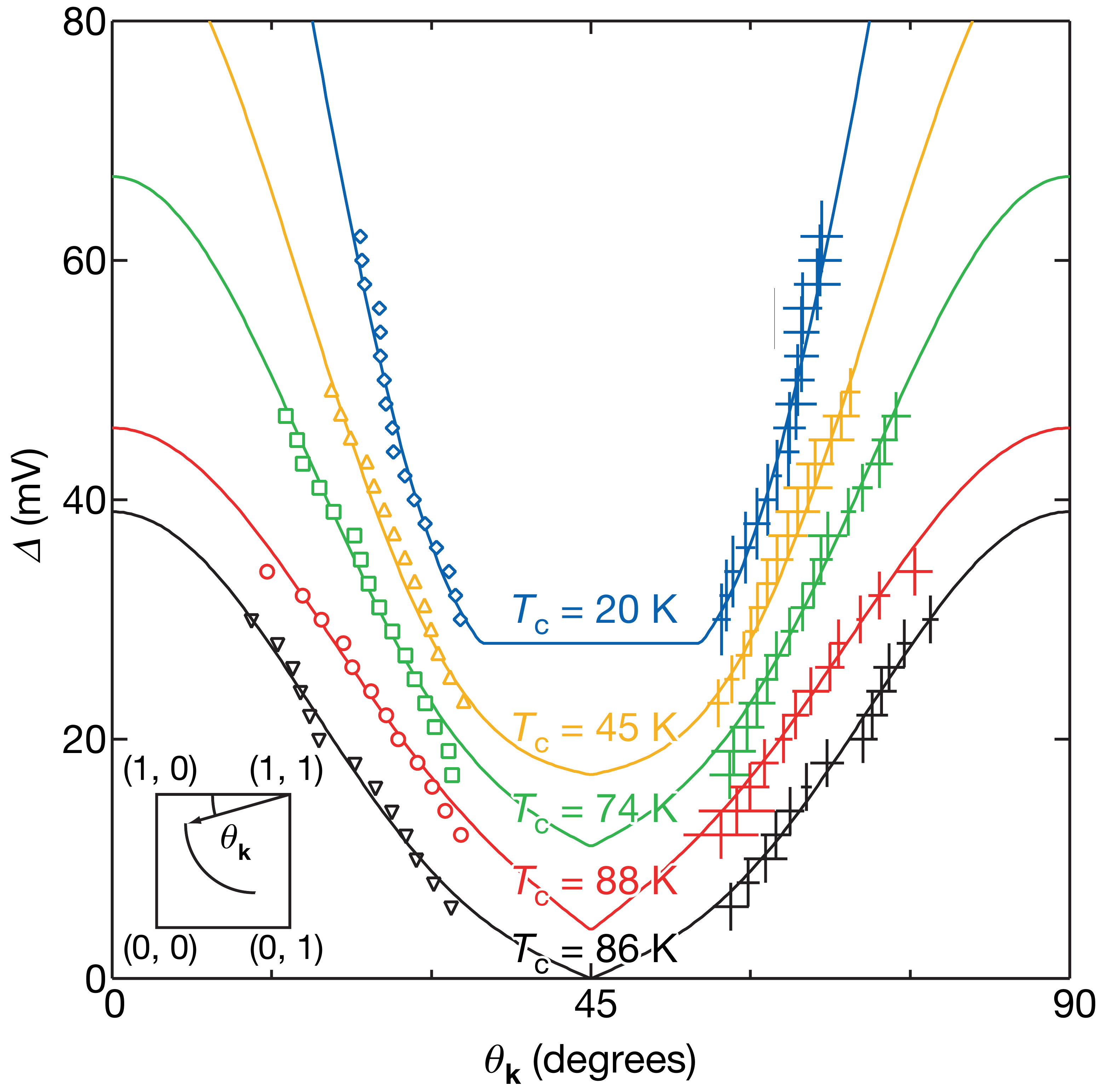}
    \caption{\label{fg:qpi} {\newr Superconducting gap $\Delta$ as a function of angle $\theta_{\rm k}$ around the nominal Ferm surface (inset) obtained from analysis Bogoliubov quasiparticle interference results of STM measurements on a series of Bi2212 crystals (labeled by $T_c$, starting from very underdoped at top) from \cite{kohs08}.  Open symbols on left are data points, corresponding error bars are on right, and lines are fitted curves; results have been offset for clarity.  Reprinted with permission from \cite{kohs08}, MacMillan \copyright2008.} }
\end{figure}

\subsection{Quantum oscillations}

The discovery of quantum oscillations in measurements of transverse and longitudingal resistance \cite{doir07}, magnetization \cite{seba08}, and specific heat \cite{rigg11} in underdoped YBCO as a function of magnetic field created a sensation.  In order for quantum oscillations to be detectable, quasiparticles near the Fermi level must be sufficiently coherent that they can complete an orbit on the Fermi surface.  Thus, their observation seemed to suggest that the state obtained by suppressing the superconducting order is a Fermi liquid, which would be a challenge to relate to the zero-field normal state at $T>T_c$.  

From the oscillation frequency it is possible to estimate the size of the orbit in reciprocal space, and that yields an associated density of carriers.  It quickly became clear that only a fraction of the doped carriers contributes to the oscillations \cite{lebo07}.  A wide variety of proposals was made for various mechanisms that could cause a modification of the nominal Fermi surface, including stripe order  and  biaxial CDW order \cite{tail09,seba12}.  In the latter case, a small pocket is effectively generated that includes the Fermi arcs, with scattering between arc ends at points connected by one of the CDW wave vectors.  The presence of a spin gap \cite{baek12,wu11} helps to minimize the scattering rate for states on the arcs.

The concept of a small pocket created from Fermi arcs due to biaxial CDW order is certainly intriguing, but it leaves some open questions.  For instance, what is the role of the uniaxial CDW order induced in high fields, and what happens to the antinodal states?  Regarding the missing antinodal states, a recent study indicates that they are likely gapped \cite{hart20}, as mentioned above.

Besides considerations of reciprocal space, it is also necessary to consider real space.  In particular, successful detection of quantum oscillations requires that the electronic states be coherent over the cyclotron radius, which is largest at the lowest fields at which the oscillations first appear.  Gannot {\it et al.} \cite{gann19} pointed out that, in comparing the cyclotron radius with the typical correlation lengths for the zero-field and high-field-induced CDWs, only the high-field CDW has a correlation length consistent with the onset field.  %Such a connection is intriguing, but more details must be sorted out before a complete understanding is reached.

{\newr
The gapless arc associated with the quantum oscillations has generally been considered to be associated with a non-superconducting state; however, gapless arcs are also compatible with PDW order, as discussed above.  There is now a proposal to explain the quantum oscillations in terms of a bidirectional PDW state \cite{capl21}.  While the bidirectionality is inconsistent with the evidence for unidirectional character of the field-induced CDW, the possibility of a PDW solution is intriguing.  This approach ties in well with new experimental work.  High-field measurements of underdoped YBCO \cite{hsu21a} suggest that, at very low temperature, there is a fragile superconducting state \cite{yu19} with low critical-current density that survives to fields well above the commonly-cited $H_{c2}$; it is bounded by an unusual vortex liquid phase \cite{hsu21b}.  This is in the regime where weak diamagnetism had previously been reported \cite{yu15,yu16}.  Clearly, we do not have all of the answers yet.
}

{\newr 
\subsection{Why are stripe-ordered nickelates not superconducting?}

I have argued that charge stripes in cuprates such as \lsco\ are good for driving pairing correlations.  There are also charge stripes in \lsno, but no superconductivity.  Why?  Well, there are a number of distinctions, but an important one is that Ni$^{2+}$ has $S=1$, in contrast to the $S=1/2$ of Cu$^{2+}$.  We have seen that antiferromagnetically-coupled $S=1/2$ moments provide a challenge for the motion of doped holes.  The situation becomes even more difficult with $S=1$, as moving a hole through a Ni site cannot simply flip the total spin.  A consequence is that the charge stripes in nickelates run diagonally (relative to a square lattice) \cite{tran13a}, similar to spin stripe order (and likely charge stripe order) in LSCO with $x\lesssim0.05$ \cite{birg06}. 

The diagonal orientation of a charge stripe is less conducive to hopping, consistent with the significant charge excitation gap \cite{ido91,kats96}.  As we have seen, the decoupled spin excitations along a diagonal charge stripe are gapless \cite{boot03b}.  There is no sign of the singlet correlations that are important for pairing in cuprates.  (Of course, it is relevant to note that superconductivity has been discovered in a different nickelate system, the ``infinite layer'' compound Nd$_{1-x}$Sr$_x$NiO$_2$ \cite{li19c}, which is the subject of intense interest and debate \cite{norm20}.)

 While there is a common tendency for hole-doped antiferromagnets to develop charge stripes, the ability to form singlet correlations relevant to pairing is special to the cuprates.
}

\subsection{Strange and bad metals}

Anderson noted that the in-plane resistivity is approximately linear in $T$ down to low $T$ near optimal doping \cite{ande92}.  This deviates from the $T^2$ behavior expected in the $T\rightarrow0$ limit of Fermi liquid theory, and it came to be known as ``strange''-metal behavior.  The possibility that it might be explained by quantum critical behavior \cite{daml97} provided one early motivation for discussions of a QCP under the superconducting dome.  Later, noting that the scattering rate associated with the in-plane resistivity is $\sim \hbar/k_{\rm B}T$, Zaanen suggested that the associated phenomenon be termed ``Planckian dissipation'' \cite{zaan04}.

Another unusual aspect of transport in cuprates occurs at high temperature.  Resistivity in metals is normally observed to saturate when it rises to a level of $\sim 0.15$~m$\Omega$-cm \cite{fisk76}.  This is presumed to be roughly the point at which the mean-free-path of an electron becomes comparable to the interatomic spacing, which has come to be known as the Mott-Ioffe-Regel (MIR) limit \cite{ioff60,mott72}.  In contrast, the resistivity of underdoped cuprates rises through this limit without any change of behavior.  Emery and Kivelson identified this as ``bad''-metal behavior \cite{emer95b}.

As we have seen from the temperature dependence of the Hall coefficient, the density of charge carriers grows at high temperature.  Hussey {\it et al.} \cite{huss04,huss08} have noted that the electronic scattering rate saturates in a fashion roughly consistent with the MIR limit, and that the carriers at high temperature are incoherent, as indicated by the loss of the Drude peak in optical conductivity.  From the perspective of the spin correlations, one can imagine the completely incoherent state corresponding to a soup of local singlet and triplet correlations with rapid fluctuations due to the lack of segregation between charges and moments.

The conventional view of resistivity is in terms of diffusive motion of independent electrons.  To describe the anomalous behavior in cuprates, a picture of hydrodynamic transport, involving a collective motion of a viscous electron fluid, has been proposed \cite{hart15}.   In particular, hydrodynamic effects associated with fluctuating charge density waves have been analyzed \cite{dela17}.  Results from related modelling have been applied to describe a range of experimental results for a Bi2201 sample \cite{amor20}.  There has also been progress on modelling Planckian dissipation by inclusion of random interactions among electrons through a variation of the Sachdev-Ye-Kitaev model \cite{pate19}. 

From the perspective of intertwined charge and spin correlations, the concept of collective, viscous transport is appealing.  It also suggests that realistic models will need to take account of the AF correlations, which are a dominant source of scattering, as we have seen.  Recent quantum Monte Carlo calculations on the Hubbard model show promising signs of this \cite{huan19}.

\subsection{Tri-layer cuprates}

Within a given family, $T_c$ varies with the number of consecutive CuO$_2$ layers, showing a maximum for three layers \cite{chak04}.  Analysis of NMR Knight shift data on 3- and 4-layer cuprates indicates that the hole concentrations are different on the inner and outer layers, with $p$ being larger on the outer planes by $\sim0.04$ \cite{shim11}.  The results shown in Figs.~\ref{fg:p} and \ref{fg:small_dope} suggest why the distinct values of $p$ could be beneficial for $T_c$.  The superfluid density reaches a maximum for $p\sim0.2$, whereas pairing strength appears to be optimized for $p\sim0.12$.  The trilayer cuprates can simultaneously benefit from both of these features, consistent with a proposed mechanism for enhancing $T_c$ \cite{kive02b}.

{\newr
\subsection{Pseudogap confirmation bias}

%delicate issue

%We have seen that the pseudogap issue is a complicated one.  

%multiple distinct phenomena that get labelled with the same name, which can be both confusing and misleading

As we saw in Sec.~\ref{sc:phenom}, there is a large pseudogap associated with scattering of holes by antiferromagnetic correlations associated with local Cu moments.  This effect evolves continuously with doping.  For underdoped samples, there is an electronically incoherent phase at high temperature; the spin correlations and electronic coherence develop as the temperature decreases, and it is only as the coherence of near-nodal electronic states develops that the strong damping of antinodal states stands out.  It is natural to try to identify a characteristic temperature $T^*(p)$ for this evolution; the challenge is that, without any sharp change in the measured properties and without an accepted, parametrized formula to describe the temperature and doping dependence of a given quantity, the problem is not well defined.  The identification of $T^*$ in particular measurements has been, at times, somewhat arbitrary.  Of course, once values of $T^*$ are plotted vs.\ $p$ in phase diagrams, it can provide motivation to identify similar temperature-dependent changes in other properties and associate them with previous $T^*$ results.  While such associations can be a way of identifying patterns in an unknown terrain, there is also a risk of confirmation bias.

One type of bias is the idea that there is a single pseudogap phenomenon.  In Sec.~\ref{sc:disorder}, I have made the case that the more sharply defined antinodal gap, with associated coherence peaks, that appears above $T_c$ for $p\gtrsim0.1$ is a distinct effect.  It is associated with locally coherent superconductivity, where long-range superconducting order is inhibited by large-scale charge disorder and regions with low-energy spin fluctuations.  Besides the distinct variation with $p$, the local superconductivity appears at a lower temperature scale than typically identified for the large pseudogap from quantities such as magnetic susceptibility and resistivity.  The finite $p$ range for the ``small pseudogap'' associated with local superconductivity has largely been ignored because of the bias that there is a single pseudogap phase that extrapolates all the way to $p=0$.

Much of condensed-matter theory is based on the concept of a Fermi liquid of quasiparticles associated with Bloch states.  From that perspective, the experimental evidence for an effective decrease in the density of charge carriers in the pseudogap phase provides motivation for considering possible types of competing order.  One such proposal was for $d$-density-wave order \cite{chak01}, with associated broken symmetries.  A different proposal, from Varma \cite{varm97,varm06}, was for a form of loop current order, involving charge hopping from Cu to O neighbors in intra-unit-cell loops that break time reversal symmetry without breaking translational symmetry. 

Varma's proposed loop-current order should be detectable by neutron scattering with spin-polarization analysis \cite{varm97,varm06}.  Because the order would not break translation symmetry, it is necessary to measure at reciprocal lattice vectors, where Bragg peaks from nuclear scattering occur.  So one has to look for a weak magnetic signal on top of significant non-magnetic intensity.  With the right scattering geometry, a neutron of known polarization will have a probability to flip its spin due to magnetic scattering while nuclear scattering does not coherently change the spin orientation.  Because the spin-flip signal is expected to be very weak and the instrumental spin-polarization sensitivity is limited, the measurements are only practical at a reciprocal lattice vector where the nuclear structure factor is small.   An early attempt to observe the predicted signal in LSCO and YBCO samples was unsuccessful \cite{lee99}.  After Varma proposed a new loop-current geometry \cite{varm06}, Bourges, Sidis, and coworkers began a series of experiments to test this possibility, starting with YBCO \cite{fauq06}

The measurements require relatively long counting times and experimental time is limited, so the measurements extend to a temperature only slightly above the anticipated $T^*$ \cite{ito93}.  Both the spin-flip and non-spin-flip signals were plotted vs.\ $T$.  The instrumental sensitivity to the spin-flip signal (the flipping ratio) was assumed to be temperature independent and, further, it was assumed that there is no magnetic response at the highest temperatures; the intensities of the two signals were plotted so as to be superimposed at high temperature, with deviations that grow with decreasing temperature interpreted as the development of magnetic response.

A complication is that the measurements were done with the sample orientation on the spectrometer held constant; this minimizes possible noise due to imperfect orientation adjustments, but it does not compensate for the small temperature-dependent rotation of the diffracted neutron beam due to thermal variation of the lattice parameters.  In a later study by Hayden and coworkers on different (and much smaller) crystals \cite{crof17}, it was demonstrated that the instrumental flipping ratio can be quite sensitive to small changes in orientation of the diffracted beam.  When the sample orientation was optimized at each temperature, the results indicated no sign of a temperature-dependent magnetic signal.  These results raise a concern of confirmation bias in the original study \cite{fauq06}.  This is obviously a delicate issue, and the arguments on each side have been debated \cite{bour18,crof18} (for a recent review  by Bourges {\it et al.}, see \cite{bour21}); nevertheless, my conclusion is that the evidence for loop currents is unconvincing.  Of course, my bias is that the pseudogap phenomena do not involve a competing order.

This is not the only controversial case.  Measurements of the polar Kerr effect (rotation of polarization of reflected light) on YBCO crystals \cite{xia08}  found that the polarization rotation angle became finite (but very small) at onset temperatures that parallel those from the polarized neutron study \cite{fauq06}, leading to further discussions of possible time-reversal symmetry breaking.   Evidence that this behavior is not universal (at least in a simple way) comes from a related study on LBCO $x=1/8$ \cite{kara12}, which found that a finite response appeared only below the structural transition at 56~K, where charge stripe order develops. That observation motivated discussions of possible broken chiral symmetry, but no clear conclusion was reached.  

Shekhter {\it et al.} \cite{shek13} reported results of resonant ultrasound spectroscopy on two crystals of YBCO, one underdoped and one near optimal doping.  In each, they saw a relatively sharp anomalous feature at $T^*$ similar to that identified in \cite{fauq06}, which they proposed was evidence of a phase transition associated with whatever type of order might be associated with the pseudogap.  This interpretation was questioned by Cooper {\it et al.} \cite{coop14} on the basis of the absence of any corresponding feature in specific heat data.  They suggested that a more plausible cause of the observed features might be resonant relaxation effects associated the oxygen in the Cu-O chain layer.

There are more examples of small surprising effects that could be cited, but they would not provide much enlightenment.  We already have big experimental effects that are a challenge for theorists to calculate.  I have presented a plausible interpretation of the pseudogap effects and the nature of the superconductivity.  Each new surprise is worthy of contemplation, but it should not require that we restart our analysis from scratch.

}

\section{Conclusion}
\label{sc:conc}

From the beginning, there have been attempts by theorists to extrapolate phenomena realized in 1D to 2D.  Anderson's proposal of a 2D quantum spin liquid, in the form of an RVB state \cite{ande87}, was made in the light of the known quantum-spin-liquid state of the 1D Heisenberg spin-$\frac12$ AF.   When AF order in 2D at $T=0$ was confirmed by theory, considerable effort was devoted to models with frustration, such as second-neighbor AF exchange, to get a spin liquid or bond order prior to hole doping \cite{vojt99}.  

A key feature of a quantum spin liquid is the absence of a spin gap.  Nevertheless, early mean-field analyses associated with the RVB model recognized that a spin gap (taken to be spatially uniform) could be associated with pairing correlations \cite{lee92}.  The spin gap was assumed to be large at low doping and to decrease linearly toward zero at large $p$.  An alternative way to get a spin gap is to consider even-leg spin ladders; however, as one extrapolates to 2D, the gap goes to zero \cite{chak96}.  

\begin{figure}
 \centering
    \includegraphics[width=5cm]{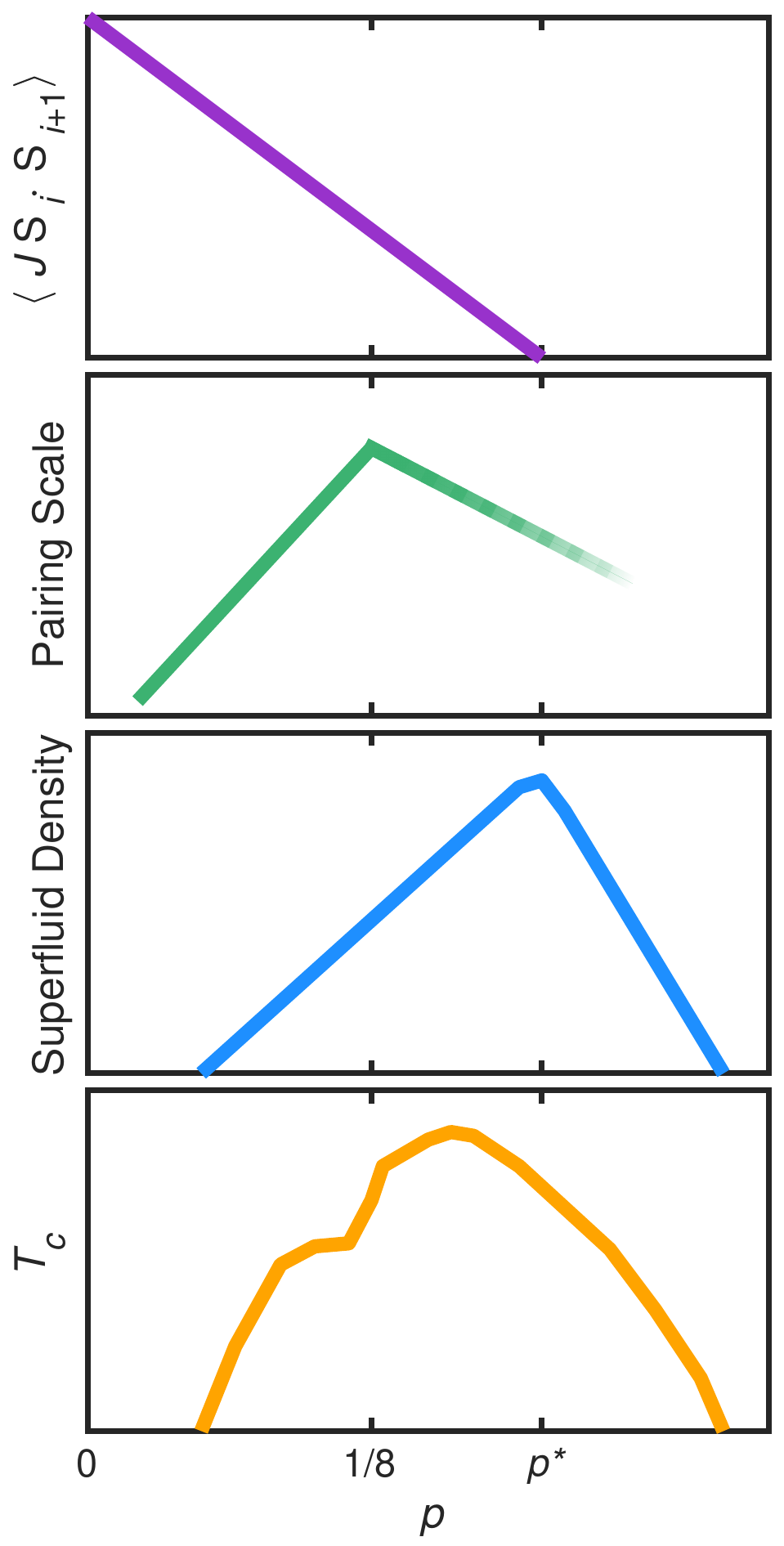}
    \caption{\label{fg:summ}  Schematic figure of hole-density dependence of (top to bottom) characteristic energy of local AF correlations, pairing scale, superfluid density, and $T_c$.}
\end{figure}

Experimentally, it is true that the average strength of local AF correlations decreases with $p$, eventually reaching a percolation transition at $p^*$, as indicated schematically in the top panel of Fig.~\ref{fg:summ}.  However, the singlet gap that sets the pairing scale is absent in the undoped system; it develops with doping due to the development of intertwined orders as a consequence of competition between the hole kinetic energy with the superexchange interactions that drive the AF correlations \footnote{There is some connection here with the ``spin-bag'' model \cite{schr88}, which assumed the frustrated interactions of doped holes and AF spins would lead to inhomogeneity and pairing.}  I have identified this gap with the energy $E_{\rm cross}$ corresponding to the neck of the typical hourglass spin excitation spectrum found in neutron scattering studies.  

It is the doping that causes the planes to segregate into quasi-1D stripes.  Emery, Kivelson, and Zachar \cite{emer97} were the first to propose that stripes could be the source of pairing and superconductivity.  I have proposed a slight variation on this, arguing that the charge stripes should be viewed as doped two-leg ladders (which are known to develop strong pairing \cite{dago96}); they are isolated by that antiphase order of the neighboring spin stripes.  Of course, those spin correlations obstruct spatially-uniform $d$-wave superconductivity, and they must be gapped to achieve it.  The pairing scale on the ladders, as measured by $E_{\rm cross}$, reaches its maximum at $p\approx\frac18$, where stripe order is optimized.  It decays gradually at higher doping and gets diluted as the regions that support stripe correlations are gradually replaced by uniformly-doped domains.  

The connection between pairing within charge stripes and spatially-uniform superconductivity is clearest where the stripe fluctuations are slow compared to the pairing scale, as in optimally-doped LSCO.  Obviously, the picture becomes fuzzier as the incommensurate spin gap approaches $E_{\rm cross}$.  Nevertheless, the doping dependence of $E_{\rm cross}$ is similar in all cuprates studied so far, and the self-organization of doped holes and local AF correlations provides a consistent interpretation.

Besides the stripe order/correlations, we also have variations in average charge density over a larger length scale due to dopant disorder and poor screening.  Because of this granular character, the development of superconducting phase order is dependent on the superfluid density \cite{emer95a,imry12}.  It follows that $T_c$ is determined by a balance between the pairing scale and the magnitude of the superfluid density.  As indicated in Fig.~\ref{fg:summ}, $T_c$ is maximized midway between the peaks in the pairing scale and the superfluid density.

Applications of advanced algorithms to simplified model Hamiltonians provide results supportive of pieces of my story, but they also demonstrate why reaching a conclusion has been challenging.  There are many competing states with similar energies, and establishing evidence for pairing order is a numerical challenge.  Rapid progress is being made, and it will be interesting to see to what extent future results may support and justify the story I have presented here.

{\newr 
Let me end with some speculations.\footnote{\newr I have been encouraged to make some predictions based on the picture painted here.  I am uncomfortable with the ``p'' word, but am happy to speculate, which is more in line with the title metaphor of a striped lens.  Of course, one scientist's speculation may be another's prediction.}  In particular, I will start with a recent speculation that has been at least partially confirmed (with pieces of the story already presented above).  In 2017, I proposed (in a talk at KITP, UC Santa Barbara) that spectroscopic-imaging STM studies had (inadvertently) measured the PDW superconducting gap in a study of Bi2212 \cite{kohs08}.  The underlying argument is the following.  In order to measure the shape of the superconducting gap in reciprocal space, STM must measure the interference pattern of Bogoliubov quasiparticles induced by defects, such as Zn.  Zn is known to enhance spin stripe order and depress uniform superfluid density.  If Zn locally induces PDW order, then it would explain the fact that the gap structure determined by STM in underdoped Bi2212 appears to have a gapless arc.

\begin{figure}
 \centering
    \includegraphics[width=0.55\textwidth]{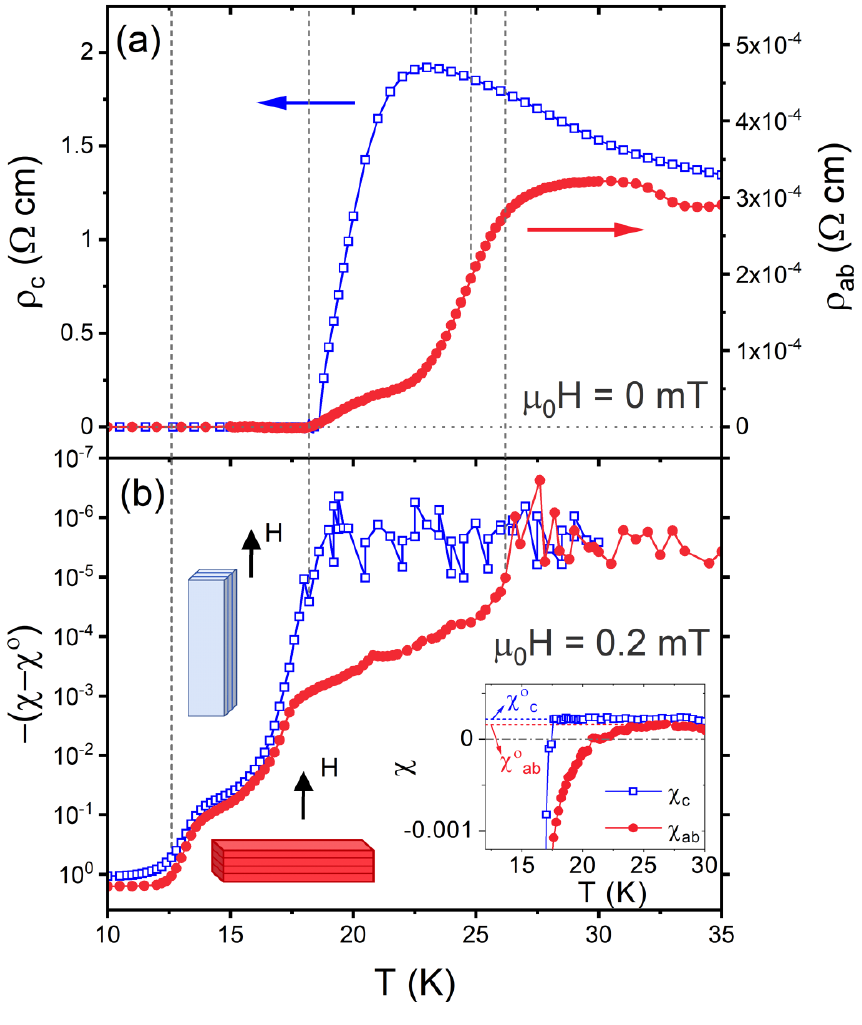}
    \caption{\label{fg:Zn} {\newr (a) $T$ dependence (in zero field) of the resistivity for current along the $c$ axis (blue squares) and parallel to the $ab$ planes (red circles) for 1\%\ Zn in LBCO $x=0.095$ \cite{loza21}. (b) $T$ dependence of the volume magnetic susceptibility with a field of $\mu_0H = 0.2$~mT applied parallel to the planes, $\chi_c$ (blue squares), and perpendicular to the planes, $\chi_{ab}$ (red circles). Inset shows the susceptibilities on a linear scale near the initial transitions; linear fits to the normal state, and extrapolated to low $T$ (indicated by dashed lines), were subtracted to enable the logarithmic-scale plots of the main panel. Vertical dashed lines indicate transition features in the data.   Reprinted with permission from \cite{loza21}, \copyright2021 by the American Physical Society.} }
\end{figure}

In locally depressing superconductivity, a Zn impurity is similar to a magnetic vortex \cite{nach96}, and the best evidence for PDW order comes from STM measurements of the vortex halo region \cite{edki19}.  The latter is consistent with the observation that application of a $c$-axis magnetic field induces 2D superconductivity, along with enhancing stripe order, in LBCO $x=0.095$ \cite{wen12b,steg13}, presumably due to favoring PDW order over spatially-uniform superconductivity.  So Steven Kivelson asked me whether we could close the loop on this: do Zn impurities induce 2D superconductivity as they depress bulk superconductivity?  My colleagues, Qiang Li and Pedro Lozano, did the anisotropic resistivity and susceptibility measurements shown in Fig.~\ref{fg:Zn}, finding strong evidence that 1\%\ Zn in LBCO $x=0.095$ results in 2D superconductivity, before transitioning to 3D superconductivity at lower temperature \cite{loza21}.  That provides support for the original speculation that the STM study on Bi2212 \cite{kohs08} selectively detected a gap associated with local PDW order.

The case of PDW order induced locally about a vortex in slightly-underdoped Bi2212 \cite{edki19} raises an interesting issue.  Such a sample has a large spin gap in zero field, and there is no evidence that the applied field is sufficient to induce any spin order.  This appears to conflict with my claim that spin stripe order is necessary to establish PDW phase coherence in LBCO $x=1/8$.  I speculate that a defect such as a vortex core, where the superconducting order parameter must go to zero, causes PDW superconducting order to be energetically favorable in its vicinity because the PDW superconducting wave function already accommodates zero-amplitude lines \cite{tran21}.\footnote{\newr P. A. Lee \cite{lee04} made the argument that, because the energetic cost of creating a vortex is empirically small, there must be some sort of competing order induced locally with a vortex.}  This idea should be tested through numerical modeling of the Hubbard or $t$-$J$ model.  A challenge is that, as discussed above, such calculations are typically performed on atomic clusters with anisotropic boundary conditions.  A new opportunity is provided by Jiang and Kivelson \cite{jian21}, who found that frustrating the nearest-neighbor superexchange with a next-nearest-neighbor coupling results in strong spatially-uniform superconducting correlations.  How would this be modified by adding a site where holes are excluded?

In terms of interesting new measurements, the field-induced CDW in underdoped YBCO is very intriguing.   For $p\approx 0.12$, it becomes readily apparent for a $c$-axis magnetic field above 15~T \cite{choi20,jang18}, below the 20-T regime where reentrant 2D superconductivity (coexisting with stripe order) has been detected in LBCO $x=1/8$ \cite{li19a}.  I am convinced that the field-induced CDW must involve modulated pair correlations.  Whether these have superconducting coherence close to the onset temperature is an interesting question, but I would bet that the superconductivity at high field and low temperature \cite{yu16,hsu21a,hsu21b} has PDW character.

I have speculated in Sec.~\ref{sc:sc_stripes} that the high-resistivity ultra-quantum-metal phase of LBCO $x=1/8$ found at low temperature and high magnetic field could be a Bose metal.  A pre-condition for such a state is that the charge carriers be bosonic pairs.  This possibility can be tested by measurements of quantum shot noise, as has been applied to identify pairing in the normal state of LSCO thin films \cite{zhou19}.  Such measurements can certainly be performed at high magnetic field, as demonstrated by the original measurement of fractional carriers associated with the fractional quantum Hall effect \cite{depi98}.  The UQM phase in LBCO extends continuously to lower fields at higher temperatures (see Fig.~\ref{fg:HF}), so it could be probed at 10~K with a field of 10~T.

An intriguing superconductor that provides a potential challenge to this perspective is Ba$_2$CuO$_{3+\delta}$, a superconductor with $T_c=73$~K that is obtained by forcing O ions into a parent compound consisting of Cu-O chains using high pressure \cite{li19b}.  (Note that the related chain compound Sr$_2$CuO$_3$ has an extremely large superexchange energy of $J\approx240$~meV \cite{walt09}.)   Because of the challenging synthesis conditions, only small polycrystalline samples have been produced, which has made it difficult to uniquely determine the crystal structure of the superconducting phase.  I believe that the underlying character of the pairing mechanism in this compound should be similar to that of the planar cuprates.  While there has been a lot of theoretical speculation on this material, we need to know the actual crystal structure and the nature of magnetic correlations before meaningful analysis is possible.

Finally, Scalapino  \cite{scal12a} has emphasized that spin fluctuations provide a ``common thread'' between cuprates and iron-based superconductors.  The latter have a more complicated starting point, with multiple Fe $3d$ orbitals playing a significant role, along with Coulomb and Hund's interactions among those orbitals \cite{mazi09,basc16,si16}.  A similarity with cuprates is that the magnetic spectral weight tends to be large at high energy and reduced at low energy, with a spin gap developing in the superconducting state \cite{dai12,dai15,tran20}.  Perhaps, as in cuprates, the superconductivity is driven by repulsive interactions associated with the magnetism.  In that case, the charge carriers should pair up in a fashion that minimizes disruption of the magnetic correlations.  This is a possibility that deserves future investigation.

}

\section*{Acknowledgments}

I have benefited from discussions with and comments by I. Bozovi\v{c}, E. Dagotto, M. P. M. Dean, E. Fradkin, K. Fujita, C. C. Homes, M. H\"ucker, P. D. Johnson, S. A. Kivelson,  P. A. Lee, Q. Li, Y. Li, H. Miao, N. J. Robinson, D. J. Scalapino, Senthil, A. Tsvelik, T. Valla, I. Zaliznyak, and many others.  I am especially grateful to T. Egami for challenging me to tell an interesting story.  This work was supported at Brookhaven National Lab by the U.S. Department of Energy, Office of Science, Basic Energy Sciences, Materials Sciences and Engineering Division under Contract No.\ DE-SC0012704.

\bibliographystyle{tADP}
\bibliography{LNO,theory,fe_sc}

\end{document}